\documentclass[11pt,a4paper]{article}
\usepackage[T1]{fontenc}
\usepackage{lmodern}
\usepackage[margin=24mm]{geometry}
\usepackage{microtype}
\usepackage{amsmath,amssymb,amsthm}
\usepackage{graphicx,pgfplots}
\pgfplotsset{compat=1.16}
\usepackage{booktabs,array,needspace}
\usepackage[font=small,labelfont=bf]{caption}
\usepackage[ruled,linesnumbered]{algorithm2e}
\SetAlFnt{\small}
\SetAlCapFnt{\small}
\SetAlCapNameFnt{\small}
\usepackage[round,authoryear]{natbib}
\usepackage{aliascnt}
\usepackage[unicode,hidelinks]{hyperref}
\usepackage[capitalise,noabbrev]{cleveref}

\newcommand{\fraction}[2]{\frac{#1}{#2}}
\theoremstyle{plain}
\newtheorem{theorem}{Theorem}
\newtheorem{proposition}{Proposition}[section]
\newaliascnt{lemma}{proposition}
\newtheorem{lemma}[lemma]{Lemma}
\aliascntresetthe{lemma}
\newtheorem{corollary}{Corollary}
\theoremstyle{definition}
\newaliascnt{definition}{proposition}
\newtheorem{definition}[definition]{Definition}
\aliascntresetthe{definition}
\newaliascnt{remark}{proposition}
\newtheorem{remark}[remark]{Remark}
\aliascntresetthe{remark}
\newaliascnt{example}{proposition}
\newtheorem{example}[example]{Example}
\aliascntresetthe{example}
\crefname{lemma}{Lemma}{Lemmas}
\crefname{definition}{Definition}{Definitions}
\crefname{remark}{Remark}{Remarks}
\crefname{example}{Example}{Examples}
\newtheorem{assumption}{Assumption}
\crefname{assumption}{Assumption}{Assumptions}

\newcommand{\R}{\mathbb{R}}
\newcommand{\bstar}{\beta^*}
\newcommand{\norm}[1]{\lVert #1\rVert_{\infty}}
\newcommand{\cE}{\mathcal{E}}
\newcommand{\cK}{\mathcal{K}}
\newcommand{\cB}{\mathcal{B}}
\newcommand{\cA}{\mathcal{A}}

\newcommand{\e}{\mathrm{e}}
\newcommand{\condref}[1]{\hyperref[cond:#1]{\textup{(#1)}}}
\newcommand{\sub}{\overline\beta}

\newcommand{\super}{\underline\beta}

\title{Last-Iterate Convergence of Payoff-Gradient Learning\\
in M\"obius Pricing Games\thanks{This project has received funding from the European Research Council (ERC) under the European Union's Horizon Europe research and innovation programme (grant agreement No 101198689).}}
\author{Martin Bichler\thanks{Corresponding author: \href{mailto:m.bichler@tum.de}{m.bichler@tum.de}. ORCID: \href{https://orcid.org/0000-0001-5491-2935}{0000-0001-5491-2935}.} \qquad Philip Hierhager \qquad Jan-Sebastian H\"ohener\\[0.55em]
  \small Decision Sciences \& Systems, Department of Computer Science\\
  \small School of Computation, Information and Technology\\
  \small Technical University of Munich, Germany}
\date{}
\hypersetup{
  pdftitle={Last-Iterate Convergence of Payoff-Gradient Learning in M\"obius Pricing Games},
  pdfauthor={Martin Bichler, Philip Hierhager, Jan-Sebastian H\"ohener}
}

\begin{document}
\maketitle

\begin{abstract}
We prove global last-iterate convergence of payoff-gradient learning in every typewise Riemannian metric, which for scalar actions is the same as typewise mirror descent, with explicit rates, in a class of Bayesian bidding and pricing games that includes first-price auctions, unit-demand pay-as-bid auctions, rank-based generalized first-price auctions, and homogeneous-good price competition with inelastic demand. The result holds for symmetric self-play: all players use a common strategy and apply the same payoff-gradient update at each type. Convergence is from every continuous weakly increasing feasible initial strategy, including pooling, and does not rely on the monotonicity or Minty conditions that underlie existing guarantees in other games. Instead, the class is characterized by two structural properties: a direct payoff effect that is independent of the action, and a payoff difference at the allocation threshold that is affine in the action. These properties reduce the learning dynamics to a singular transport–damping equation for the common strategy, which, under the stated coefficient and metric assumptions, admits a unique viscosity solution. Comparison barriers make the uniform distance to the unique rest point a strict Lyapunov functional and yield convergence rates with optimal algebraic exponents. In allocation games, the rest point is the unique weakly increasing symmetric equilibrium under the stated tie-breaking conditions, and a positive lower slope bound converts strategy-error bounds into explicit approximate-equilibrium guarantees. A projected grid scheme with a slope floor makes the approximation constructive, with error bounds under deterministic perturbations and almost-sure convergence on a fixed grid under martingale gradient noise.
\end{abstract}

\begingroup
\small
\makeatletter
\renewcommand*{\l@section}{\@dottedtocline{1}{0pt}{1.8em}}
\makeatother
\tableofcontents
\endgroup
\clearpage

% MOR: DE: Mertikopoulos - Perchet, Candogan, Zhou (NYU)
% OR: Candogan, Zhengyuan Zhou, Yonatan Gur
% TE: Takahashi, Peski, Krishna
% ECTA: Co-Editor - Siniscalchi; Annie Liang, Vasilis Syrgkanis, Zapeckelnyuk, Ryota Iijima, 
%=====================================================================
\section{Introduction}\label{sec:intro}
%=====================================================================

First-price auctions, procurement, and price competition are central models
of price formation and resource allocation under private information. They
share one strategic trade-off: a more aggressive bid or price raises the
probability of receiving an allocation but lowers the surplus conditional on
receiving it. Whether learning reaches equilibrium in these environments
matters both for equilibrium computation and for the behavioral foundations
of predictions about prices, revenues, and allocations.

Numerical studies show that learning algorithms approximate Bayes--Nash
equilibria with remarkable accuracy in a range of auction environments with
continuous types and actions, including settings in which analytical
equilibrium characterizations are difficult to obtain
\citep{bichler2021learning,bichler2025computing}. The theoretical explanation
remains incomplete. The classical sufficient conditions for convergence of
gradient methods, monotonicity of the game \citep{rosen1965existence} and
Minty-type variational conditions
\citep{minty1962monotone,mertikopoulos2019learning,cai2022finite}, fail
already in the first-price auction \citep{bichler2025beyond}. Their failure
does not preclude convergence, but it leaves a gap between the empirical
success of learning algorithms and the available guarantees.

We close this gap, under symmetric self-play, for a class of Bayesian bidding
and pricing games that we call \emph{M\"obius pricing games}. The class
contains first-price auctions, unit-demand pay-as-bid auctions, scalar-bid
generalized first-price auctions, and homogeneous-good Bertrand competition
with inelastic demand. All players use one common strategy, a map from types
to actions, and the action of each type moves in the direction of its
marginal interim payoff. We show that this payoff-gradient dynamics converges
uniformly to its unique rest point from every continuous, weakly increasing
initial strategy within natural bounds, including strategies that pool types.
In allocation games the rest point is the symmetric Bayes--Nash equilibrium.
The guarantee is a last-iterate guarantee: it concerns the current strategy
itself, not its time average or empirical distribution.

% EC27 revision: a motivating example, without changing result numbering.
\paragraph{A motivating example: the first-price auction.}
Consider $n\ge2$ risk-neutral bidders with independent private values drawn
from a common distribution $F_V$ on $[0,1]$ with continuous positive density
$f_V$. Suppose that the rivals use a continuous strictly increasing bid
function $\alpha$ with positive derivative in the interior. A bidder with
value $v$ who submits an interior bid $b$ has interim payoff
\[
U(b;\alpha,v)=(v-b)\,F_V\bigl(\alpha^{-1}(b)\bigr)^{n-1}.
\]
At the current bid $b=\alpha(v)$, the derivative of this payoff in $b$ is
\[
D[\alpha](v)
=-F_V(v)^{n-1}
+(n-1)f_V(v)F_V(v)^{n-2}\,
 \frac{v-\alpha(v)}{\fraction{\mathrm{d} \alpha}{\mathrm{d} v}(v)}.
\]
The first term is the additional payment conditional on winning. The second
is the gain in winning probability times the surplus at the margin: a small
increase in the bid raises the highest rival value that the bidder can beat,
by an amount inversely proportional to the slope of the rivals' strategy.

Now let the common strategy evolve through learning. Write $\alpha(t,v)$ for
the bid of a bidder with value $v$ at learning time $t$; the value is fixed,
and $t$ tracks changes in the strategy. For two bidders with uniform values,
a gradient update with step size $h>0$ reads
\[
\alpha^{k+1}(v)
=\alpha^k(v)
+h\left(-v+\frac{v-\alpha^k(v)}{\fraction{\mathrm{d} \alpha^k}{\mathrm{d} v}(v)}\right),
\]
where $k$ indexes learning rounds. Writing $t=kh$ and formally letting
$h\to0$ gives the continuous-time payoff-gradient dynamics
\[
\fraction{\partial \alpha}{\partial t}(t,v)
=-v+\frac{v-\alpha(t,v)}{\fraction{\partial \alpha}{\partial v}(t,v)}.
\]
The symmetric equilibrium $\alpha^*(v)=v/2$ makes the right-hand side zero:
at equilibrium, no type changes its bid.

To study convergence, we track the deviation of each type's bid from its
equilibrium bid, $y(t,v)=\alpha(t,v)-v/2$. Since
$\fraction{\partial y}{\partial t}=\fraction{\partial \alpha}{\partial t}$ and
$\fraction{\partial \alpha}{\partial v}=1/2+\fraction{\partial y}{\partial v}$, the learning equation
becomes
\[
\fraction{\partial y}{\partial t}
+\frac{v}{\fraction{\partial \alpha}{\partial v}}\,
 \fraction{\partial y}{\partial v}
=-\frac{y}{\fraction{\partial \alpha}{\partial v}}.
\]
This equation separates two effects. The term in $\fraction{\partial y}{\partial v}$ is a
\emph{transport} term: it moves the error profile across types, here toward
higher values. The right-hand side is a \emph{damping} term: it pulls the
error toward zero whenever the strategy has positive slope. At a smooth
interior maximum or minimum of the error, $\fraction{\partial y}{\partial v}=0$, so only
damping remains and the largest error shrinks.
% TODO: at such a point d\alpha/dv = 1/2, so dy/dt = -2y, whereas
% \cref{cor:auction} gives e^{-f_- t} = e^{-t} for the uniform case. Say in
% the corollary (or here) where the constant comes from and whether the
% exponential rate for n=2 is tight.

This observation is the core of our convergence proof. Making it rigorous
requires treating boundary types, nonsmooth strategies, and pooling, where
the slope vanishes and the equation is singular.

\paragraph{M\"obius pricing games.}
The same transport--damping structure appears in procurement, multi-unit
allocation, and Bertrand competition. To treat these settings jointly we use
procurement coordinates, in which types are costs $c\in[0,1]$, strategies
$\beta$ map types to scalar actions, and lower actions are favored. The
selling auction above is translated by
\[
c=1-v,\qquad \beta(c)=1-\alpha(1-c),
\]
so that high values correspond to low costs, and transport toward higher
values becomes transport toward lower costs.
In these coordinates each game is a pricing problem: a player with cost $c$
quotes a price, lower prices win, and the anchor $\mu(c)$ below is the
break-even price. We keep the neutral terms \emph{type} and \emph{action},
since in the selling auction the action is $1-\alpha$.

A \emph{M\"obius pricing game} is a symmetric game in which each type
$c\in[0,1]$ chooses a scalar action, interpreted as a bid or a price, and
for which there are functions $A$, $B$, and $\mu$ on $[0,1]$ such that,
whenever the rivals use a continuous strictly increasing $\beta$ with
$\fraction{\mathrm{d} \beta}{\mathrm{d} c}(c)>0$ for $0<c<1$, the derivative of the payoff of type $c$ at
its action $\beta(c)$ is
\begin{equation}\label{eq:intro-field}
G[\beta](c)=A(c)-B(c)\,\frac{\beta(c)-\mu(c)}{\fraction{\mathrm{d} \beta}{\mathrm{d} c}(c)} .
\end{equation}
We call $G$ the \emph{payoff field}. The coefficient $A$ is the direct
marginal effect of the action on the payoff. The second term is the effect
through the allocation threshold: $B$ collects the threshold density and any
payment scaling, and $\mu(c)$, the \emph{anchor}, is the action at which type
$c$ is indifferent about crossing the threshold. For the first-price auction,
\[
A(c)=F_V(1-c)^{n-1},\qquad
B(c)=(n-1)f_V(1-c)F_V(1-c)^{n-2},\qquad
\mu(c)=c,
\]
and $G[\beta](c)=-D[\alpha](1-c)$; the sign reflects that a higher action
$\beta$ is a lower bid $\alpha$. \Cref{prop:membership} characterizes
membership through the payoff primitives: the direct effect must not depend
on the action, and the density-weighted payoff difference across the
threshold must be affine in the action.

The name refers to the slope: for fixed $c$ and $z=\beta(c)$, the field is
the M\"obius transformation
\[
u\longmapsto\frac{A(c)\,u-B(c)\bigl(z-\mu(c)\bigr)}{u}
\]
of the slope $u=\fraction{\mathrm{d} \beta}{\mathrm{d} c}(c)$ \citep[Section~II.7]{gamelin2001complex},\footnote{Unrelated
to the M\"obius transform of a cooperative game.} degenerate when $B(c)(z-\mu(c))=0$". Any threshold game is M\"obius in the slope; the class is defined by
coefficients in which $u$ enters with an action-independent factor and the
constant term is affine in the action (\cref{prop:membership}).

The class includes first-price auctions, unit-demand pay-as-bid auctions for
multiple units, scalar-bid generalized first-price auctions with
rank-dependent allocation weights, and homogeneous-good Bertrand competition
with inelastic demand. Proportional fees and cost sharing change the margin
but not the structure, and the interdependent- and affiliated-cost
specifications in \cref{tab:members} extend the information structure. One
structural test therefore covers different allocation rules, transfers, and
information assumptions. Elastic demand, risk aversion, and all-pay contests,
in the specifications of \cref{prop:nonmembers}, fall outside the class.

\paragraph{Results.}
Three notions are needed to state the results. An action is \emph{feasible}
for type $c$ if it lies between the anchor $\mu(c)$ and a common cap $C$. A
\emph{feasible strategy} is continuous, weakly increasing, and takes feasible
values; pooling is allowed. In the first-price auction, $\mu(c)=c$ and $C=1$,
so feasibility reads $c\le\beta(c)\le1$, that is, $0\le\alpha(v)\le v$: bids
are nonnegative and no type bids above its value. In procurement, the cap is
a reserve price at the highest cost. A \emph{rest point} is a strategy
$\bstar$ with $G[\bstar]=0$. Finally, a learner may measure changes of
actions in any \emph{typewise Riemannian metric}, which weights the change of
each type's action by a positive function of the type and the action. Because
actions are scalar, every such metric is the Hessian of a typewise mirror map,
so Riemannian gradient play and mirror descent
\citep{nemirovski1983problem,beck2003mirror} give the same continuous-time
dynamics.

Our main result holds for every M\"obius pricing game whose coefficients are
continuous and positive below the top type and whose anchor is weakly
increasing, and for every typewise Riemannian
metric that satisfies two regularity conditions in the action. It has
three parts.
\begin{enumerate}
\item \emph{Convergence.} From every feasible initial strategy the dynamics
has exactly one solution. The uniform distance of this solution to the unique
rest point is a strict Lyapunov functional: it decreases strictly, at an
explicit rate determined by the metric and the coefficients, and the
resulting power rates are optimal (\cref{thm:main}).
% TODO: state the solution concept here in one clause (viscosity solution?),
% since \cref{prop:blowup} rules out global classical solutions.
\item \emph{Equilibrium.} In every allocation game, in which the payoff is
the margin times a winning probability, the rest point is the unique weakly
increasing symmetric equilibrium. If $\bstar$ has slope bounded below, the
learned strategy is an approximate equilibrium whose gap is at most
proportional to its uniform distance to $\bstar$. More generally, in every
M\"obius pricing game with continuous payoffs and a strictly increasing rest
point, learning reaches approximate equilibria if and only if the rest point
is an equilibrium (\cref{thm:selection}).
\item \emph{Computation.} A projected grid algorithm attains these rates up
to explicit error terms, uniformly in the grid, under bounded deterministic
errors. With stochastic gradient estimates on a fixed grid it converges
almost surely to its grid rest point. Its iterates converge to the continuum
solution as the discretization parameters and the error bound vanish
(\cref{thm:grid,thm:scheme,thm:stochastic}). The scheme accepts every initial
action vector; a nonmonotone initial profile is interpreted through the lower
monotone envelope that the scheme selects.
\end{enumerate}
In first-price auctions with $n$ bidders whose continuous value density lies
in $[f_-,f_+]$ with $f_->0$, all constants are explicit: in the time $t$ of
the dynamics, the uniform distance and the equilibrium gap decay like
$\e^{-f_-t}$ for $n=2$ and like $t^{-1/(n-2)}$ for $n\ge3$
(\cref{cor:auction,cor:rounds}).
% Every numbered result of this paper is
% formally verified in Lean~4 (\cref{app:lean}).
% TODO: add one clause on what the Lean development takes as axioms, if
% anything; referees will ask.

\paragraph{Method: comparison instead of monotonicity.}
The state of the dynamics is a strategy, a function of the type, and the
velocity of each type depends on the slope of this function. The dynamics is
therefore a nonlinear partial differential equation that is singular wherever
the strategy pools types, and classical solutions need not exist globally:
\cref{prop:blowup} gives smooth initial data whose solution becomes spatially
nondifferentiable at a positive time.
The standard route to last-iterate convergence is not available either. The
$L^2$ distance to the rest point, the natural Lyapunov candidate of
variational-inequality methods, can increase along the flow
(\cref{prop:minty}).

We replace monotonicity by comparison. A rest point satisfies
$A\,\fraction{\mathrm{d} \bstar}{\mathrm{d} c}=B(\bstar-\mu)$, so every strategy $\beta=\bstar+y$ satisfies
\begin{equation}\label{eq:intro-identity}
G[\beta]=\frac{A\,\fraction{\partial y}{\partial c}-B\,y}{\fraction{\partial \beta}{\partial c}} .
\end{equation}
This is the transport--damping representation of the example in general form:
the error $y$ is transported toward lower types and damped, and both effects
are scaled by the positive factor $1/(\fraction{\partial \beta}{\partial c})$. The coefficients depend on
the evolving strategy, so the dynamics remains nonlinear. A comparison
principle for the singular equation makes the clipped shifts
$\max\{\mu,\bstar-e(t)\}$ and $\min\{C,\bstar+e(t)\}$ barriers for suitable
strictly decreasing scalar envelopes $e$. Initializing the envelope at the
current uniform error and restarting the comparison at any time yields the
strict Lyapunov property of the uniform distance to $\bstar$. Existence comes
from the algorithm: under a step-size condition each update is monotone in
the actions, and the limits of the iterates are identified by the comparison
principle, as in the method of \citet{barles1991convergence}.

\paragraph{Scope: symmetric self-play.}
Throughout, all players use one common strategy, updated with one estimate
per type and round, as in the symmetry-preserving equilibrium-learning
procedures of \citet{bichler2021learning,bichler2025computing}. This mirrors
the analytical literature, which studies symmetric equilibria of symmetric
models
\citep{vickrey1961counterspeculation,riley1981optimal,krishna2010auction};
in first-price auctions the symmetric equilibrium is the unique equilibrium
\citep{maskin2003uniqueness,lebrun2006uniqueness}, so the restriction loses
no equilibria. The restriction is, however, dynamic as well as static:
symmetric primitives alone do not preserve a common strategy when players use
different initializations, learning rates, or independent estimation noise,
and the analysis of asymmetric strategies requires coupled equations
\citep{marshall1994numerical,fibich2012asymmetric,hubbard2014numerical}.
Whether the symmetric dynamics is stable under asymmetric perturbations, that
is, under independent learning, is a separate question that we leave open.

\paragraph{Related work.} Equilibria of auctions have been computed by ODE methods
\citep{marshall1994numerical,fibich2012asymmetric,hubbard2014numerical}, by best responses on restricted strategy
spaces \citep{reeves2004computing,armantier2008approximation,rabinovich2013computing}, by fixed-point methods
\citep{hoang2019return}, and by learning: gradient ascent with neural strategies
\citep{bichler2021learning,bichler2023learning,heidekruger2021equilibrium}, dual averaging on distributional strategies
\citep{bichler2025computing}, and derivative-free methods \citep{martin2023finding,martin2025joint}; see the surveys of
\citet{bichler2025differentiable,bichler2025potentialness}. These methods parametrize strategies by finitely many numbers
or by distributions. We analyze a pure-strategy dynamics in function space. Neural parameterizations
generally couple updates across types, and distributional methods evolve different state variables;
transferring the present guarantees to those procedures requires a separate argument.

\Citet{bichler2025beyond,bichler2026stability} show that monotonicity and the Minty condition fail in first-price
auctions and in Bertrand competition. Many last-iterate guarantees in games rely on monotonicity or Minty-type
conditions \citep{daskalakis2019last,golowich2020tight,diakonikolas2021efficient,pethick2022escaping,bohm2023solving,gorbunov2023convergence},
and without them gradient dynamics may cycle
\citep{mertikopoulos2018cycles,mazumdar2020gradient,milionis2023impossibility,heymann2026tiebreaking}; related results concern
complete information or discrete bids
\citep{ahunbay2025semicoarse,deng2022nash,ahunbay2026gradient,ipsen2026revenue,feng2021convergence,ahunbay2025uniqueness,bichler2023bargaining,banchio2022artificial}.
We use order and comparison to obtain convergence without those conditions. Order-based methods have
a substantial history in games with strategic complementarities and in monotone dynamical systems
\citep{milgrom1990rationalizability,vives1990nash,hirsch1988stability,smith1995monotone,athey2001single};
the broader theory also includes infinite-dimensional settings. Our analysis uses viscosity
solutions \citep{crandall1983viscosity,crandall1992users} and monotone schemes
\citep{barles1991convergence}. Uniqueness of equilibrium in first-price auctions is classical
\citep{maskin2003uniqueness,lebrun2006uniqueness}.

The contribution is the convergence analysis enabled by the particular singular payoff field of this
class of Bayesian games. The membership characterization identifies its economic primitives;
comparison and shrinking barriers establish well-posedness and global last-iterate convergence even
from pooling, with explicit dissipation and matching algebraic exponents. Additional payoff conditions
identify the limit as an equilibrium and bound profitable deviations, while the monotone grid scheme
provides a constructive approximation and error guarantees. Thus the novelty is the combination of
these results for the specified learning dynamics, rather than comparison or viscosity methods
themselves.

%=====================================================================
\section{M\"obius pricing games}\label{sec:model}
%=====================================================================

This section defines the games, states the assumptions, introduces the learning dynamics, and shows why
the classical tools fail.

\subsection{The payoff field}\label{sec:field}

Each of $n\ge2$ players has a privately known type $c\in[0,1]$ and chooses a scalar action $z\in\mathbb{R}$. A strategy is a function $\beta:[0,1]\to\mathbb{R}$. We call a strategy \emph{regular} if it is continuous and strictly increasing on $[0,1]$, differentiable on $(0,1)$, and satisfies $\fraction{\mathrm{d} \beta}{\mathrm{d} c}>0$ on $(0,1)$. All players use the same strategy and apply a common update at each type; this is the symmetric
self-play restriction used throughout.

A pricing game specifies the interim payoff $\pi(z;\beta,c)$ of type $c$ from action $z$ when the rivals use strategy $\beta$.

Let $S:[0,1]^{n-1}\to[0,1]$ be a statistic of the rival types that determines the payoff regime. We write $s=S(c_{-i})$ for its realization and let $g(s\mid c)$ be its density conditional on the player's type $c$. We orient $S$ so that the losing payoff $V(z,s;c)$ applies when $s<\beta^{-1}(z)$ and the winning payoff $W(z,s;c)$ applies when $s>\beta^{-1}(z)$.

To derive the payoff field, suppose that $\beta$ is regular.
The inverse $\beta^{-1}$ is defined on $[\beta(0),\beta(1)]$. We extend it to $\mathbb{R}$ by setting $\beta^{-1}(z)=0$ for $z\le\beta(0)$ and $\beta^{-1}(z)=1$ for $z\ge\beta(1)$. At $s=\beta^{-1}(z)$, the rival statistic is exactly at the threshold, corresponding to a tie in the auction and price-competition examples. Since $S$ has a density, this event has probability zero.
Hence
\begin{equation}\label{eq:payoff}
\pi(z;\beta,c)
=
\int_0^{\beta^{-1}(z)} V(z,s;c)g(s\mid c)\,ds
+
\int_{\beta^{-1}(z)}^1 W(z,s;c)g(s\mid c)\,ds .
\end{equation}

Differentiating \eqref{eq:payoff} at $z=\beta(c)$, where $\beta^{-1}(\beta(c))=c$ and $\fraction{\mathrm{d} \beta^{-1}}{\mathrm{d} z}=\frac{1}{\fraction{\mathrm{d} \beta}{\mathrm{d} c}(c)}$, gives
\begin{align}
\fraction{\partial \pi}{\partial z}(z;\beta,c)\Big|_{z=\beta(c)}
={}&
\int_0^c \fraction{\partial V}{\partial z}(\beta(c),s;c)g(s\mid c)\,ds
+
\int_c^1 \fraction{\partial W}{\partial z}(\beta(c),s;c)g(s\mid c)\,ds
\nonumber\\
&-
g(c\mid c)
\bigl(
W(\beta(c),c;c)-V(\beta(c),c;c)
\bigr)
\frac{1}{\fraction{\mathrm{d} \beta}{\mathrm{d} c}(c)} .
\label{eq:generic-field}
\end{align}
The first line is the direct effect of changing \(z\) while holding the threshold fixed: it captures how \(V\) and \(W\) vary with the player's action. The second line is the threshold effect: changing \(z\) moves \(\beta^{-1}(z)\), transferring probability mass between the two payoff regimes at a rate weighted by the payoff difference \(W-V\).

If $\beta$ pools types, ties can have positive probability, and the pointwise payoff derivative may fail to exist or depend on the tie-breaking rule. If $\beta$ is nonmonotone, the representation through the single inverse $\beta^{-1}$ can also fail. We therefore identify the payoff gradient through \eqref{eq:generic-field} on regular strategies. The learning dynamics for pooled and nonsmooth feasible strategies is defined later as a viscosity continuation of this field, and \cref{thm:scheme} identifies the continuation selected by the grid approximation. Arbitrary tie-breaking rules are treated when studying equilibrium in \cref{sec:main}.

\begin{definition}[M\"obius pricing game]\label{def:mobius}
Let $A,B,\mu:[0,1]\to\mathbb{R}$. A pricing game is a \emph{M\"obius pricing game with coefficients $(A,B,\mu)$} if, for every regular strategy $\beta$ and every $c\in(0,1)$, the interim payoff $\pi(z;\beta,c)$ is differentiable in $z$ at $z=\beta(c)$ and
\begin{equation}\label{eq:field}
G[\beta](c)
:=
\left.
\fraction{\partial \pi}{\partial z}(z;\beta,c)
\right|_{z=\beta(c)}
=
A(c)
-
B(c)
\frac{\beta(c)-\mu(c)}
{\fraction{\mathrm{d} \beta}{\mathrm{d} c}(c)}.
\end{equation}
We call $G[\beta]$ the \emph{payoff field} and $\mu$ the \emph{anchor}.
\end{definition}

\begin{proposition}[Membership]\label{prop:membership}
Let $W$ and $V$ be differentiable in $z$, and suppose that $W(z,s;c)g(s\mid c)$, $V(z,s;c)g(s\mid c)$, $\fraction{\partial W}{\partial z}(z,s;c)g(s\mid c)$, and $\fraction{\partial V}{\partial z}(z,s;c)g(s\mid c)$ are continuous in $(z,s)$ for every $c\in(0,1)$. The game \eqref{eq:payoff} is a M\"obius pricing game with coefficients $(A,B,\mu)$ if and only if, for every $c\in(0,1)$ and every $z\in\mathbb{R}$,
\begin{gather*}
\int_0^c \fraction{\partial V}{\partial z}(z,s;c)g(s\mid c)\,ds
+
\int_c^1 \fraction{\partial W}{\partial z}(z,s;c)g(s\mid c)\,ds
=
A(c),
\\
g(c\mid c)\bigl(W(z,c;c)-V(z,c;c)\bigr)
=
B(c)\bigl(z-\mu(c)\bigr).
\end{gather*}
\end{proposition}

The proposition gives a simple test for membership. A game is M\"obius exactly when the direct effect depends only on the type $c$ and the cutoff payoff gap is affine in the player's action $z$. The function $\mu(c)$ is the zero of this affine cutoff term.

An important subclass consists of allocation games. Their payoff has the form
\begin{equation}\label{eq:allocation}
\pi(z;\beta,c)
=
\kappa\bigl(z-m(c)\bigr)Q\bigl(\beta^{-1}(z)\bigr),
\end{equation}
where $\kappa>0$, $m$ is continuous and strictly increasing, and $Q(c)$ is the allocation probability of type $c$. We assume that $Q$ has a continuous derivative $\fraction{\mathrm{d} Q}{\mathrm{d} c}=-q\le0$ on $[0,1]$, that $Q(1)=0$, and that $Q(c)>0$ for $c<1$. Differentiating \eqref{eq:allocation} at $z=\beta(c)$ shows that allocation games are M\"obius with $A=\kappa Q$, $B=\kappa q$, and $\mu=m$.

\subsection{Assumptions}\label{sec:dynamics}

\begin{assumption}\label{ass:main}
The coefficients $(A,B,\mu)$ satisfy:
\begin{enumerate}
\item[(A1)]\phantomsection\label{cond:A1} $A$, $B$ and $\mu$ are continuous on $[0,1]$;
\item[(A2)]\phantomsection\label{cond:A2} $A(c)>0$ and $B(c)>0$ for $0\le c<1$;
\item[(A3)]\phantomsection\label{cond:A3} $\mu$ is weakly increasing on $[0,1]$.
\end{enumerate}
\end{assumption}

We write
\[
C=\mu(1),\qquad w(c)=C-\mu(c),\qquad R=w(0).
\]
The \emph{cap} $C$ is the anchor of the top type, the \emph{cost gap} $w(c)$ is the length of the interval $[\mu(c),C]$,
and $R$ is the largest cost gap.

\Cref{tab:members} lists the main members of the class: first-price auctions \citep{vickrey1961counterspeculation},
Bertrand competition with private costs \citep{spulber1995bertrand}, pay-as-bid auctions
\citep{vickrey1961counterspeculation,krishna2010auction}, generalized first-price auctions \citep{edelman2007internet},
auctions with a random number of rivals \citep{mcafee1987auctions}, procurement with fees or cost sharing, auctions
with interdependent or affiliated costs \citep{milgrom1982theory}, and a nonlinear family that \cref{sec:main} uses as a counterexample. The
columns $p$ and $a$ give the bound $A\ge a\,w^p$, which determines the rates of \cref{thm:main}.

\begin{proposition}[Members]\label{prop:examples}
Let $f$ be continuous with $\int_0^1f=1$ and $f\ge f_->0$, and let $\bar m(c,\cdot)$ and $g_c$ be continuous on $[0,1]$ for every $c$. Each row of \cref{tab:members} is a M\"obius pricing
game with the stated coefficients, and the columns $p$ and $a$ are correct. The coefficients satisfy \cref{ass:main}, in the
interdependent row if $c\mapsto\bar m(c,c)$ is weakly increasing, and in the affiliated row if $(c,s)\mapsto g_c(s)$ is jointly
continuous with $g_c(c)>0$ for $c<1$; the only
exception is $B(0)=0$ for pay-as-bid auctions with $K\ge2$ and generalized first-price auctions with $\lambda_1=\lambda_2$,
which a change of type removes (\cref{rem:density}).
\end{proposition}

Elastic demand, risk aversion and all-pay contests fall outside the class (\cref{prop:nonmembers} in \cref{app:model}).

\begin{table}[tbp]
\centering
\footnotesize
\caption{Members of the class. Types are independent with distribution $F$ and continuous density $f$ with $f_-\le f\le f_+$,
and $S=1-F$; there are $n$ players, low costs win, and $x=x(z)$. The allocation $Q_K$ is the probability that at most $K-1$
of $n-1$ rivals have lower costs; the allocation $Q^\lambda=\sum_{j\le K}(\lambda_j-\lambda_{j+1})Q_j$ has click rates
$\lambda_1\ge\dots\ge\lambda_K>\lambda_{K+1}=0$; the allocation $Q^\varrho=\sum_{r=1}^{r_{\max}}\varrho_rS^r$ has the probability $\varrho_r$ of $r$ rivals, and $r_0=\min\{r:\varrho_r>0\}$; the seller keeps the
share $\kappa\in(0,1]$ of the price; $\bar m(c,s)$ is the expected own cost given the own type $c$ and the lowest rival
type $s$, whose density is $g$; the function $\Sigma_c(x)=\int_x^1g_c$ is the probability that the lowest rival type exceeds $x$ given the
own type $c$; and $\zeta_c(s)=(s-c)(3s-c-2)$. For an allocation, $q_K=-\fraction{\mathrm{d} Q_K}{\mathrm{d} c}$, $q^\lambda=-\fraction{\mathrm{d} Q^\lambda}{\mathrm{d} c}$ and $q^\varrho=-\fraction{\mathrm{d} Q^\varrho}{\mathrm{d} c}$; in every row except the affiliated one, $B=-\fraction{\mathrm{d} A}{\mathrm{d} c}$. The columns $p$ and $a$ give the bound $A\ge a\,w^p$ on $[0,1)$. In the interdependent row
$L_m$ is a Lipschitz constant of $c\mapsto\bar m(c,c)$; in the affiliated row $\min_{c,s}g_c(s)>0$.}
\label{tab:members}
\setlength{\tabcolsep}{3pt}
\resizebox{\linewidth}{!}{%
\begin{tabular}{@{}lllllll@{}}
\toprule
Game & Interim payoff & $A$ & $B$ & $\mu$ & $p$ & $a$\\
\midrule
First price; Bertrand & $(z-c)\,S(x)^{n-1}$ & $S^{n-1}$ & $(n-1)S^{n-2}f$ & $c$ & $n-1$ & $f_-^{\,n-1}$\\
Pay-as-bid, $K<n$ units & $(z-c)\,Q_K(x)$ & $Q_K$ & $q_K$ & $c$ & $n-K$ & $f_-^{\,n-K}$\\
Generalized first price, $K<n$ slots & $(z-c)\,Q^\lambda(x)$ & $Q^\lambda$ & $q^\lambda$ & $c$ & $n-K$ & $\lambda_Kf_-^{\,n-K}$\\
Random number of rivals & $(z-c)\,Q^\varrho(x)$ & $Q^\varrho$ & $q^\varrho$ & $c$ & $r_0$ & $\varrho_{r_0}f_-^{\,r_0}$\\
Platform fee $1-\kappa$ & $(\kappa z-c)\,S(x)^{n-1}$ & $\kappa S^{n-1}$ & $\kappa(n-1)S^{n-2}f$ & $c/\kappa$ & $n-1$ & $\kappa^nf_-^{\,n-1}$\\
Cost sharing, share $\chi$ & $(z-\chi c)\,S(x)^{n-1}$ & $S^{n-1}$ & $(n-1)S^{n-2}f$ & $\chi c$ & $n-1$ & $(f_-/\chi)^{n-1}$\\
Interdependent costs & $\int_x^1(z-\bar m(c,s))\,g(s)\,ds$ & $S^{n-1}$ & $(n-1)S^{n-2}f$ & $\bar m(c,c)$ & $n-1$ & $(f_-/L_m)^{n-1}$\\
First price, affiliated costs & $(z-c)\,\Sigma_c(x)$ & $\Sigma_c(c)$ & $g_c(c)$ & $c$ & $1$ & $\min_{c,s}g_c(s)$\\
Nonlinear profit, $n=2$, $f\equiv1$ & $\int_x^1(z-c+\theta z^2\zeta_c(s))\,ds$ & $1-c$ & $1$ & $c$ & $1$ & $1$\\
\bottomrule
\end{tabular}}
\end{table}

\subsection{Learning dynamics}

A type $c$ chooses an action $z$ in its feasible interval $[\mu(c),C]$. A \emph{feasible strategy} is a continuous weakly increasing function $b:[0,1]\to\mathbb{R}$ with $\mu\le b\le C$. Since $\mu(1)=C$, every feasible strategy satisfies $b(1)=C$. Let $\mathcal K=\{(c,z):0\le c\le1,\ \mu(c)\le z\le C\}$ be the set of feasible type--action pairs. For bounded $f:[0,1]\to\mathbb{R}$, write $\|f\|=\sup_{0\le c\le1}|f(c)|$.

A \emph{rest point} is a continuous function $\beta^*:[0,1]\to\mathbb{R}$, differentiable on $[0,1)$ with a one-sided derivative at $0$, that satisfies
\begin{equation}\label{eq:rest}
A(c)\fraction{\mathrm{d} \beta^*}{\mathrm{d} c}(c)
=
B(c)\bigl(\beta^*(c)-\mu(c)\bigr)
\quad (0\le c<1),
\qquad
\beta^*(1)=C.
\end{equation}
Thus $G[\beta^*](c)=0$ wherever $\fraction{\mathrm{d} \beta^*}{\mathrm{d} c}>0$.

In symmetric self-play, the common action of each type is moved in the direction of its interim-payoff
derivative against the current common rival strategy. The same update is used by every player at a
given type. The learning geometry is a \emph{typewise Riemannian metric}, a continuous function $g:\mathcal K\to(0,\infty)$: at $(c,z)$, an action change $dz$ has length $g(c,z)^{1/2}|dz|$. The Riemannian gradient of the interim payoff is the payoff field scaled by the \emph{weight} $\varphi=1/g$. For a regular iterate $\beta^k$ and step size $h>0$, the explicit Euler step is
\[
\beta^{k+1}(c)
=
\beta^k(c)
+
h\,\varphi(c,\beta^k(c))G[\beta^k](c).
\]
Formally letting $h\to0$ and $kh\to t$ gives the \emph{payoff-gradient dynamics}
\begin{equation}\label{eq:GF}
\fraction{\partial \beta}{\partial t}(t,c)
=
\varphi(c,\beta(t,c))
\left(
A(c)
-
B(c)
\frac{\beta(t,c)-\mu(c)}
{\fraction{\partial \beta}{\partial c}(t,c)}
\right),
\quad 0\le c<1,
\qquad
\beta(t,1)=C.
\tag{GF}
\end{equation}
The geometry is \emph{typewise}: the velocity of each type is rescaled using only that type and its
current action. Equation \eqref{eq:GF} agrees with ordinary payoff-gradient ascent where the strategy
is regular. For feasible pooled and nonsmooth strategies, \cref{def:solution} specifies a viscosity
continuation. Section~\ref{sec:grid} constructs a grid approximation and states the conditions on
deterministic or stochastic errors in its update. Both formulations maintain one common strategy;
independently updated players require a coupled analysis.

Since actions are one-dimensional, every typewise Riemannian metric is a Hessian metric, so Riemannian gradient flow and mirror descent coincide. Indeed, for each $c$ choose a strictly convex function $\Phi(c,\cdot)$ with $\fraction{\partial^2 \Phi}{\partial z^2}=g=1/\varphi$. The continuous-time mirror flow generated by $\int_0^1\Phi(c,\beta(c))\,dc$ satisfies
\[
\fraction{\partial}{\partial t}
\fraction{\partial \Phi}{\partial z}(c,\beta(t,c))
=
G[\beta(t,\cdot)](c),
\]
and therefore coincides with \eqref{eq:GF}. Conversely, every typewise mirror map $\Phi(c,\cdot)$ whose second derivative is continuous and positive on $\mathcal K$ defines the metric $g=\fraction{\partial^2 \Phi}{\partial z^2}$. Thus \eqref{eq:GF} covers every typewise Riemannian metric and every typewise mirror map.

\begin{assumption}\label{ass:metric}
The weight $\varphi=1/g$ is continuous and positive on $\mathcal K$, and there is a constant $L_\varphi\ge0$ such that, for every $0\le c\le1$ and $\mu(c)\le z'\le z\le C$,
\begin{enumerate}
\item[(G1)]\phantomsection\label{cond:G1} $|\varphi(c,z)-\varphi(c,z')|\le L_\varphi(z-z')$;
\item[(G2)]\phantomsection\label{cond:G2} $\varphi(c,z')(z'-\mu(c))<\varphi(c,z)(z-\mu(c))$ if $z'<z$.
\end{enumerate}
\end{assumption}

Let $\varphi_-$ and $\varphi_+$ be the minimum and maximum of $\varphi$ on $\mathcal K$; they are positive and finite because $\mathcal K$ is compact. Condition \condref{G1} bounds how fast the weight changes with the action; since $\varphi_-\le\varphi\le\varphi_+$, it is equivalent to the same condition on the metric $g$, with another constant. Condition \condref{G2} makes the weighted restoring term $z\mapsto\varphi(c,z)(z-\mu(c))=(z-\mu(c))/g(c,z)$ strictly increasing, so it depends on the anchor of the game. Every weight that is weakly increasing in $z$, in particular every weight that depends only on the type, satisfies \condref{G2}. Decreasing weights are allowed: for differentiable $\varphi$, \condref{G2} holds if $\varphi+(z-\mu(c))\,\fraction{\partial \varphi}{\partial z}>0$ on $\mathcal K$, that is, if the elasticity of $\varphi$ with respect to the distance $z-\mu(c)$ from the anchor stays above $-1$. Apart from these two regularity conditions, every typewise Riemannian metric is allowed. For example, Euclidean gradient flow has $\varphi\equiv1$, so $\varphi_\pm=1$ and $L_\varphi=0$; type-dependent learning rates give a weight $\varphi(c,z)$ that does not depend on $z$, so $L_\varphi=0$; and the shifted-entropy mirror map $\Phi(c,z)=(z-\underline z)\log(z-\underline z)$ with $\underline z<\mu(0)$ gives $\varphi(c,z)=z-\underline z$, so $\varphi_-=\mu(0)-\underline z$, $\varphi_+=C-\underline z$ and $L_\varphi=1$.

\begin{remark}[On the assumptions]\label{rem:assumptions}
For $c<1$, the conditions $A(c)>0$ and $B(c)>0$ ensure that both the direct effect and the threshold effect are present. At the top type only continuity is assumed: \eqref{eq:GF} is imposed for $c<1$, and the boundary condition $\beta(t,1)=C$ fixes the only feasible action of the top type, since $\mu(1)=C$. In the main examples $A(1)=0$, so the field is stationary at the top anchor. The cap is essential when $A(1)=B(1)=0$, because the field then does not determine the top action, and when $A(1)>0$, as with a reserve price (\cref{app:model}), because the field at the cap is then positive. Monotonicity of $\mu$ makes the feasible intervals $[\mu(c),C]$ nested.
\end{remark}

\subsection{Two obstructions}\label{sec:obstructions}

The first obstruction is that the $L^2$ distance to the rest point can increase along the flow.

\begin{proposition}[Minty failure]\label{prop:minty}
Consider the first-price procurement auction with $n=2$ and uniform costs, so that $A=1-c$, $B=1$, $\mu=c$, and the
rest point is $\bstar=(1+c)/2$. The feasible strategy $\beta$ with slope $\frac15$ on $[0,\frac12]$, slope $\frac45$ on $[\frac12,1]$ and
$\beta(1)=1$ satisfies $\int_0^1(\beta-\bstar)\,G[\beta]\,dc=\frac3{320}>0$, so the rest point violates the Minty condition that
$\int_0^1(\beta-\bstar)\,G[\beta]\,dc\le0$ for all feasible $\beta$ \citep{minty1962monotone}, as \citet{bichler2025beyond} observed. Hence, if a family $(\beta_t)$ of measurable strategies, defined for $t$ near a time $t_0$, is
differentiable in $t$, with $\beta_t$ and $\fraction{\partial \beta_t}{\partial t}$ bounded uniformly near $t_0$, passes through $\beta$ at $t_0$, and moves along the field there,
$\fraction{\partial \beta_t}{\partial t}=G[\beta_t]$ at $t_0$, then half the squared $L^2$ distance, $\frac12\int_0^1(\beta_t-\bstar)^2dc$, has the derivative $\frac3{320}>0$
at $t_0$.
\end{proposition}

The second obstruction concerns regularity. A strategy that pools an interval of types has slope zero
there, so the reciprocal-slope expression in \eqref{eq:field} is not defined. Even from a smooth
strictly increasing initial strategy, the solution can lose spatial differentiability in finite time.

\begin{proposition}[Breakdown of classical solutions]\label{prop:blowup}
In the auction of \cref{prop:minty} with $\varphi\equiv1$, the strategy
\[
b(c)=\frac{1+c}{2}
+\frac{\sqrt{40001}-\sqrt{1+160000\,(c-\frac12)^2}}{1600}
\]
is $C^\infty$ and feasible, with $b(1)=1$ and $\frac14<\fraction{\mathrm{d} b}{\mathrm{d} c}(c)<\frac34$. Its unique solution
$\beta$ in the sense of \cref{def:solution} remains continuous, but for some $0<t_0<\frac1{200}$ and
$c_0\in(0,1)$ the function $c\mapsto\beta(t_0,c)$ is not differentiable at $c_0$. Consequently,
there is no feasible $C^1$ classical solution with positive slope on $[0,\frac1{200}]\times[0,1]$
from this datum.
\end{proposition}

The viscosity notion in \cref{def:solution} therefore provides a continuous continuation after spatial
differentiability is lost. The proof of \cref{thm:main} allows such nonsmooth strategies and pooling,
and measures progress by the uniform distance instead of the $L^2$ distance.

\Cref{app:model} proves the results of this section.

%=====================================================================
\section{Global convergence}\label{sec:solutions}
%=====================================================================

This section defines solutions of \eqref{eq:GF} and proves that every solution converges to the rest point at an explicit rate.

At a positive slope $u$ the right-hand side of \eqref{eq:GF} is $-\varphi(c,z)\,H(c,z,u)$ with the Hamiltonian
\[
H(c,z,u)=-A(c)+B(c)\,\frac{z-\mu(c)}u .
\]
For $c<1$ and $z>\mu(c)$, $H(c,z,u)$ tends to $+\infty$ as $u\downarrow0$, while at $z=\mu(c)$ it equals $-A(c)$ for every $u>0$. We therefore extend $H$ to pooling slopes by setting
\[
H(c,z,0)=+\infty \quad\text{for }z>\mu(c),
\qquad
H(c,\mu(c),0)=-A(c).
\]
Thus $H$ is extended-real-valued on feasible strategies. In an allocation game, pooling above cost can create a payoff jump from a slight undercut, whose size and one-sided behavior depend on the allocation and tie-breaking rule; hence an ordinary payoff derivative need not exist at pooling. We do not extend $H$ to negative slopes: feasible strategies are weakly increasing, and the single-inverse derivation of \eqref{eq:field} generally fails there.

The continuation is specified by the test-function conditions in \cref{def:solution}. Above the anchor,
a nonpositive upper-test slope is incompatible with the subsolution condition; at the anchor a
separate clause applies. We use viscosity solutions following \citet{crandall1983viscosity} and
\citet{crandall1992users}, with the anchor treated as a state constraint
\citep{soner1986optimal}. \Cref{thm:scheme} identifies this continuation as the limit of the prescribed
regularized grid scheme. Let $\cB_T=[0,T]\times[0,1]$.

\begin{definition}[Solution]\label{def:solution}
Let $T>0$, and extend $\varphi$ continuously to $[0,1]\times\R$. A function $\sub:\cB_T\to\R$ is a \emph{subsolution} of \eqref{eq:GF} if, for every continuously
differentiable $\psi:\R^2\to\R$ and every $(t_{\star},c_{\star})\in(0,T]\times[0,1)$ at which $\sub-\psi$ attains its maximum over $\cB_T$: if
$\sub(t_{\star},c_{\star})>\mu(c_{\star})$, then $\fraction{\partial \psi}{\partial c}(t_{\star},c_{\star})>0$ and $\fraction{\partial \psi}{\partial t}+\varphi(c_{\star},\sub)\,H(c_{\star},\sub,\fraction{\partial \psi}{\partial c})\le0$ at $(t_{\star},c_{\star})$; if
$\sub(t_{\star},c_{\star})\le\mu(c_{\star})$, then $\fraction{\partial \psi}{\partial t}(t_{\star},c_{\star})\le\varphi(c_{\star},\mu(c_{\star}))\,A(c_{\star})$. A function $\super:\cB_T\to\R$ is a
\emph{supersolution} if $\fraction{\partial \psi}{\partial t}+\varphi(c_{\star},\super)\,H(c_{\star},\super,\fraction{\partial \psi}{\partial c})\ge0$ at every $(t_{\star},c_{\star})\in(0,T]\times[0,1)$ at which
$\super-\psi$ attains its minimum over $\cB_T$ and $\fraction{\partial \psi}{\partial c}(t_{\star},c_{\star})>0$. A function $\beta:[0,\infty)\times[0,1]\to\R$ is a
\emph{solution from} a feasible strategy $b$ if $\beta(0,\cdot)=b$ and, for every $T>0$, the function $\beta$ is continuous on $\cB_T$, satisfies
$\mu\le\beta\le C$ and $\beta(t,1)=C$, and is a subsolution and a supersolution on $\cB_T$.
\end{definition}

The subsolution condition excludes contact from above by a test function that is weakly decreasing in
the type at a point where the strategy lies above its anchor. A function with positive slope and
continuous partial derivatives that solves \eqref{eq:GF} pointwise and satisfies $\mu\le\beta\le C$ is a solution. Away from the anchor the
conditions are those of the semicontinuous envelopes of $\varphi H$ \citep{crandall1992users}; at the anchor with a nonpositive
test slope they ask less, so every solution in the envelope sense is a solution. \Cref{thm:scheme} shows that
\cref{alg:grid} of \cref{sec:grid} converges to this solution. Thus the definition describes the limit
of the prescribed regularization; it does not assert that every tie-breaking or payoff-estimation
procedure selects the same limit.

A \emph{Lyapunov functional} vanishes at the rest point, is positive elsewhere, and has a negative time derivative along
every solution wherever it is positive \citep{lyapunov1992general}; see \citet[Theorem~2.0.1]{wiggins2003introduction} for this
criterion on $\R^n$. The uniform distance to the rest point need not be differentiable in time, so we use the upper right
Dini derivative of a function $f$ of time, the limit superior of $(f(t+s)-f(t))/s$ as $s\downarrow0$. The local damping rate at the rest point is
\[
\frac{B(c)}{\fraction{\mathrm{d} \beta^*}{\mathrm{d} c}(c)}
=
\frac{A(c)}{\beta^*(c)-\mu(c)}
\ge
\frac{A(c)}{w(c)},
\]
where the equality follows from the rest-point equation. For an error level $0<E\le R$, let
\[
\Psi(E)=\min\Bigl\{\frac{A(c)}{w(c)}:\ 0\le c<1,\ w(c)\ge E\Bigr\}
\]
be the resulting guaranteed damping rate among types whose cost gap is at least $E$, and let $\Psi(E)=0$ for $E>R$.

\begin{theorem}[Global last-iterate convergence]\label{thm:main}
Let $(A,B,\mu)$ satisfy \cref{ass:main} and $\varphi$ \cref{ass:metric}. There is a unique rest point $\bstar$; it is weakly increasing, satisfies $\mu\le\bstar\le C$, and does not depend on $\varphi$. Let $b$ be a feasible strategy.
\begin{enumerate}
\item[(i)] \emph{Solution.}
There is a unique solution $\beta$ from $b$. It is continuous on $[0,\infty)\times[0,1]$, and $\beta(t,\cdot)$ is a feasible strategy for every $t\ge0$.

\item[(ii)] \emph{Lyapunov functional.}
Let
\[
\cE(t)=\norm{\beta(t,\cdot)-\bstar}.
\]
The function $\Psi$ is positive on $(0,R]$, and whenever $\cE(t)>0$,
\[
D^+\cE(t)
\le
-\frac12\varphi_-\Psi\bigl(\cE(t)/2\bigr)\cE(t)
<0.
\]
Hence $\cE$ is a Lyapunov functional and $\cE(t)\to0$ as $t\to\infty$.

If, in addition, $A\ge a\,w^p$ on $[0,1)$ with $a>0$ and $p\ge1$, then
\[
D^+\cE(t)\le-\varphi_-a\,\cE(t)^p,
\]
and, for every $t\ge t_0\ge0$,
\begin{equation}\label{eq:power}
\cE(t)\le
\begin{cases}
\cE(t_0)\,\e^{-\varphi_-a(t-t_0)}, & p=1,\\[1mm]
\displaystyle
\frac{\cE(t_0)}
{\bigl(1+(p-1)\varphi_-a\,\cE(t_0)^{p-1}(t-t_0)\bigr)^{1/(p-1)}},
& p>1.
\end{cases}
\end{equation}

\item[(iii)] \emph{Optimality.}
For every $p\ge1$, there are coefficients satisfying \cref{ass:main} with $A=w^p$, and a feasible initial strategy, for which the rate in \eqref{eq:power} is attained in order: for $p=1$ the error decays exponentially, while for $p>1$,
\[
\cE(t)\asymp (1+t)^{-1/(p-1)}.
\]
Hence the exponential rate for $p=1$ and the exponent $1/(p-1)$ for $p>1$ are optimal in order. Moreover, for every $\nu>0$, there are coefficients satisfying \cref{ass:main} for which $\cE(t)$ is not $O((1+t)^{-\nu})$.
\end{enumerate}
\end{theorem}

\Cref{thm:main} holds from every feasible initial strategy, pooling included; it needs no bound on slopes and no
regularity of $A$ and $\mu$ beyond continuity. The metric changes the speed of learning, not its limit. Part (ii)
identifies the Lyapunov functional: the uniform distance decreases at the rate of the slowest type whose cost gap
exceeds the current error, whereas by \cref{prop:minty} the $L^2$ distance can increase. For an integer $p$ the family
of (iii) is the first-price auction with $p+1$ bidders and uniform costs, written in the type $1-(1-c)^p$.

The next example shows how a pooled strategy separates.

\begin{example}[Complete pooling in the uniform duopoly]\label{ex:pooled}
Let $\varphi\equiv1$, $A(c)=1-c$, $B=1$, $\mu(c)=c$ and $b\equiv1$. Let $u(t)$ be the inverse of the increasing bijection
$\mathcal T(u)=-\frac u2-\frac14\log(1-2u)$ of $[0,\frac12)$ onto $[0,\infty)$. The solution is $\beta(t,c)=1-u(t)(1-c)$; it has slope
$u(t)\ge\min\{\sqrt{t/2},\frac14\}$ for $t>0$, and $\cE(t)=\frac12-u(t)=\frac12\e^{-4t-2u(t)}\le\frac12\e^{-4t}$.
\end{example}

The pooled strategy thus has a positive slope at every positive time, at least of order $\sqrt t$; \cref{fig:pooled}
compares it with the grid algorithm, and \cref{app:example} verifies the example.

\paragraph{Proof outline.}
\Cref{app:main} proves \cref{thm:main}. The rest point is an explicit weighted average of the anchors of
higher types (\cref{lem:rest}). The central tool is the comparison principle of \cref{lem:comparison}:
\begin{quote}\itshape
if $\sub$ is an upper semicontinuous subsolution with $\sub\le C$, $\super$ is a lower semicontinuous supersolution with
$\mu\le\super\le C$, and $\sub\le\super$ at $t=0$ and at $c=1$, then $\sub\le\super$ on $\cB_T$.
\end{quote}
It orders strategies type by type, so progress is measured in the uniform norm, not through the inner product
$\int(\beta-\bstar)G[\beta]$ that monotonicity arguments control; in its proof, \condref{G1} controls the direct term and \condref{G2}
the singular term. Comparison gives uniqueness of solutions; existence follows from the grid-approximation lemma \cref{lem:scheme-feasible}, based on the monotone scheme of \cref{sec:grid}.
For (ii), the error identity \eqref{eq:intro-identity} shows that the clipped shifts $\min\{C,\bstar+e(t)\}$
and $\max\{\mu,\bstar-e(t)\}$ are a supersolution and a subsolution whenever $-\fraction{\mathrm{d} e}{\mathrm{d} t}\le\varphi_-\Psi(e)\,e$
(\cref{lem:barriers}). Comparison traps the solution between them, and restarting the comparison at an
arbitrary time $t_0$ (\cref{lem:shift}) gives the bound on the Dini derivative. For (iii), an explicit
subsolution bounds the error from below (\cref{lem:lowerbarrier}).

%=====================================================================
\section{The limit is the equilibrium}\label{sec:main}
%=====================================================================

This section states the additional payoff conditions under which the rest point is a Bayes--Nash
equilibrium and the learned strategies are approximate equilibria. The distinction matters because
the coefficients $(A,B,\mu)$ specify derivatives along regular symmetric profiles, not all gains from
unilateral deviations. The nonlinear family in \cref{thm:selection}(d) illustrates the distinction:
its payoff field is unchanged while the equilibrium status of its rest point varies.

The feasible actions of type $c$ are $[\mu(c),C]$. A feasible strategy $\beta$ that pools types makes players tie.
For an action $z$ let $x^-(z)=\sup(\{s\in[0,1]:\beta(s)<z\}\cup\{0\})$ and $x^+(z)=\inf(\{s\in[0,1]:\beta(s)>z\}\cup\{1\})$; the rival
types strictly between $x^-(z)$ and $x^+(z)$ choose $z$, and $x^-=x^+=x$ if $\beta$ is strictly increasing. A \emph{tie-breaking rule}
decides, possibly at random, which of the tied players receives the allocation; under it, the payoff $\pi_c(z;\beta)$ is \eqref{eq:payoff} with $x(z)$
replaced by $x^-(z)$ and, for $x^-(z)<s<x^+(z)$, with $W_c(z,s)$ replaced by a convex combination of $W_c(z,s)$ and $V_c(z,s)$.
Let $\varepsilon\ge0$. A feasible strategy $\beta$ is an
\emph{$\varepsilon$-equilibrium} under a tie-breaking rule if $\pi_c(z;\beta)\le\pi_c(\beta(c);\beta)+\varepsilon$ for every type $c\in[0,1]$ and every feasible
action $z$, and a \emph{symmetric Bayes--Nash equilibrium} if this holds with $\varepsilon=0$.
We say that the equilibrium inequality \emph{holds globally} if it holds for every $z\in\R$.

\begin{theorem}[Equilibrium]\label{thm:selection}
Let a M\"obius pricing game have coefficients $(A,B,\mu)$ that satisfy \cref{ass:main}, and let $\bstar$ be its rest point.
\begin{enumerate}
\item[(a)] Every regular symmetric Bayes--Nash equilibrium equals $\bstar$.

\item[(b)] In allocation games \eqref{eq:allocation}, the rest point is a symmetric Bayes--Nash equilibrium, and the equilibrium inequality holds globally. Under some tie-breaking rule
in which a tied player does not surely receive the allocation ahead of every tied rival it is the only feasible strategy that is a symmetric Bayes--Nash equilibrium . If $B\le\bar B$ and $\bstar(c')-\bstar(c)\ge u_-(c'-c)$ for $c\le c'$, with $u_->0$, every feasible $\beta$ is a $2\norm{\beta-\bstar}\bigl(A(0)+R\bar B/u_-\bigr)$-equilibrium, against every action, under every tie-breaking rule.

\item[(c)] In the game with interdependent costs of \cref{tab:members}, with $\bar m(c,\cdot)$ continuous for every $c$ and $\bar m(c,c)<\bar m(1,1)$ for
$c<1$, the rest point $\bstar$ is a symmetric Bayes--Nash equilibrium against every action if $\bar m(\cdot,s)$ is weakly increasing for every
$s$; in the first-price auction with affiliated costs of \cref{tab:members} \citep{milgrom1982theory}, with $\Sigma_c>0$ on $[0,1)$, it is one if $g_c(x)/\Sigma_c(x)$
is weakly decreasing in $c$.

\item[(d)] In the nonlinear family of \cref{tab:members}, $\bstar(c)=(1+c)/2$ for every $\theta\ge0$, and it is a Bayes--Nash equilibrium if and only if
$\theta\le27/16$. For $\theta>27/16$ the type $0$ gains by choosing the action $\bstar(1/3)$.

\item[(e)] Let the payoffs \eqref{eq:payoff} have $W_cg_c$ and $V_cg_c$ jointly continuous in $(c,z,s)$ and $g_c\ge0$, and let $\bstar$ be
strictly increasing. If $\bstar$ is a symmetric Bayes--Nash equilibrium, then for every $\varepsilon>0$ there is $\delta>0$ such that
every feasible $\beta$ with $\norm{\beta-\bstar}\le\delta$ is an $\varepsilon$-equilibrium under every tie-breaking rule. If instead some type $c$
and feasible action $z$ have $\pi_c(z;\bstar)-\pi_c(\bstar(c);\bstar)=\varepsilon_0>0$, then there is $\delta>0$ such that
$\pi_c(z;\beta)-\pi_c(\beta(c);\beta)\ge\varepsilon_0/2$ under every tie-breaking rule for every feasible $\beta$ with $\norm{\beta-\bstar}\le\delta$.
\end{enumerate}
\end{theorem}

Together with \cref{thm:main}, \cref{thm:selection} decides whether learning reaches equilibrium. Along the learning path
$\cE(t)\to0$, so by (e) the learned strategy is an $\varepsilon(t)$-equilibrium with $\varepsilon(t)\to0$ exactly when $\bstar$ is an equilibrium. In
allocation games this always holds, with $\varepsilon(t)=2\cE(t)(A(0)+R\bar B/u_-)$ by (b). In the nonlinear family with $\theta>27/16$ it
never holds, although the dynamics does not depend on $\theta$. The action sets matter in (d): with unrestricted
actions, every type $c>0$ would gain by a very negative action.

\paragraph{Proof idea.} \Cref{app:equilibrium} proves \cref{thm:selection}. For (a), at a regular equilibrium the
first-order condition $G[\beta](c)=0$ at an interior action is the rest-point equation; an action at the anchor would
need $A(c)\le0$, and the rest-point equation has one solution with $\beta(1)=C$. For (b), the rest point of an
allocation game is a weighted average of the costs $m$ of higher types, so type $c$ earns
$\Pi_c(x)=\kappa\int_x^1(m(s)-m(c))q(s)\,ds$ by posting the action of type $x$; this is largest at $x=c$ because $m$ is
increasing, and actions outside the range of $\bstar$ do no better. Uniqueness follows from revealed preference, as
in the integral form of equilibrium payoffs \citep{myerson1981optimal}, and a pooling interval is broken by a slight
undercut under the stated tie-breaking rules. For the approximate equilibria, the rivals' implied types under $\beta$ and
$\bstar$ differ by at most $\norm{\beta-\bstar}/u_-$, and each unit of implied type changes the winning probability by at
most $\bar B/\kappa$. Part (c) uses the same mimicking payoff, whose derivative in $x$ has the sign of $c-x$ under the
stated monotonicity conditions, and (d) computes the deviation. Part (e) holds because payoffs depend on the rivals'
strategy only through implied types, which are uniformly close when strategies are; this is why the uniform norm of
\cref{thm:main} is the right one.

\paragraph{First-price auctions.} In the selling first-price auction, $n\ge2$ bidders have independent private
values with distribution $F_V$ and density $f_V$ on $[0,1]$, and a bid function $\alpha$ is feasible if it is continuous and weakly increasing
with $0\le\alpha(v)\le v$. In the coordinates $c=1-v$ and $\beta(c)=1-\alpha(1-c)$ the auction is the procurement auction of
\cref{tab:members} with $f(c)=f_V(1-c)$, and the cap $C=1$ is the constraint $\alpha\ge0$. A solution $\alpha$ from $\alpha_0$ is the
reflection of the solution of \eqref{eq:GF} with $\varphi\equiv1$ from the reflection of $\alpha_0$.

\begin{corollary}[First-price auctions]\label{cor:auction}
Let $F_V$ have a continuous density with $0<f_-\le f_V\le f_+$, and let $\alpha^*(v)=v-F_V(v)^{1-n}\int_0^vF_V(s)^{n-1}ds$ ($\alpha^*(0)=0$) be the
equilibrium bid function of \citet{vickrey1961counterspeculation} and \citet{riley1981optimal}.
\begin{enumerate}
\item[(a)] The bid function $\alpha^*$ is a symmetric Bayes--Nash equilibrium against every bid, and it is the only weakly increasing
symmetric Bayes--Nash equilibrium under every tie-breaking rule of \cref{thm:selection}(b)
\citep[cf.][]{maskin2003uniqueness,lebrun2006uniqueness}.
\item[(b)] For every feasible initial strategy $\alpha_0$, there exists a unique solution $\alpha$ with $\alpha(0,\cdot)=\alpha_0$; moreover, $\mathcal E(t)=\|\alpha(t,\cdot)-\alpha^*\|_\infty$ is a strict Lyapunov functional, and
\[
\cE(t)\le\e^{-f_-t}\quad(n=2),\qquad \cE(t)\le\bigl(1+(n-2)f_-^{\,n-1}t\bigr)^{-1/(n-2)}\quad(n\ge3).
\]
\item[(c)] Under every tie-breaking rule, $\alpha(t,\cdot)$ is an $\varepsilon(t)$-equilibrium against every bid, where
\[
\varepsilon(t)=2\left(1+\frac{nf_+^2}{f_-}\right)\mathcal E(t).
\]
\item[(d)] These rates are sharp in order. Let $u_-=(n-1)f_-/(nf_+)$ and $\Gamma=2nf_+^{\,n-1}/u_-$. From the feasible initial strategy $\alpha_0(v)=\alpha^*(v)-\frac12u_-v$,
\[
\cE(t)\ge\tfrac12u_-\e^{-\Gamma t}\quad(n=2),\qquad
\cE(t)\ge\tfrac12u_-\bigl(\e\,(n-2)\Gamma t\bigr)^{-1/(n-2)}\quad\bigl(n\ge3,\ (n-2)\Gamma t\ge1\bigr).
\]
\end{enumerate}
\end{corollary}

For $n\ge3$ the decay is slow because bidders with low values win with probability of order $F_V(v)^{n-1}$
and therefore receive a weak gradient. \Cref{app:equilibrium} proves the corollary.

%=====================================================================
\section{Learning on a grid}\label{sec:grid}
%=====================================================================

This section gives a finite grid approximation of \eqref{eq:GF} and establishes its convergence and
error guarantees; \cref{app:grid} contains the proofs. The algorithm uses the game coefficients and
the common strategy vector. Projection and a slope floor specify its behavior at boundaries and
singular grid configurations.

For computation, we discretize the type space. The \emph{loss coordinate} $X(c)=\int_0^cB(s)\,ds$ measures types by the
mass of ties below them; in a first-price auction $X(c)=1-A(c)$ is the probability that type $c$ loses. Fix a grid
$0=c_0<c_1<\dots<c_N=1$ and write
\[
A_i=A(c_i),\quad \mu_i=\mu(c_i),\quad w_i=C-\mu_i,\quad \Delta_i=X(c_{i+1})-X(c_i)\quad(i<N),\quad \varphi_i(z)=\varphi(c_i,z).
\]
A grid strategy $(\beta_0,\dots,\beta_N)$ is \emph{feasible} if $\mu_i\le\beta_i\le C$ for all $i$ and $\beta_N=C$; unlike a feasible strategy on
$[0,1]$, it need not be weakly increasing. The algorithm uses the forward action difference, floored at
$\ell\Delta_i$, to approximate the slope in the loss coordinate, scales the step by the metric weight,
and projects onto the feasible actions. The error $\xi_i^k$ is measured relative to this regularized
discrete field. In symmetric self-play, the same estimate and update at a given type are applied
to the common strategy.

\begin{algorithm}[H]
\KwIn{grid $c_0<\dots<c_N$, weights $\varphi_i$, floor $\ell>0$, step $h>0$, initial actions $\beta^0_0,\dots,\beta^0_N$}
\For{$k=0,1,2,\dots$}{
  \For{$i=0,\dots,N-1$ (in parallel)}{
    $G_i\gets A_i-\Delta_i\,(\beta^k_i-\mu_i)/\max\{\beta^k_{i+1}-\beta^k_i,\ \ell\Delta_i\}$\tcp*{regularized discrete payoff field}
    $\beta^{k+1}_i\gets\min\bigl\{C,\max\{\mu_i,\ \beta^k_i+h\,\varphi_i(\beta^k_i)\,(G_i+\xi^k_i)\}\bigr\}$\tcp*{$\xi^k_i$: gradient error}
  }
  $\beta^{k+1}_N\gets C$\;
}
\caption{Projected grid learning with a slope floor}\label{alg:grid}
\end{algorithm}

In \eqref{eq:field}, $(\fraction{\mathrm{d} \beta}{\mathrm{d} c})/B$ is the slope of $\beta$ in the loss coordinate; the estimate $G_i$ replaces it by the difference
quotient $(\beta^k_{i+1}-\beta^k_i)/\Delta_i$, floored at $\ell$. In an allocation game
$\Delta_i=A_i-A_{i+1}$. For an increasing grid vector and an inactive floor, $G_i$ is the marginal
profit of type $c_i$ using the corresponding piecewise-linear interpolation of the rivals' winning
probability, with the one-sided derivative selected by the scheme at a knot. When the floor is active,
it modifies that update. For a nonmonotone vector, the neighboring-action formula is a numerical
extension and need not equal the actual interim-payoff derivative. 
The \emph{grid rest point} $\hat\beta$ is defined by $\hat\beta_N=C$ and, for $i=N-1,\dots,0$,
\begin{equation}\label{eq:gridrest}
\hat\beta_i
=
\max\Bigl\{
\frac{A_i\hat\beta_{i+1}+\Delta_i\mu_i}{A_i+\Delta_i},
\ \min\{\mu_i+\ell A_i,C\}
\Bigr\}.
\end{equation}
Setting $\ell=0$ removes the slope-floor regularization and defines $\check\beta$ by the same recursion. Equivalently, $\check\beta$ is obtained by replacing the derivative in the continuum rest-point equation by its one-sided forward difference, following the upwind principle of \citet{courant1952solution}.
With $\hat\varphi=\varphi_++L_\varphi R$ and $\norm A=\max A$, we impose the following \emph{CFL-type condition} \citep{courant1928partiellen}: for the error bound $\eta\ge0$,
\begin{equation}\label{eq:cfl}
h\Bigl(L_\varphi(\norm A+\eta)+\frac{\hat\varphi}{\ell}\Bigr)\le1
\qquad\text{and}\qquad
h\Bigl(L_\varphi(\norm A+\eta)+\frac{\hat\varphi R}{\ell^2\Delta_i}\Bigr)\le1
\quad(i<N).
\end{equation}
For $\varphi\equiv1$, this reduces to $h\le\ell$ and $hR\le\ell^2\Delta_i$.

\begin{theorem}[Grid learning]\label{thm:grid}
Let $(A,B,\mu)$ satisfy \cref{ass:main} and $\varphi$ \cref{ass:metric}, let $\ell>0$, let the errors satisfy $|\xi^k_i|\le\eta$ and be
otherwise arbitrary, let $h$ satisfy \eqref{eq:cfl}, and let $E_k=\max_i|\beta^k_i-\hat\beta_i|$.
\begin{enumerate}
\item[(a)] \emph{Rest point.} The grid rest point is the only feasible fixed point of the update without errors, and
$0\le\hat\beta_i-\check\beta_i\le\ell\norm A$. If $\mu_{i+1}-\mu_i+\ell(A_{i+1}-A_i)\ge\ell\Delta_i$ for all $i<N$, with $A_N$ read as $0$ (the algorithm does not use $A_N$), then $\hat\beta=\check\beta$; if moreover
$\varphi\equiv1$, $\eta=0$, $\beta^0$ is feasible and $\beta^0_{i+1}-\beta^0_i\ge\ell\Delta_i$ for all $i<N$, the floor and the projection are never active, and the iterates
are those of plain gradient play $\beta^{k+1}_i=\beta^k_i+h\bigl(A_i-\Delta_i(\beta^k_i-\mu_i)/(\beta^k_{i+1}-\beta^k_i)\bigr)$.
\item[(b)] \emph{Rates.} After one round the iterate is feasible, whatever the initial vector. Let $\beta^0$ be feasible and assume $R>0$.
% PH: the statement with $\bar w>0$ follows directly from A>0 on continuum interior since discretization introduces compactness and $\bar w>0$ implies $\bar c < 1$.
If $\bar w\ge0$ and $a>0$ satisfy $\ell a\le R$ and $A_i\ge a$ whenever $w_i>\bar w$, then
\[
E_k\le\bar w+(1-h\varphi_-a/R)^kR+R\varphi_+\eta/(\varphi_-a).
\]
If $A_i\ge a\,w_i^p$ for all $i<N$, with $a>0$, real $p\ge1$, $\ell aR^{p-1}\le1$ and $h\varphi_-apR^{p-1}\le1$, then
\[
E_k\le\begin{cases}
(1-h\varphi_-a)^kR+\varphi_+\eta/(\varphi_-a), & p=1,\\[1mm]
\max\Bigl\{\bigl(2\varphi_+\eta/(\varphi_-a)\bigr)^{1/p},\ R\bigl(1+(p-1)\tfrac{\varphi_-a}2R^{p-1}\,kh\bigr)^{-1/(p-1)}\Bigr\}, & p>1 .
\end{cases}
\]
\item[(c)] \emph{Consistency.} For every $\varepsilon>0$ there is $\delta>0$ with $\max_i|\check\beta_i-\bstar(c_i)|\le\varepsilon$ on every grid of mesh
$\max_i(c_{i+1}-c_i)\le\delta$.
\end{enumerate}
\end{theorem}

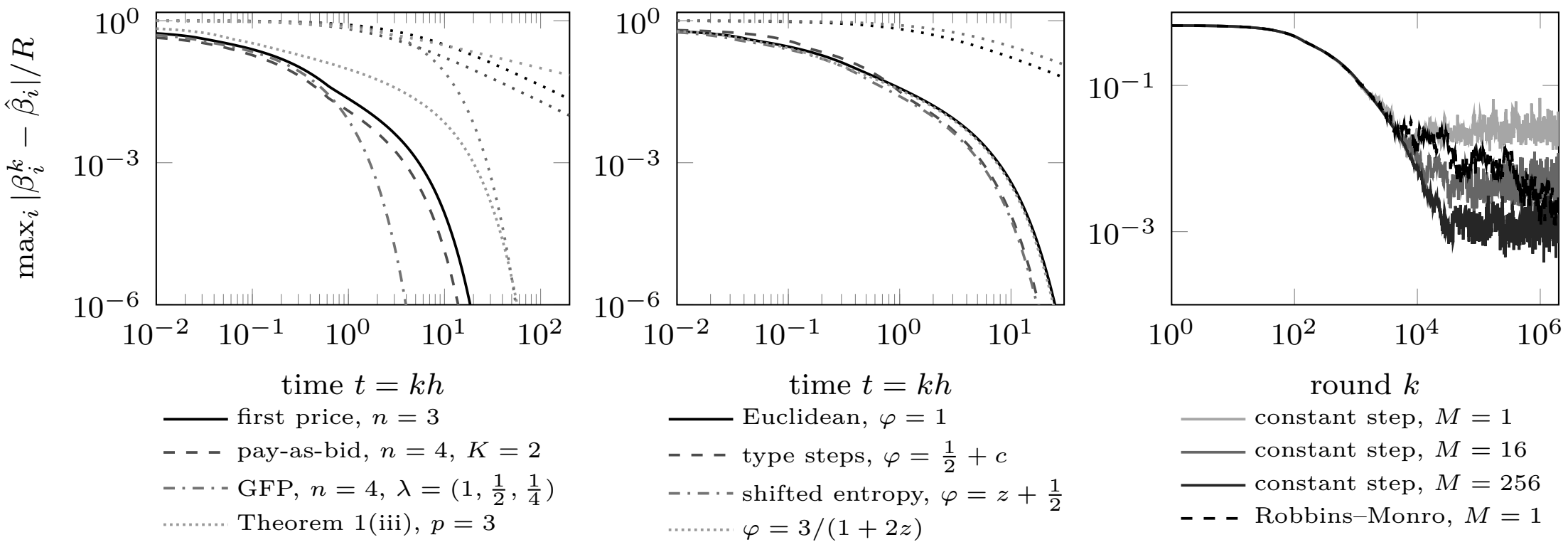
\begin{figure}[t]
\centering
\begin{tikzpicture}
\begin{axis}[
  name=left, width=0.335\linewidth, height=4.3cm,
  xmode=log, ymode=log, xmin=0.01, xmax=200, ymin=1e-6, ymax=1.5,
  xlabel={time $t=kh$}, ylabel={$\max_i|\beta^k_i-\hat\beta_i|/R$},
  tick label style={font=\scriptsize}, label style={font=\scriptsize},
  legend style={font=\tiny, draw=none, fill=none, cells={anchor=west}, at={(0.5,-0.3)}, anchor=north, row sep=-1pt},
  every axis plot/.append style={line width=0.7pt},
]
\addplot[black] table[x=t, y=fpa3] {figures/general_rates.dat};
\addlegendentry{first price, $n=3$}
\addplot[black!70, dashed] table[x=t, y=payasbid] {figures/general_rates.dat};
\addlegendentry{pay-as-bid, $n=4$, $K=2$}
\addplot[black!55, dashdotted] table[x=t, y=gfp] {figures/general_rates.dat};
\addlegendentry{GFP, $n=4$, $\lambda=(1,\frac12,\frac14)$}
\addplot[black!40, densely dotted] table[x=t, y=optimal3] {figures/general_rates.dat};
\addlegendentry{\cref{thm:main}(iii), $p=3$}
\addplot[black, dotted, domain=0.01:200, samples=150, forget plot] {1/(1+(2/9)*x)};
\addplot[black!70, dotted, domain=0.01:200, samples=150, forget plot] {1/(1+0.5*x)};
\addplot[black!55, dotted, domain=0.01:200, samples=150, forget plot] {exp(-0.25*x)};
\addplot[black!40, dotted, domain=0.01:200, samples=150, forget plot] {1/sqrt(1+x)};
\end{axis}
\begin{axis}[
  name=mid, at={(left.east)}, anchor=west, xshift=1.0cm, width=0.32\linewidth, height=4.3cm,
  xmode=log, ymode=log, xmin=0.01, xmax=30, ymin=1e-6, ymax=1.5,
  xlabel={time $t=kh$},
  tick label style={font=\scriptsize}, label style={font=\scriptsize},
  legend style={font=\tiny, draw=none, fill=none, cells={anchor=west}, at={(0.5,-0.3)}, anchor=north, row sep=-1pt},
  every axis plot/.append style={line width=0.7pt},
]
\addplot[black] table[x=t, y=euclid] {figures/geometry_rates.dat};
\addlegendentry{Euclidean, $\varphi=1$}
\addplot[black!70, dashed] table[x=t, y=type] {figures/geometry_rates.dat};
\addlegendentry{type steps, $\varphi=\frac12+c$}
\addplot[black!55, dashdotted] table[x=t, y=entropy] {figures/geometry_rates.dat};
\addlegendentry{shifted entropy, $\varphi=z+\frac12$}
\addplot[black!40, densely dotted] table[x=t, y=decr] {figures/geometry_rates.dat};
\addlegendentry{$\varphi=3/(1+2z)$}
\addplot[black, dotted, forget plot] table[x=t, y=bound_euclid] {figures/geometry_rates.dat};
\addplot[black!55, dotted, forget plot] table[x=t, y=bound_entropy] {figures/geometry_rates.dat};
\end{axis}
\begin{axis}[
  at={(mid.east)}, anchor=west, xshift=1.0cm, width=0.32\linewidth, height=4.3cm,
  xmode=log, ymode=log, xmin=1, xmax=2e6, ymin=1e-4, ymax=1,
  xlabel={round $k$},
  tick label style={font=\scriptsize}, label style={font=\scriptsize},
  legend style={font=\tiny, draw=none, fill=none, cells={anchor=west}, at={(0.5,-0.3)}, anchor=north, row sep=-1pt},
  every axis plot/.append style={line width=0.7pt},
]
\addplot[black!35] table[x=k, y=const1] {figures/stochastic_rates.dat};
\addlegendentry{constant step, $M=1$}
\addplot[black!60] table[x=k, y=const16] {figures/stochastic_rates.dat};
\addlegendentry{constant step, $M=16$}
\addplot[black!85] table[x=k, y=const256] {figures/stochastic_rates.dat};
\addlegendentry{constant step, $M=256$}
\addplot[black, dashed] table[x=k, y=rm] {figures/stochastic_rates.dat};
\addlegendentry{Robbins--Monro, $M=1$}
\end{axis}
\end{tikzpicture}
\caption{\Cref{alg:grid} from complete pooling on grids with $N$ equal steps of the loss coordinate. Left:
four members of \cref{tab:members} without noise ($N=100$, floor $\frac18$, largest CFL step; the first-price auction
has the cost density $2(1+c)/3$, so $a=\frac49$), normalized error $\max_i|\beta^k_i-\hat\beta_i|/R$ and the bounds of \cref{thm:grid}(b) (dotted). Middle and right: error $\max_i|\beta^k_i-\hat\beta_i|$. Middle:
first-price auction with $n=3$ uniform bidders under four metric weights (same grid, floor and step rule); the grid rest
point is the same for every weight, and the dotted lines are the bounds $(1+\varphi_-t/2)^{-1}$ of \cref{thm:grid}(b) for
$\varphi_-=1$ and $\varphi_-=\frac12$. Right: the same auction ($N=50$) under Gaussian gradient noise with standard
deviation $1$ per sample: constant step with mini-batches of $M$ samples, and steps $h_k=h(1+k/10^4)^{-0.7}$
(\cref{thm:stochastic}).}
\label{fig:general}
\end{figure}

The bounds in part~(b), expressed in physical time $kh$, contain no explicit dependence on the mesh and match the rates of \cref{thm:main}(ii), up to the factor $1/2$ in $\varphi_-a$ for $p>1$ and terms that vanish with the error bound. \Cref{fig:general} illustrates these rates for four members of \cref{tab:members} (left) and four metric weights (middle). The slope floor and the positive weight are safeguards of this guarantee.

Both safeguards are needed (\cref{prop:safeguards} in \cref{app:grid}): without the floor, gradient errors of order $1/N$ move the top slope out of every prescribed window, whereas with it \cref{thm:grid}(b) holds for every error bound; and a weight that vanishes on the boundary of $\cK$, that is, a metric that blows up there, such as the Hessian of the entropy of the feasible interval, freezes learning there.
\paragraph{Proof idea.} The CFL condition makes the one-step update monotone (\cref{lem:monotone-update}): raising an
action, or the action of the next type, never lowers the updated action. This is a discrete comparison principle. The
grid rest point solves the backward recursion \eqref{eq:gridrest}, and the clipped shifts $\min\{C,\hat\beta+E\}$ and
$\max\{\mu,\hat\beta-E\}$ are discrete barriers: by a discrete form of the error identity, the field at a type with cost
gap $w_i>E$ restores at rate at least $A_iE/w_i$. Monotonicity keeps the iterates between the barriers, so $E_k$ obeys a
scalar recursion, which gives (b). The floor bounds the singular term, so a fixed step suffices. For (c),
\eqref{eq:gridrest} with $\ell=0$ is the forward-difference discretization of the rest-point equation, and the mean
value theorem bounds its local error.

Parts (a)--(c) of \cref{thm:grid} combine into a guarantee that does not depend on the grid.

\begin{corollary}[Learning the equilibrium]\label{cor:learn}
Let $(A,B,\mu)$ satisfy \cref{ass:main} and $\varphi$ \cref{ass:metric}. For every $\varepsilon>0$ there are $\ell>0$, $\delta>0$, $\eta_0>0$ and $T\ge0$
such that the iterates of \cref{alg:grid} satisfy $\max_i|\beta^k_i-\bstar(c_i)|\le\varepsilon$ whenever $kh\ge T$,
for every grid of mesh at most $\delta$, every step that satisfies \eqref{eq:cfl} for $\eta_0$, all errors with $|\xi^k_i|\le\eta_0$, and
every feasible initial vector.
\end{corollary}

The next theorem connects the algorithm with \cref{thm:main}: as step, mesh, floor, and error bound vanish, the
iterates converge to the solution. A profile $b$ that is not weakly increasing is first replaced by its lower monotone
envelope $b^\flat(c)=\min_{s\in[c,1]}b(s)$; the scheme selects this adjustment, which is not the payoff gradient at $b$
(\cref{app:grid}), so convergence is then uniform only away from time zero.

\begin{theorem}[Convergence to the solution]\label{thm:scheme}
Let $(A,B,\mu)$ satisfy \cref{ass:main} and $\varphi$ \cref{ass:metric}, and let $b$ be continuous with $\mu\le b\le C$. Then $b^\flat$ is
feasible; let $\beta$ be the solution from $b^\flat$. For $N\ge1$ let $0=c^N_0<\dots<c^N_N=1$ be grids whose mesh tends to $0$, let
$\ell_N\to0$ and $\eta_N\to0$, let $|\xi^k_i|\le\eta_N$, and let $h_N$ satisfy \eqref{eq:cfl} for $\eta_N$. The iterates of \cref{alg:grid} from
$\beta^{N,0}_i=b(c^N_i)$ satisfy, for all $0<T_0\le T$,
\[
\max\bigl\{|\beta^{N,k}_i-\beta(kh_N,c^N_i)|:\ T_0\le kh_N\le T,\ 0\le i\le N\bigr\}\to0\qquad(N\to\infty),
\]
and also for $T_0=0$ if $b$ is weakly increasing.
\end{theorem}

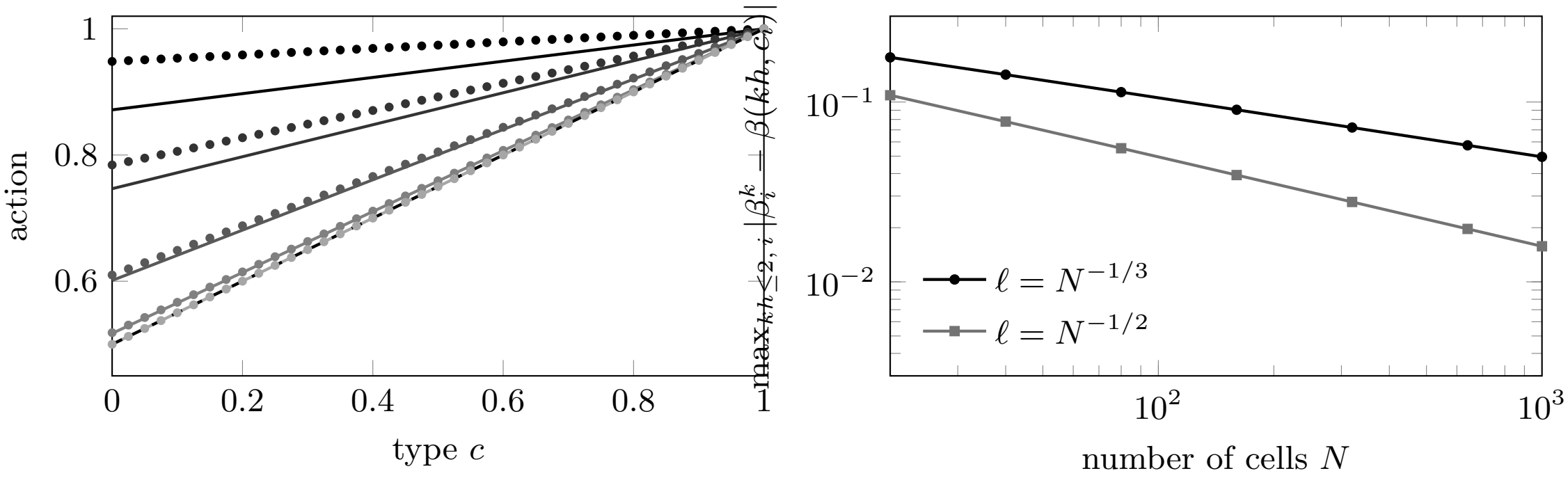
\begin{figure}[t]
\centering
\begin{tikzpicture}
\begin{axis}[
  name=left, width=0.48\linewidth, height=5cm,
  xmin=0, xmax=1, ymin=0.45, ymax=1.02,
  xlabel={type $c$}, ylabel={action},
  tick label style={font=\footnotesize}, label style={font=\footnotesize},
  every axis plot/.append style={line width=0.8pt},
]
\addplot[black] table[x=c, y=ex0] {figures/pooled_profiles.dat};
\addplot[black!80] table[x=c, y=ex1] {figures/pooled_profiles.dat};
\addplot[black!65] table[x=c, y=ex2] {figures/pooled_profiles.dat};
\addplot[black!50] table[x=c, y=ex3] {figures/pooled_profiles.dat};
\addplot[black!35] table[x=c, y=ex4] {figures/pooled_profiles.dat};
\addplot[black, only marks, mark=*, mark size=0.8pt] table[x=c, y=gr0] {figures/pooled_grid.dat};
\addplot[black!80, only marks, mark=*, mark size=0.8pt] table[x=c, y=gr1] {figures/pooled_grid.dat};
\addplot[black!65, only marks, mark=*, mark size=0.8pt] table[x=c, y=gr2] {figures/pooled_grid.dat};
\addplot[black!50, only marks, mark=*, mark size=0.8pt] table[x=c, y=gr3] {figures/pooled_grid.dat};
\addplot[black!35, only marks, mark=*, mark size=0.8pt] table[x=c, y=gr4] {figures/pooled_grid.dat};
\addplot[black, dashed, domain=0:1] {(1+x)/2};
\end{axis}
\begin{axis}[
  at={(left.east)}, anchor=west, xshift=1.2cm, width=0.48\linewidth, height=5cm,
  xmode=log, ymode=log, xmin=20, xmax=1000, ymin=3e-3, ymax=0.3,
  xlabel={number of cells $N$}, ylabel={$\max_{kh\le2,\,i}|\beta^k_i-\beta(kh,c_i)|$},
  legend pos=south west, legend style={font=\footnotesize, draw=none, fill=none},
  tick label style={font=\footnotesize}, label style={font=\footnotesize},
  every axis plot/.append style={line width=0.8pt},
]
\addplot[black, mark=*, mark size=1pt] table[x=N, y=err13] {figures/pooled_convergence.dat};
\addlegendentry{$\ell=N^{-1/3}$}
\addplot[black!55, mark=square*, mark size=1pt] table[x=N, y=err12] {figures/pooled_convergence.dat};
\addlegendentry{$\ell=N^{-1/2}$}
\end{axis}
\end{tikzpicture}
\caption{Complete pooling in the uniform duopoly (\cref{ex:pooled}). Left: the solution at $t=0.01$, $0.05$, $0.2$, $0.6$,
$2$ (lines, dark to light), \cref{alg:grid} with $N=40$, floor $\ell=N^{-1/2}$ and step $\ell^2/N$ (dots), and the equilibrium
$(1+c)/2$ (dashed). Right: largest error of \cref{alg:grid} against the solution on $[0,2]$ as the floor vanishes
(\cref{thm:scheme}).}
\label{fig:pooled}
\end{figure}

For weakly increasing initial data the proof follows \citet{barles1991convergence} (\cref{lem:scheme-feasible}): by
monotonicity of the update, the upper and lower half-relaxed limits of the iterates are a subsolution and a
supersolution; barriers at $t=0$ show that both start from $b$, and \cref{lem:comparison} makes them equal. This also
gives existence in \cref{thm:main}(i). For a profile $b$ that is not weakly increasing, the iterates from $b^\flat\le b$
give a lower bound by monotonicity, and a barrier on the higher types shows that within a time of order $\ell$ the scheme
lowers each action to at most the initial actions of higher types, which gives the upper bound away from $t=0$. The floor must vanish: with a fixed floor the iterates lag behind the solution while its slope is below the floor (\cref{fig:pooled}). The convergence holds on every sequence of grids and for every weight that satisfies \cref{ass:metric}.

% EC27 revision: unnumbered illustration; all original theorem numbers are retained.

\begin{remark}[Densities with jumps]\label{rem:density}
Continuity of $B$ and $B(0)>0$ are not needed. Let $B$ be integrable with $\int_{c'}^cB>0$ for $c'<c$, as for type densities with
jumps or for $B(0)=0$. The change of type $c\mapsto X(c)/X(1)$ gives coefficients that satisfy \cref{ass:main}, with $B$ replaced by
the constant $X(1)$ and the same values of $A$, $w$ and $R$ at corresponding types, and it changes neither \eqref{eq:GF} nor
\cref{alg:grid}. For continuous $B$ the solutions in the two coordinates agree. For all such $B$, therefore, the solution
exists and is unique, its error is weakly decreasing and satisfies \eqref{eq:power}, \cref{thm:grid,thm:scheme} hold, and in
allocation games the rest point is a symmetric Bayes--Nash equilibrium against every action.
\end{remark}
\paragraph{Stochastic gradients.}
The grid scheme is deterministic when the regularized field $G_i(\beta^k)$ is evaluated exactly. We now allow stochastic gradient estimates
\[
\widehat G_i^k=G_i(\beta^k)+\xi_i^k
\]
and study convergence on a fixed grid and floor. There is one estimate per grid type and round, applied to the common strategy. The next theorem gives almost-sure convergence under martingale noise and finite-horizon guarantees under tail bounds.

\begin{theorem}[Stochastic gradients]\label{thm:stochastic}
Let $(A,B,\mu)$ satisfy \cref{ass:main} and $\varphi$ \cref{ass:metric}, fix a grid and a floor, let the deterministic steps $h_k\ge0$ satisfy
\eqref{eq:cfl} for $\eta=0$, and let $\beta^0$ be feasible.
\begin{enumerate}
\item[(a)] If $\sum_kh_k=\infty$ and the series $\sum_kh_k\varphi_i(\beta^k_i)\xi^k_i$ converges for every $i<N$, then $\beta^k\to\hat\beta$.
\item[(b)] \emph{Robbins--Monro steps} \citep{robbins1951stochastic}. Let $\xi^k_i$ be square integrable and $\mathcal F_{k+1}$-measurable for a filtration $(\mathcal F_k)$, with $\mathbb E[\xi^k_i\mid\mathcal F_k]=0$ and
$\mathbb E[(\xi^k_i)^2]\le\sigma^2$ for a constant $\sigma$. If $\sum_kh_k=\infty$ and $\sum_kh_k^2<\infty$, then $\beta^k\to\hat\beta$ almost surely.
\item[(c)] Let $h_k=h$ satisfy \eqref{eq:cfl} for $\eta$. If $\Pr[|\xi^k_i|>\eta]\le\varpi$ for all $k<k_1$ and $i<N$, then with probability at least
$1-Nk_1\varpi$ the bounds of \cref{thm:grid}(b) hold for all $k\le k_1$.
\end{enumerate}
\end{theorem}

Thus unbiased estimates with bounded variance and decreasing steps learn the grid rest point almost surely (b); reaching the continuum rest point also requires the floor and mesh to vanish, as in \cref{thm:grid}(a),(c). With a constant step, the deterministic guarantee survives with high probability (c): for averages of $M$ independent sub-Gaussian samples with variance proxy $\sigma^2$, \citet{chernoff1952measure} gives $\varpi=2\e^{-M\eta^2/(2\sigma^2)}$, so $M=\lceil2\sigma^2\eta^{-2}\log(2Nk_1/\vartheta)\rceil$ samples per estimate give confidence $1-\vartheta$. \Cref{fig:general} (right) illustrates both regimes.

\paragraph{First-price auctions.} For first-price auctions the constants of \cref{thm:grid} are explicit, and they bound
the number of rounds.

\begin{corollary}[Rounds in first-price auctions]\label{cor:rounds}
In the setting of \cref{cor:auction}, run \cref{alg:grid} with $\varphi\equiv1$ in procurement coordinates from a feasible initial vector.
\begin{enumerate}
\item[(a)] On a grid of mesh $\delta$, with $\ell\le1/(2(n-1)f_+)$, errors $|\xi^k_i|\le\eta$ and a step that satisfies \eqref{eq:cfl} for $\eta$,
\[
\max_i|\beta^k_i-\bstar(c_i)|\le\frac{f_+}{f_-}\,\delta+
\begin{cases}
(1-hf_-)^k+\eta/f_-, & n=2,\\[1mm]
\max\Bigl\{\bigl(2\eta/f_-^{\,n-1}\bigr)^{1/(n-1)},\ \bigl(1+\tfrac{n-2}2f_-^{\,n-1}kh\bigr)^{-1/(n-2)}\Bigr\}, & n\ge3 .
\end{cases}
\]
\item[(b)] Let $0<\varepsilon\le1$, $N=\lceil3f_+/(f_-\varepsilon)\rceil$, $c_i=i/N$, $\ell=1/(2(n-1)f_+)$, $h=\ell^2(f_-/N)^{n-1}$, and
$\eta\le\frac12f_-^{\,n-1}(\varepsilon/3)^{n-1}$ (with $\eta\le f_-\varepsilon/3$ if $n=2$). Then there is $k_\varepsilon$ with $\max_i|\beta^k_i-\bstar(c_i)|\le\varepsilon$ for all
$k\ge k_\varepsilon$ and
\[
\begin{aligned}
k_\varepsilon&\le16\Bigl(\frac{f_+}{f_-}\Bigr)^3\frac{\log(3/\varepsilon)}\varepsilon+1\quad(n=2),\\
k_\varepsilon&\le\frac{2\cdot3^{n-2}4^n(n-1)^2f_+^{\,n+1}}{(n-2)\,f_-^{\,3(n-1)}}\,\varepsilon^{-(2n-3)}+1\quad(n\ge3).
\end{aligned}
\]
\end{enumerate}
\end{corollary}

The three factors of $k_\varepsilon$ have distinct sources: the continuous rate of order $\varepsilon^{-(n-2)}$, the mesh of
order $\varepsilon$, and the step of order $N^{-(n-1)}$ imposed by the CFL condition at the top type.

%=====================================================================
\section{Conclusion}\label{sec:conclusion}
%=====================================================================

The membership characterization identifies a shared economic structure across the bidding and
pricing games considered here. Its singular transport--damping field admits comparison barriers
that contract the uniform strategy error. Together with the equilibrium conditions and the grid
analysis, this yields explicit last-iterate guarantees for the specified symmetric learning dynamics.

The main open question is learning by players who estimate their payoff fields independently: their strategies
may differ, each velocity depends on the rivals' slopes, and \cref{lem:comparison} does not apply.
Symmetric primitives alone do not remove this difficulty when initialization, noise, or learning rates
differ across players. A further direction is to connect the stated error assumptions to concrete
payoff-observation models, controlling statistical bias, discretization, and regularization together.

\clearpage
\phantomsection
\addcontentsline{toc}{section}{References}
\bibliographystyle{plainnat}
\bibliography{references}

\begin{thebibliography}{70}
\providecommand{\natexlab}[1]{#1}
\providecommand{\url}[1]{\texttt{#1}}
\expandafter\ifx\csname urlstyle\endcsname\relax
  \providecommand{\doi}[1]{doi: #1}\else
  \providecommand{\doi}{doi: \begingroup \urlstyle{rm}\Url}\fi

\bibitem[Ahunbay and Bichler(2025{\natexlab{a}})]{ahunbay2025semicoarse}
Mete~{\c{S}}eref Ahunbay and Martin Bichler.
\newblock Semicoarse correlated equilibria and {LP}-based guarantees for
  gradient dynamics in normal-form games.
\newblock In \emph{Proceedings of the 26th ACM Conference on Economics and
  Computation (EC '25)}, page~91. ACM, 2025{\natexlab{a}}.
\newblock \doi{10.1145/3736252.3742500}.

\bibitem[Ahunbay and Bichler(2025{\natexlab{b}})]{ahunbay2025uniqueness}
Mete~{\c{S}}eref Ahunbay and Martin Bichler.
\newblock On the uniqueness of {Bayesian} coarse correlated equilibria in
  standard first-price and all-pay auctions.
\newblock In \emph{Proceedings of the 2025 Annual ACM-SIAM Symposium on
  Discrete Algorithms (SODA)}, pages 2491--2537. SIAM, 2025{\natexlab{b}}.
\newblock \doi{10.1137/1.9781611978322.82}.

\bibitem[Ahunbay et~al.(2026)Ahunbay, Zheng, and Lin]{ahunbay2026gradient}
Mete~{\c{S}}eref Ahunbay, Weiqiang Zheng, and Tao Lin.
\newblock Gradient dynamics in first-price auctions: Iterative strategy
  elimination via cubic potentials, 2026.
\newblock Accepted to the 27th ACM Conference on Economics and Computation
  (EC'26).

\bibitem[Armantier et~al.(2008)Armantier, Florens, and
  Richard]{armantier2008approximation}
Olivier Armantier, Jean-Pierre Florens, and Jean-Francois Richard.
\newblock Approximation of {Nash} equilibria in {Bayesian} games.
\newblock \emph{Journal of Applied Econometrics}, 23\penalty0 (7):\penalty0
  965--981, 2008.
\newblock \doi{10.1002/jae.1040}.

\bibitem[Athey(2001)]{athey2001single}
Susan Athey.
\newblock Single crossing properties and the existence of pure strategy
  equilibria in games of incomplete information.
\newblock \emph{Econometrica}, 69\penalty0 (4):\penalty0 861--889, 2001.
\newblock \doi{10.1111/1468-0262.00223}.

\bibitem[Banchio and Skrzypacz(2022)]{banchio2022artificial}
Martino Banchio and Andrzej Skrzypacz.
\newblock Artificial intelligence and auction design.
\newblock In \emph{Proceedings of the 23rd ACM Conference on Economics and
  Computation (EC '22)}, pages 30--31. ACM, 2022.
\newblock \doi{10.1145/3490486.3538244}.

\bibitem[Barles and Perthame(1987)]{barles1987discontinuous}
Guy Barles and Beno{\^\i}t Perthame.
\newblock Discontinuous solutions of deterministic optimal stopping time
  problems.
\newblock \emph{ESAIM: Mathematical Modelling and Numerical Analysis},
  21\penalty0 (4):\penalty0 557--579, 1987.

\bibitem[Barles and Souganidis(1991)]{barles1991convergence}
Guy Barles and Panagiotis~E. Souganidis.
\newblock Convergence of approximation schemes for fully nonlinear second order
  equations.
\newblock \emph{Asymptotic Analysis}, 4\penalty0 (3):\penalty0 271--283, 1991.
\newblock \doi{10.3233/ASY-1991-4305}.

\bibitem[Beck and Teboulle(2003)]{beck2003mirror}
Amir Beck and Marc Teboulle.
\newblock Mirror descent and nonlinear projected subgradient methods for convex
  optimization.
\newblock \emph{Operations Research Letters}, 31\penalty0 (3):\penalty0
  167--175, 2003.
\newblock \doi{10.1016/S0167-6377(02)00231-6}.

\bibitem[Bichler and Hoehener(2026)]{bichler2026stability}
Martin Bichler and Jan-Sebastian Hoehener.
\newblock Stability beyond {Minty}: Numerical {Lyapunov} certification for
  equilibrium learning in {Bayesian} {Bertrand} competition, 2026.
\newblock v1 17 May 2026; v2 5 October 2026.

\bibitem[Bichler and Parkes(2025)]{bichler2025differentiable}
Martin Bichler and David~C. Parkes.
\newblock Differentiable economics: Strategic behavior, mechanisms, and machine
  learning.
\newblock \emph{Communications of the ACM}, 68\penalty0 (9):\penalty0 80--88,
  2025.
\newblock \doi{10.1145/3725809}.

\bibitem[Bichler et~al.(2021)Bichler, Fichtl, Heidekr{\"u}ger, Kohring, and
  Sutterer]{bichler2021learning}
Martin Bichler, Maximilian Fichtl, Stefan Heidekr{\"u}ger, Nils Kohring, and
  Paul Sutterer.
\newblock Learning equilibria in symmetric auction games using artificial
  neural networks.
\newblock \emph{Nature Machine Intelligence}, 3\penalty0 (8):\penalty0
  687--695, 2021.
\newblock \doi{10.1038/s42256-021-00365-4}.

\bibitem[Bichler et~al.(2023{\natexlab{a}})Bichler, Kohring, and
  Heidekr{\"u}ger]{bichler2023learning}
Martin Bichler, Nils Kohring, and Stefan Heidekr{\"u}ger.
\newblock Learning equilibria in asymmetric auction games.
\newblock \emph{INFORMS Journal on Computing}, 35\penalty0 (3):\penalty0
  523--542, 2023{\natexlab{a}}.
\newblock \doi{10.1287/ijoc.2023.1281}.

\bibitem[Bichler et~al.(2023{\natexlab{b}})Bichler, Kohring, Oberlechner, and
  Pieroth]{bichler2023bargaining}
Martin Bichler, Nils Kohring, Matthias Oberlechner, and Fabian~R. Pieroth.
\newblock Learning equilibrium in bilateral bargaining games.
\newblock \emph{European Journal of Operational Research}, 311\penalty0
  (2):\penalty0 660--678, 2023{\natexlab{b}}.
\newblock \doi{10.1016/j.ejor.2022.12.022}.

\bibitem[Bichler et~al.(2025{\natexlab{a}})Bichler, Fichtl, and
  Oberlechner]{bichler2025computing}
Martin Bichler, Max Fichtl, and Matthias Oberlechner.
\newblock Computing {Bayes}--{Nash} equilibrium strategies in auction games via
  simultaneous online dual averaging.
\newblock \emph{Operations Research}, 73\penalty0 (2):\penalty0 1102--1127,
  2025{\natexlab{a}}.
\newblock \doi{10.1287/opre.2022.0287}.

\bibitem[Bichler et~al.(2025{\natexlab{b}})Bichler, Legacci, Mertikopoulos,
  Oberlechner, and Pradelski]{bichler2025potentialness}
Martin Bichler, Davide Legacci, Panayotis Mertikopoulos, Matthias Oberlechner,
  and Bary Pradelski.
\newblock Characterizing the convergence of game dynamics via potentialness.
\newblock \emph{Transactions on Machine Learning Research}, 2025{\natexlab{b}}.

\bibitem[Bichler et~al.(2025{\natexlab{c}})Bichler, Lunowa, Oberlechner,
  Pieroth, and Wohlmuth]{bichler2025beyond}
Martin Bichler, Stephan~B. Lunowa, Matthias Oberlechner, Fabian~R. Pieroth, and
  Barbara Wohlmuth.
\newblock Beyond monotonicity: On the convergence of learning algorithms in
  standard auction games.
\newblock In \emph{Proceedings of the AAAI Conference on Artificial
  Intelligence}, volume~39, pages 13649--13657, 2025{\natexlab{c}}.
\newblock \doi{10.1609/aaai.v39i13.33491}.

\bibitem[B{\"o}hm(2023)]{bohm2023solving}
Axel B{\"o}hm.
\newblock Solving nonconvex-nonconcave min-max problems exhibiting weak {Minty}
  solutions.
\newblock \emph{Transactions on Machine Learning Research}, 2023.

\bibitem[Cai et~al.(2022)Cai, Oikonomou, and Zheng]{cai2022finite}
Yang Cai, Argyris Oikonomou, and Weiqiang Zheng.
\newblock Finite-time last-iterate convergence for learning in multi-player
  games.
\newblock In \emph{Advances in Neural Information Processing Systems},
  volume~35, pages 33904--33919, 2022.
\newblock \doi{10.52202/068431-2457}.

\bibitem[Chernoff(1952)]{chernoff1952measure}
Herman Chernoff.
\newblock A measure of asymptotic efficiency for tests of a hypothesis based on
  the sum of observations.
\newblock \emph{The Annals of Mathematical Statistics}, 23\penalty0
  (4):\penalty0 493--507, 1952.

\bibitem[Courant et~al.(1928)Courant, Friedrichs, and
  Lewy]{courant1928partiellen}
Richard Courant, Kurt Friedrichs, and Hans Lewy.
\newblock {\"U}ber die partiellen {Differenzengleichungen} der mathematischen
  {Physik}.
\newblock \emph{Mathematische Annalen}, 100:\penalty0 32--74, 1928.

\bibitem[Courant et~al.(1952)Courant, Isaacson, and Rees]{courant1952solution}
Richard Courant, Eugene Isaacson, and Mina Rees.
\newblock On the solution of nonlinear hyperbolic differential equations by
  finite differences.
\newblock \emph{Communications on Pure and Applied Mathematics}, 5\penalty0
  (3):\penalty0 243--255, 1952.

\bibitem[Crandall and Lions(1983)]{crandall1983viscosity}
Michael~G. Crandall and Pierre-Louis Lions.
\newblock Viscosity solutions of {Hamilton}-{Jacobi} equations.
\newblock \emph{Transactions of the American Mathematical Society},
  277\penalty0 (1):\penalty0 1--42, 1983.
\newblock \doi{10.1090/S0002-9947-1983-0690039-8}.

\bibitem[Crandall et~al.(1992)Crandall, Ishii, and Lions]{crandall1992users}
Michael~G. Crandall, Hitoshi Ishii, and Pierre-Louis Lions.
\newblock User's guide to viscosity solutions of second order partial
  differential equations.
\newblock \emph{Bulletin of the American Mathematical Society}, 27\penalty0
  (1):\penalty0 1--67, 1992.
\newblock \doi{10.1090/S0273-0979-1992-00266-5}.

\bibitem[Daskalakis and Panageas(2019)]{daskalakis2019last}
Constantinos Daskalakis and Ioannis Panageas.
\newblock Last-iterate convergence: Zero-sum games and constrained min-max
  optimization.
\newblock In \emph{10th Innovations in Theoretical Computer Science Conference
  (ITCS 2019)}, volume 124 of \emph{Leibniz International Proceedings in
  Informatics (LIPIcs)}, pages 27:1--27:18. Schloss Dagstuhl--Leibniz-Zentrum
  f{\"u}r Informatik, 2019.
\newblock \doi{10.4230/LIPIcs.ITCS.2019.27}.

\bibitem[Deng et~al.(2022)Deng, Hu, Lin, and Zheng]{deng2022nash}
Xiaotie Deng, Xinyan Hu, Tao Lin, and Weiqiang Zheng.
\newblock Nash convergence of mean-based learning algorithms in first price
  auctions.
\newblock In \emph{Proceedings of the ACM Web Conference 2022 (WWW '22)}, pages
  141--150. ACM, 2022.
\newblock \doi{10.1145/3485447.3512059}.

\bibitem[Diakonikolas et~al.(2021)Diakonikolas, Daskalakis, and
  Jordan]{diakonikolas2021efficient}
Jelena Diakonikolas, Constantinos Daskalakis, and Michael~I. Jordan.
\newblock Efficient methods for structured nonconvex-nonconcave min-max
  optimization.
\newblock In \emph{Proceedings of The 24th International Conference on
  Artificial Intelligence and Statistics (AISTATS)}, volume 130 of
  \emph{Proceedings of Machine Learning Research}, pages 2746--2754. PMLR,
  2021.

\bibitem[Doob(1953)]{doob1953stochastic}
Joseph~L. Doob.
\newblock \emph{Stochastic Processes}.
\newblock Wiley, New York, 1953.

\bibitem[Edelman et~al.(2007)Edelman, Ostrovsky, and
  Schwarz]{edelman2007internet}
Benjamin Edelman, Michael Ostrovsky, and Michael Schwarz.
\newblock Internet advertising and the generalized second-price auction:
  Selling billions of dollars worth of keywords.
\newblock \emph{American Economic Review}, 97\penalty0 (1):\penalty0 242--259,
  2007.

\bibitem[Feng et~al.(2021)Feng, Guruganesh, Liaw, Mehta, and
  Sethi]{feng2021convergence}
Zhe Feng, Guru Guruganesh, Christopher Liaw, Aranyak Mehta, and Abhishek Sethi.
\newblock Convergence analysis of no-regret bidding algorithms in repeated
  auctions.
\newblock In \emph{Proceedings of the AAAI Conference on Artificial
  Intelligence}, volume~35, pages 5399--5406, 2021.
\newblock \doi{10.1609/aaai.v35i6.16680}.

\bibitem[Fibich and Gavish(2012)]{fibich2012asymmetric}
Gadi Fibich and Nir Gavish.
\newblock Asymmetric first-price auctions---a dynamical-systems approach.
\newblock \emph{Mathematics of Operations Research}, 37\penalty0 (2):\penalty0
  219--243, 2012.
\newblock \doi{10.1287/moor.1110.0535}.

\bibitem[Gamelin et~al.(2001)Gamelin, Gehring, and Halmos]{gamelin2001complex}
Theodore~W Gamelin, FW~Gehring, and PR~Halmos.
\newblock \emph{Complex analysis}, volume~1.
\newblock Springer, New York, 2001.

\bibitem[Golowich et~al.(2020)Golowich, Pattathil, and
  Daskalakis]{golowich2020tight}
Noah Golowich, Sarath Pattathil, and Constantinos Daskalakis.
\newblock Tight last-iterate convergence rates for no-regret learning in
  multi-player games.
\newblock In \emph{Advances in Neural Information Processing Systems},
  volume~33, 2020.

\bibitem[Gorbunov et~al.(2023)Gorbunov, Taylor, Horv{\'a}th, and
  Gidel]{gorbunov2023convergence}
Eduard Gorbunov, Adrien Taylor, Samuel Horv{\'a}th, and Gauthier Gidel.
\newblock Convergence of proximal point and extragradient-based methods beyond
  monotonicity: the case of negative comonotonicity.
\newblock In \emph{Proceedings of the 40th International Conference on Machine
  Learning (ICML)}, volume 202 of \emph{Proceedings of Machine Learning
  Research}, 2023.

\bibitem[Heidekr{\"u}ger et~al.(2021)Heidekr{\"u}ger, Sutterer, Kohring,
  Fichtl, and Bichler]{heidekruger2021equilibrium}
Stefan Heidekr{\"u}ger, Paul Sutterer, Nils Kohring, Maximilian Fichtl, and
  Martin Bichler.
\newblock Equilibrium learning in combinatorial auctions: Computing approximate
  {Bayesian} {Nash} equilibria via pseudogradient dynamics, 2021.

\bibitem[Heymann(2026)]{heymann2026tiebreaking}
Benjamin Heymann.
\newblock On the role of tie-breaking rules in the convergence of fictitious
  play for symmetric first-price auctions, 2026.

\bibitem[Hirsch(1988)]{hirsch1988stability}
Morris~W. Hirsch.
\newblock Stability and convergence in strongly monotone dynamical systems.
\newblock \emph{Journal f{\"u}r die reine und angewandte Mathematik},
  383:\penalty0 1--53, 1988.

\bibitem[Hoang et~al.(2019)Hoang, Soumis, and Zaccour]{hoang2019return}
L{\^e}~Nguy{\^e}n Hoang, Fran{\c{c}}ois Soumis, and Georges Zaccour.
\newblock The return function: A new computable perspective on
  {Bayesian}--{Nash} equilibria.
\newblock \emph{European Journal of Operational Research}, 279\penalty0
  (2):\penalty0 471--485, 2019.
\newblock \doi{10.1016/j.ejor.2019.05.036}.

\bibitem[Hubbard and Paarsch(2014)]{hubbard2014numerical}
Timothy~P. Hubbard and Harry~J. Paarsch.
\newblock On the numerical solution of equilibria in auction models with
  asymmetries within the private-values paradigm.
\newblock In \emph{Handbook of Computational Economics}, volume~3, pages
  37--115. Elsevier, 2014.
\newblock \doi{10.1016/B978-0-444-52980-0.00002-5}.

\bibitem[Ipsen and Skoulakis(2026)]{ipsen2026revenue}
Anders~Bo Ipsen and Stratis Skoulakis.
\newblock Revenue guarantees of no-swap-regret dynamics in first price
  auctions, 2026.
\newblock To appear in ICML 2026 (arXiv comment).

\bibitem[Krishna(2010)]{krishna2010auction}
Vijay Krishna.
\newblock \emph{Auction Theory}.
\newblock Academic Press, Burlington, MA, 2nd edition, 2010.
\newblock ISBN 978-0-12-374507-1.

\bibitem[Lebrun(2006)]{lebrun2006uniqueness}
Bernard Lebrun.
\newblock Uniqueness of the equilibrium in first-price auctions.
\newblock \emph{Games and Economic Behavior}, 55\penalty0 (1):\penalty0
  131--151, 2006.
\newblock \doi{10.1016/j.geb.2005.01.006}.

\bibitem[Lyapunov(1992)]{lyapunov1992general}
Aleksandr~Mikhailovich Lyapunov.
\newblock The general problem of the stability of motion.
\newblock \emph{International Journal of Control}, 55\penalty0 (3):\penalty0
  531--773, 1992.
\newblock Translated by A.~T. Fuller from the 1892 Russian original.

\bibitem[Marshall et~al.(1994)Marshall, Meurer, Richard, and
  Stromquist]{marshall1994numerical}
Robert~C. Marshall, Michael~J. Meurer, Jean-Francois Richard, and Walter
  Stromquist.
\newblock Numerical analysis of asymmetric first price auctions.
\newblock \emph{Games and Economic Behavior}, 7\penalty0 (2):\penalty0
  193--220, 1994.
\newblock \doi{10.1006/game.1994.1045}.

\bibitem[Martin and Sandholm(2023)]{martin2023finding}
Carlos Martin and Tuomas Sandholm.
\newblock Finding mixed-strategy equilibria of continuous-action games without
  gradients using randomized policy networks.
\newblock In \emph{Proceedings of the Thirty-Second International Joint
  Conference on Artificial Intelligence (IJCAI)}, pages 2844--2852, 2023.
\newblock \doi{10.24963/ijcai.2023/317}.

\bibitem[Martin and Sandholm(2025)]{martin2025joint}
Carlos Martin and Tuomas Sandholm.
\newblock Joint-perturbation simultaneous pseudo-gradient.
\newblock In \emph{Proceedings of the Thirty-Fourth International Joint
  Conference on Artificial Intelligence (IJCAI)}, pages 3996--4004, 2025.
\newblock \doi{10.24963/ijcai.2025/445}.

\bibitem[Maskin and Riley(2003)]{maskin2003uniqueness}
Eric Maskin and John Riley.
\newblock Uniqueness of equilibrium in sealed high-bid auctions.
\newblock \emph{Games and Economic Behavior}, 45\penalty0 (2):\penalty0
  395--409, 2003.
\newblock \doi{10.1016/S0899-8256(03)00150-7}.

\bibitem[Mazumdar et~al.(2020)Mazumdar, Ratliff, and
  Sastry]{mazumdar2020gradient}
Eric Mazumdar, Lillian~J. Ratliff, and S.~Shankar Sastry.
\newblock On gradient-based learning in continuous games.
\newblock \emph{SIAM Journal on Mathematics of Data Science}, 2\penalty0
  (1):\penalty0 103--131, 2020.
\newblock \doi{10.1137/18M1231298}.

\bibitem[McAfee and McMillan(1987)]{mcafee1987auctions}
R.~Preston McAfee and John McMillan.
\newblock Auctions with a stochastic number of bidders.
\newblock \emph{Journal of Economic Theory}, 43\penalty0 (1):\penalty0 1--19,
  1987.

\bibitem[Mertikopoulos and Zhou(2019)]{mertikopoulos2019learning}
Panayotis Mertikopoulos and Zhengyuan Zhou.
\newblock Learning in games with continuous action sets and unknown payoff
  functions.
\newblock \emph{Mathematical Programming}, 173\penalty0 (1--2):\penalty0
  465--507, 2019.
\newblock \doi{10.1007/s10107-018-1254-8}.

\bibitem[Mertikopoulos et~al.(2018)Mertikopoulos, Papadimitriou, and
  Piliouras]{mertikopoulos2018cycles}
Panayotis Mertikopoulos, Christos Papadimitriou, and Georgios Piliouras.
\newblock Cycles in adversarial regularized learning.
\newblock In \emph{Proceedings of the Twenty-Ninth Annual ACM-SIAM Symposium on
  Discrete Algorithms (SODA)}, pages 2703--2717. SIAM, 2018.
\newblock \doi{10.1137/1.9781611975031.172}.

\bibitem[Milgrom and Roberts(1990)]{milgrom1990rationalizability}
Paul Milgrom and John Roberts.
\newblock Rationalizability, learning, and equilibrium in games with strategic
  complementarities.
\newblock \emph{Econometrica}, 58\penalty0 (6):\penalty0 1255--1277, 1990.

\bibitem[Milgrom and Segal(2002)]{milgrom2002envelope}
Paul Milgrom and Ilya Segal.
\newblock Envelope theorems for arbitrary choice sets.
\newblock \emph{Econometrica}, 70\penalty0 (2):\penalty0 583--601, 2002.

\bibitem[Milgrom and Weber(1982)]{milgrom1982theory}
Paul~R. Milgrom and Robert~J. Weber.
\newblock A theory of auctions and competitive bidding.
\newblock \emph{Econometrica}, 50\penalty0 (5):\penalty0 1089--1122, 1982.

\bibitem[Milionis et~al.(2023)Milionis, Papadimitriou, Piliouras, and
  Spendlove]{milionis2023impossibility}
Jason Milionis, Christos Papadimitriou, Georgios Piliouras, and Kelly
  Spendlove.
\newblock An impossibility theorem in game dynamics.
\newblock \emph{Proceedings of the National Academy of Sciences}, 120\penalty0
  (41):\penalty0 e2305349120, 2023.
\newblock \doi{10.1073/pnas.2305349120}.

\bibitem[Minty(1962)]{minty1962monotone}
George~J. Minty.
\newblock Monotone (nonlinear) operators in {Hilbert} space.
\newblock \emph{Duke Mathematical Journal}, 29\penalty0 (3):\penalty0 341--346,
  1962.

\bibitem[Myerson(1981)]{myerson1981optimal}
Roger~B. Myerson.
\newblock Optimal auction design.
\newblock \emph{Mathematics of Operations Research}, 6\penalty0 (1):\penalty0
  58--73, 1981.

\bibitem[Nemirovski and Yudin(1983)]{nemirovski1983problem}
Arkadi~S. Nemirovski and David~B. Yudin.
\newblock \emph{Problem Complexity and Method Efficiency in Optimization}.
\newblock Wiley, New York, 1983.

\bibitem[Pethick et~al.(2022)Pethick, Latafat, Patrinos, Fercoq, and
  Cevher]{pethick2022escaping}
Thomas Pethick, Puya Latafat, Panagiotis Patrinos, Olivier Fercoq, and Volkan
  Cevher.
\newblock Escaping limit cycles: Global convergence for constrained
  nonconvex-nonconcave minimax problems.
\newblock In \emph{International Conference on Learning Representations
  (ICLR)}, 2022.

\bibitem[Rabinovich et~al.(2013)Rabinovich, Naroditskiy, Gerding, and
  Jennings]{rabinovich2013computing}
Zinovi Rabinovich, Victor Naroditskiy, Enrico~H. Gerding, and Nicholas~R.
  Jennings.
\newblock Computing pure {Bayesian}-{Nash} equilibria in games with finite
  actions and continuous types.
\newblock \emph{Artificial Intelligence}, 195:\penalty0 106--139, 2013.
\newblock \doi{10.1016/j.artint.2012.09.007}.

\bibitem[Reeves and Wellman(2004)]{reeves2004computing}
Daniel~M. Reeves and Michael~P. Wellman.
\newblock Computing best-response strategies in infinite games of incomplete
  information.
\newblock In \emph{Proceedings of the 20th Conference on Uncertainty in
  Artificial Intelligence (UAI)}, pages 470--478. AUAI Press, 2004.

\bibitem[Riley and Samuelson(1981)]{riley1981optimal}
John~G. Riley and William~F. Samuelson.
\newblock Optimal auctions.
\newblock \emph{American Economic Review}, 71\penalty0 (3):\penalty0 381--392,
  1981.

\bibitem[Robbins and Monro(1951)]{robbins1951stochastic}
Herbert Robbins and Sutton Monro.
\newblock A stochastic approximation method.
\newblock \emph{The Annals of Mathematical Statistics}, 22\penalty0
  (3):\penalty0 400--407, 1951.
\newblock \doi{10.1214/aoms/1177729586}.

\bibitem[Rosen(1965)]{rosen1965existence}
J.~B. Rosen.
\newblock Existence and uniqueness of equilibrium points for concave
  {$N$}-person games.
\newblock \emph{Econometrica}, 33\penalty0 (3):\penalty0 520--534, 1965.
\newblock \doi{10.2307/1911749}.

\bibitem[Smith(1995)]{smith1995monotone}
Hal~L. Smith.
\newblock \emph{Monotone Dynamical Systems: An Introduction to the Theory of
  Competitive and Cooperative Systems}, volume~41 of \emph{Mathematical Surveys
  and Monographs}.
\newblock American Mathematical Society, Providence, RI, 1995.

\bibitem[Soner(1986)]{soner1986optimal}
H.~Mete Soner.
\newblock Optimal control with state-space constraint {I}.
\newblock \emph{SIAM Journal on Control and Optimization}, 24\penalty0
  (3):\penalty0 552--561, 1986.

\bibitem[Spulber(1995)]{spulber1995bertrand}
Daniel~F. Spulber.
\newblock {Bertrand} competition when rivals' costs are unknown.
\newblock \emph{The Journal of Industrial Economics}, 43\penalty0 (1):\penalty0
  1--11, 1995.
\newblock \doi{10.2307/2950422}.

\bibitem[Vickrey(1961)]{vickrey1961counterspeculation}
William Vickrey.
\newblock Counterspeculation, auctions, and competitive sealed tenders.
\newblock \emph{The Journal of Finance}, 16\penalty0 (1):\penalty0 8--37, 1961.
\newblock \doi{10.1111/j.1540-6261.1961.tb02789.x}.

\bibitem[Vives(1990)]{vives1990nash}
Xavier Vives.
\newblock Nash equilibrium with strategic complementarities.
\newblock \emph{Journal of Mathematical Economics}, 19\penalty0 (3):\penalty0
  305--321, 1990.

\bibitem[Wiggins(2003)]{wiggins2003introduction}
Stephen Wiggins.
\newblock \emph{Introduction to Applied Nonlinear Dynamical Systems and Chaos},
  volume~2 of \emph{Texts in Applied Mathematics}.
\newblock Springer, New York, 2 edition, 2003.

\end{thebibliography}

\clearpage
\appendix
%=====================================================================
\section{Proofs for Section~\ref{sec:model}}\label{app:model}
%=====================================================================

\paragraph{Notation.} In the appendices we abbreviate the objects of \cref{sec:field} as follows. For a type $c$ we write
$W_c(z,s)=W(z,s;c)$, $V_c(z,s)=V(z,s;c)$, $g_c(s)=g(s\mid c)$ and $\pi_c(z;\beta)=\pi(z;\beta,c)$. For a strategy $\beta$ we
write $x(z)=\beta^{-1}(z)$ for the \emph{implied type} of the action $z$, with the extension $x(z)=0$ for $z\le\beta(0)$ and
$x(z)=1$ for $z\ge\beta(1)$ of \cref{sec:field}. In this notation \eqref{eq:payoff} reads
$\pi_c(z;\beta)=\int_0^{x(z)}V_c(z,s)g_c(s)\,ds+\int_{x(z)}^1W_c(z,s)g_c(s)\,ds$.

\begin{proof}[Proof of \cref{prop:membership}]
\emph{Idea.} We first compute the derivative of the payoff along an arbitrary regular strategy. It is the direct
effect of the action minus the profit jump at a tie, weighted by the density $g_c(c)$ of tied rivals and by the speed
$1/u$ at which the implied type moves with the action. Sufficiency is then a substitution. For necessity we use that
this derivative depends on the strategy only through the action $\beta(c)$ and the slope $u$. Affine strategies realize
every pair $(\beta(c),u)$, and comparing two slopes separates the two identities.

Fix $c\in(0,1)$, and write
\[
\cA_c(z)=\int_0^c\fraction{\partial V_c}{\partial z}(z,s)g_c(s)\,ds+\int_c^1\fraction{\partial W_c}{\partial z}(z,s)g_c(s)\,ds,\qquad J_c(z)=W_c(z,c)-V_c(z,c)
\]
for the direct effect of the action and the profit jump at a tie. The proposition states that the game is M\"obius if
and only if $\cA_c(z)=A(c)$ and $g_c(c)J_c(z)=B(c)(z-\mu(c))$ for all $c\in(0,1)$ and $z\in\R$.

\emph{Step 1: the implied type moves at speed $1/u$.} Let $\beta$ be regular with $\fraction{\mathrm{d} \beta}{\mathrm{d} c}(c)=u>0$. Near the action
$\beta(c)$, which lies strictly between $\beta(0)$ and $\beta(1)$, the implied type is the inverse of $\beta$, and it is
continuous there because it is monotone and maps an interval onto an interval. Since $x(\beta(c))=c$ and $\beta$ has the nonzero derivative $u$
at $c$, the inverse function rule gives
\[
x(\beta(c))=c,\qquad \fraction{\mathrm{d} x}{\mathrm{d} z}(\beta(c))=\frac1u .
\]

\emph{Step 2: the payoff as a fixed integral plus a cutoff integral.} Writing $\int_0^{x(z)}=\int_0^1-\int_{x(z)}^1$ in the
losing term of \eqref{eq:payoff} gives
\begin{equation}\label{eq:A-split}
\pi_c(z;\beta)=\int_0^1V_c(z,s)g_c(s)\,ds+\int_{x(z)}^1\bigl(W_c(z,s)-V_c(z,s)\bigr)g_c(s)\,ds .
\end{equation}
The first integral does not involve $\beta$. The second integral is the gain from winning against the rivals whose
types exceed the implied type; only its lower limit depends on $\beta$.

\emph{Step 3: the derivative along $\beta$.} If $F$ and $\fraction{\partial F}{\partial z}$ are jointly continuous and the cutoff $x$ is
differentiable at $z_0$ with $x(z_0)=c$, the Leibniz rule for a cutoff integral gives
\[
\fraction{\mathrm{d}}{\mathrm{d} z}\int_{x(z)}^1F(z,s)\,ds\Big|_{z=z_0}=\int_c^1\fraction{\partial F}{\partial z}(z_0,s)\,ds-F(z_0,c)\,\fraction{\mathrm{d} x}{\mathrm{d} z}(z_0).
\]
The continuity hypotheses of the proposition are exactly what this rule needs, for $F=V_cg_c$ with the constant cutoff $0$
and for $F=(W_c-V_c)g_c$ with the cutoff $x$. Applying it to both terms of \eqref{eq:A-split} at $z_0=\beta(c)$, with
Step~1, gives
\begin{equation}\label{eq:A-field}
\begin{aligned}
\fraction{\partial \pi_c}{\partial z}(z;\beta)\Big|_{z=\beta(c)}
&=\int_0^1\fraction{\partial V_c}{\partial z}g_c\,ds+\int_c^1\bigl(\fraction{\partial W_c}{\partial z}-\fraction{\partial V_c}{\partial z}\bigr)g_c\,ds-\frac{g_c(c)J_c(\beta(c))}{u}\\
&=\cA_c(\beta(c))-\frac{g_c(c)J_c(\beta(c))}{u},
\end{aligned}
\end{equation}
where the integrands are evaluated at $(\beta(c),s)$.

\emph{Step 4: sufficiency.} If the two identities hold, then $\cA_c(\beta(c))=A(c)$ and
$g_c(c)J_c(\beta(c))=B(c)(\beta(c)-\mu(c))$, so \eqref{eq:A-field} is \eqref{eq:field}.

\emph{Step 5: necessity.} Fix $z\in\R$ and $u>0$. The affine strategy $\beta_u(s)=z+u(s-c)$ is regular, takes the value
$\beta_u(c)=z$ and has slope $u$. Since the game is M\"obius, its payoff derivative at $z$ is given both by \eqref{eq:field}
and by \eqref{eq:A-field}, so
\[
\cA_c(z)-\frac{g_c(c)J_c(z)}u=A(c)-\frac{B(c)(z-\mu(c))}u\qquad(u>0).
\]
Both sides are affine in $1/u$. Subtracting the equation for $u=1$ from the one for $u=2$ gives
$\frac12g_c(c)J_c(z)=\frac12B(c)(z-\mu(c))$, the second identity. Inserting it into the equation for $u=1$ gives
$\cA_c(z)=A(c)$, the first identity.
\end{proof}

\begin{proof}[Proof of \cref{prop:examples}]
We identify the coefficients of each row (Steps 1 and 2) and then verify \cref{ass:main} (Step 3) and the
columns $p$ and $a$ (Step 4). The first six rows are allocation games, whose coefficients can be read off from
\eqref{eq:allocation}; the last three rows are tested with \cref{prop:membership}. Throughout, $S(c)=\int_c^1f\ge f_-(1-c)$,
so $S>0$ on $[0,1)$ and $S(1)=0$.

\emph{Step 1: the six allocation rows.} By the computation after \eqref{eq:allocation}, an allocation game is
M\"obius with $A=\kappa Q$, $B=\kappa q$ and $\mu=m$, where $q=-\fraction{\mathrm{d} Q}{\mathrm{d} c}$. The rows specify $\kappa$, $m$ and $Q$:
\begin{itemize}
\item \emph{First price; Bertrand:} $\kappa=1$, $m(c)=c$, $Q=S^{n-1}$, hence $q=(n-1)S^{n-2}f$.
\item \emph{Pay-as-bid:} $\kappa=1$, $m(c)=c$, $Q_K=\sum_{j<K}\binom{n-1}jF^jS^{n-1-j}$. With $j\binom{n-1}j=(n-1)\binom{n-2}{j-1}$ and
$(n-1-j)\binom{n-1}j=(n-1)\binom{n-2}j$, the derivative is $(n-1)f\sum_{j<K}(a_{j-1}-a_j)$ with $a_j=\binom{n-2}jF^jS^{n-2-j}$
and $a_{-1}=0$. The sum telescopes to $-a_{K-1}$, so $\fraction{\mathrm{d} Q_K}{\mathrm{d} c}=-q_K$ with
$q_K=(n-1)\binom{n-2}{K-1}F^{K-1}S^{n-1-K}f\ge0$.
\item \emph{Generalized first price:} $\kappa=1$, $m(c)=c$, $Q^\lambda=\sum_{j\le K}(\lambda_j-\lambda_{j+1})Q_j$, hence
$q^\lambda=\sum_{j\le K}(\lambda_j-\lambda_{j+1})q_j$.
\item \emph{Random number of rivals:} $\kappa=1$, $m(c)=c$, $Q^\varrho=\sum_{r=1}^{r_{\max}}\varrho_rS^r$, hence
$q^\varrho=\sum_{r=1}^{r_{\max}}r\varrho_rS^{r-1}f$.
\item \emph{Platform fee:} $(\kappa z-c)S^{n-1}=\kappa(z-c/\kappa)S^{n-1}$, so $m(c)=c/\kappa$, $Q=S^{n-1}$ and $B=\kappa(n-1)S^{n-2}f$.
\item \emph{Cost sharing:} $\kappa=1$, $m(c)=\chi c$, $Q=S^{n-1}$.
\end{itemize}
Each $Q$ has a continuous derivative, $Q(1)=0$ and $Q>0$ on $[0,1)$. For $Q_K$, every term contains $S^{n-K}$ with
$n-K\ge1$, and $Q_K\ge Q_1=S^{n-1}$; $Q^\lambda$ and $Q^\varrho$ are combinations of these with nonnegative weights.

\emph{Step 2: the three remaining rows.} We apply \cref{prop:membership}, whose hypotheses hold because the integrands
below are continuous. With $V_c=0$ it suffices to compute the direct effect $\cA_c=\int_c^1\fraction{\partial W_c}{\partial z}(z,s)g_c(s)\,ds$
and the profit jump $J_c(z)=W_c(z,c)$; then $A(c)=\cA_c$ and $g_c(c)J_c(z)=B(c)(z-\mu(c))$.
\begin{itemize}
\item \emph{Interdependent costs:} $W_c=z-\bar m(c,s)$, and the lowest rival type has the density $g=(n-1)S^{n-2}f$, which does not
depend on the own type. Then $\fraction{\partial W_c}{\partial z}=1$ gives $\cA_c=\int_c^1g=S(c)^{n-1}$, and $J_c(z)=z-\bar m(c,c)$ gives
$B=g=(n-1)S^{n-2}f$ and $\mu(c)=\bar m(c,c)$.
\item \emph{Affiliated costs:} $W_c=z-c$. Then $\cA_c=\int_c^1g_c=\Sigma_c(c)$, and $J_c(z)=z-c$ gives $B(c)=g_c(c)$ and $\mu(c)=c$.
\item \emph{Nonlinear profit:} $W_c=z-c+\theta z^2\zeta_c(s)$ and $g\equiv1$. Then
$\fraction{\partial W_c}{\partial z}=1+2\theta z\zeta_c(s)$, and $\int_c^1\zeta_c(s)\,ds=0$ gives $\cA_c=1-c$; $\zeta_c(c)=0$ gives
$J_c(z)=z-c$, so $B=1$ and $\mu(c)=c$. The term $\theta z^2\zeta_c$ is invisible to both identities: it has zero mean over
the winning types and vanishes at a tie.
\end{itemize}
In every row except the affiliated one, $B=-\fraction{\mathrm{d} A}{\mathrm{d} c}$.

\emph{Step 3: \cref{ass:main}.}
\begin{itemize}
\item \condref{A1} holds because $f$, $c\mapsto\bar m(c,c)$ and $(c,s)\mapsto g_c(s)$ are continuous; in the affiliated row,
$A(c)=\int_c^1g_c(s)\,ds$ and $B(c)=g_c(c)$ are continuous by joint continuity.
\item $A>0$ on $[0,1)$: in the allocation rows because $Q>0$ there, in the interdependent row because $S>0$ there, in the
affiliated row because $g_c\ge0$ is continuous with $g_c(c)>0$, so that $\Sigma_c(c)=\int_c^1g_c>0$, and in the nonlinear
row because $A=1-c$.
\item $B>0$ on $[0,1)$ because $S>0$ there and $f>0$; in the affiliated row because $g_c(c)>0$ for $c<1$. For $q^\lambda$ on
$(0,1)$ this uses $q^\lambda\ge\lambda_Kq_K$, and for $q^\varrho$ the term $r_0\varrho_{r_0}S^{r_0-1}f$. The exceptions are
at $c=0$: the factor $F(0)^{K-1}=0$ in $q_K(0)$ for $K\ge2$, and $q^\lambda(0)=(\lambda_1-\lambda_2)(n-1)f(0)$, since
$q_j(0)=0$ for $j\ge2$, which vanishes if $\lambda_1=\lambda_2$.
\item \condref{A3} holds because $c$, $c/\kappa$ and $\chi c$ increase, and $c\mapsto\bar m(c,c)$ is weakly increasing by hypothesis.
\end{itemize}

\emph{Step 4: columns $p$ and $a$.} The cost gap is $w(c)=1-c$ in the rows with $\mu(c)=c$, $w(c)=(1-c)/\kappa$ for
the platform fee, $w(c)=\chi(1-c)$ for cost sharing, and $w(c)=\bar m(1,1)-\bar m(c,c)\le L_m(1-c)$ for interdependent
costs. In each row we bound $A$ below by a power of $1-c$ and then express $1-c$ through $w$. With $S\ge f_-(1-c)$:
\begin{itemize}
\item $S^{n-1}\ge f_-^{\,n-1}(1-c)^{n-1}$ gives the first-price, platform-fee ($\kappa S^{n-1}\ge\kappa^nf_-^{\,n-1}w^{n-1}$), cost-sharing
($1-c=w/\chi$) and interdependent ($1-c\ge w/L_m$) rows.
\item $Q_K\ge S^{n-K}$, because at most $K-1$ rivals have lower costs whenever $n-K$ fixed rivals have higher costs.
This gives the pay-as-bid row.
\item Summation by parts gives $Q^\lambda=\sum_{j\le K}\lambda_j(Q_j-Q_{j-1})\ge\lambda_KQ_K$, with $Q_0=0$, because $Q_j-Q_{j-1}\ge0$ and
$\lambda_j\ge\lambda_K$. This gives the generalized first-price row.
\item $Q^\varrho\ge\varrho_{r_0}S^{r_0}$ gives the random-rivals row.
\item $\Sigma_c(c)\ge(1-c)\min_{c',s}g_{c'}(s)$ gives the affiliated row, and $A=1-c=w$ gives the nonlinear row.
\end{itemize}
\end{proof}

\paragraph{Reserve prices.} In a first-price procurement auction with reserve price $r\in(0,1)$, the types above $r$
abstain. We write a participating type as $c\in[0,1]$, with cost $rc$, so that the participating types are rescaled to
$[0,1]$. A participating bidder wins when every rival has a higher cost, abstaining rivals included. Against a rival
strategy with implied type $x(z)$, the action $z$ therefore wins with probability $S(r\,x(z))^{n-1}$, and type $c$ earns
\[
\pi_c(z;\beta)=\bigl(z-rc\bigr)\,S\bigl(r\,x(z)\bigr)^{n-1}.
\]
This has the form \eqref{eq:allocation} with $\kappa=1$, $m(c)=rc$ and $Q(c)=S(rc)^{n-1}$, except that $Q(1)=S(r)^{n-1}$ is
positive. The computation after \eqref{eq:allocation} uses only $\fraction{\mathrm{d} Q}{\mathrm{d} c}=-q$, so the game is M\"obius with
\[
A(c)=S(rc)^{n-1},\qquad B(c)=r(n-1)S(rc)^{n-2}f(rc),\qquad \mu(c)=rc .
\]
These coefficients satisfy \cref{ass:main}, because $S>0$ and $f>0$ on $[0,r]$. The cap is the reserve, $C=\mu(1)=r$, and
$A(1)=S(r)^{n-1}>0$ is the probability that all rivals abstain (\cref{rem:assumptions}).

\begin{proposition}[Non-members]\label{prop:nonmembers}
Let $g_c$ be continuous. If $W_c=\upsilon(z-c)$ with $\upsilon$ strictly concave and $V_c=0$, or in the all-pay contest with prize
$\mathcal V$, where $W_c=\mathcal V-z$ and $V_c=-z$, every triple $(A,B,\mu)$ for which the game is a M\"obius pricing game has $B=0$ on
$(0,1)$. If $W_c=(z-c)D(z)$ with continuous $D$ and $V_c=0$, and the game is a M\"obius pricing game with coefficients
that have $B(c)\ne0$ at some $c\in(0,1)$, then $D$ is constant.
\end{proposition}

\begin{proof}
\emph{Idea.} In each case we use only the second identity of \cref{prop:membership}, the jump identity
\begin{equation}\label{eq:A-jump}
g_c(c)J_c(z)=B(c)\bigl(z-\mu(c)\bigr)\qquad(z\in\R),
\end{equation}
which says that the weighted profit jump is affine in the action. In the all-pay contest the profits are differentiable,
so \cref{prop:membership} applies directly. In the other two games the winning profit depends only on the action and is
merely continuous in it, so we derive \eqref{eq:A-jump} from \eqref{eq:field} by the argument of Step~5 of
\cref{prop:membership}, with difference quotients in place of derivatives.

\emph{Step 1: the jump identity for continuous winning profits.} Let $W_c(z,s)=P_c(z)$ with $P_c$ continuous, $V_c=0$, and let
$g_c$ be continuous. Then $J_c=P_c$. Fix $c\in(0,1)$ and $z\in\R$, and for $u>0$ let $\beta_u(s)=z+u(s-c)$ as in
\cref{prop:membership}, with implied type $x_u$. Against $\beta_u$ the payoff is $P_c(y)H_u(y)$ with
$H_u(y)=\int_{x_u(y)}^1g_c(s)\,ds$. By Step~1 of \cref{prop:membership} and the fundamental theorem of calculus,
\[
H_u(z)=\int_c^1g_c(s)\,ds=:\Sigma,\qquad \fraction{\mathrm{d} H_u}{\mathrm{d} y}(z)=-\frac{g_c(c)}u,
\]
and by \eqref{eq:field} the product $P_cH_u$ has the derivative $A(c)-B(c)(z-\mu(c))/u$ at $z$. The product rule cannot be
applied, because $P_c$ need not be differentiable, but its difference-quotient form can: for $y\ne z$,
\[
\Sigma\,\frac{P_c(y)-P_c(z)}{y-z}=\frac{P_c(y)H_u(y)-P_c(z)H_u(z)}{y-z}-P_c(y)\,\frac{H_u(y)-H_u(z)}{y-z}.
\]
As $y\to z$, continuity of $P_c$ gives the limit $A(c)-B(c)(z-\mu(c))/u+P_c(z)g_c(c)/u$ on the right. The left side does not
depend on $u$, so the limits for $u=1$ and $u=2$ agree:
\[
A(c)-B(c)\bigl(z-\mu(c)\bigr)+g_c(c)P_c(z)=A(c)-\tfrac12B(c)\bigl(z-\mu(c)\bigr)+\tfrac12g_c(c)P_c(z).
\]
This is \eqref{eq:A-jump}. No condition such as $\Sigma\ne0$ is needed.

\emph{Step 2: the three games.} Fix $c\in(0,1)$.
\begin{itemize}
\item \emph{Elastic demand:} $P_c(z)=(z-c)D(z)$, and suppose $B(c)\ne0$. At $z=c$, \eqref{eq:A-jump} gives $0=B(c)(c-\mu(c))$, so
$\mu(c)=c$. Then $g_c(c)\ne0$, since otherwise $z=c+1$ would give $0=B(c)$. Hence $(z-c)\bigl(g_c(c)D(z)-B(c)\bigr)=0$, so
$D(z)=B(c)/g_c(c)$ for $z\ne c$, and also at $z=c$ by continuity of $D$. Thus $D$ is constant.
\item \emph{Risk aversion:} $P_c(z)=\upsilon(z-c)$, and $\upsilon$ is continuous because it is concave on $\R$. With $t=z-c$,
\eqref{eq:A-jump} reads $g_c(c)\upsilon(t)=B(c)(t+c-\mu(c))$ for all $t$, so $g_c(c)\upsilon$ is affine. An affine function equals
the average of its values at $0$ and $1$ at the midpoint $\frac12$, whereas strict concavity gives
$\upsilon(\frac12)>\frac12(\upsilon(0)+\upsilon(1))$. Hence $g_c(c)=0$. Then $B(c)(t+c-\mu(c))=0$ for all $t$, so $B(c)=0$.
\item \emph{All-pay contest:} $W_c=\mathcal V-z$ and $V_c=-z$ are differentiable with continuous derivatives, so
\cref{prop:membership} applies, and $J_c\equiv\mathcal V$ does not depend on $z$. Then \eqref{eq:A-jump} at $z=\mu(c)$ gives
$g_c(c)\mathcal V=0$, and at $z=\mu(c)+1$ it gives $g_c(c)\mathcal V=B(c)$. Hence $B(c)=0$.
\end{itemize}
\end{proof}

\begin{proof}[Proof of \cref{prop:minty}]
\emph{Idea.} The strategy $\beta$ lies below the rest point. It is too flat on $[0,\frac12]$ (slope $\frac15$ instead of
$\frac12$) and too steep on $[\frac12,1]$ (slope $\frac45$). A flat strategy makes the reciprocal-slope term in
\eqref{eq:field} large, so on the left half the field is negative and pushes $\beta$ further away from $\bstar$; there
$(\beta-\bstar)G[\beta]>0$. On the right half the field pushes $\beta$ back toward $\bstar$, but this half contributes less. We
compute both contributions (Steps 1 and 2) and then identify the integral with the derivative of the $L^2$ distance
(Step 3).

\emph{Step 1: the strategy and its field.} The strategy is $\beta(c)=\frac12+\frac c5$ on $[0,\frac12]$ and
$\beta(c)=\frac15+\frac{4c}5$ on $[\frac12,1]$. The two pieces agree at $c=\frac12$, with value $\frac35$, and $\beta(1)=1$. Its
slopes lie in $(0,1)$, so $\beta(c)=1-\int_c^1 \fraction{\mathrm{d} \beta}{\mathrm{d} s}(s)\,\mathrm{d}s\ge c$ and $\beta\le1$; hence $\beta$ is feasible. Off the kink at
$c=\frac12$, $G[\beta](c)=(1-c)-(\beta(c)-c)/\fraction{\mathrm{d} \beta}{\mathrm{d} c}(c)$, and with $\bstar=\frac12+\frac c2$:
\begin{itemize}
\item on $(0,\frac12)$: $\beta-\bstar=-\frac{3c}{10}$ and $\beta-c=\frac12-\frac{4c}5$, so $G[\beta]=(1-c)-5(\frac12-\frac{4c}5)=3(c-\frac12)<0$;
\item on $(\frac12,1)$: $\beta-\bstar=-\frac{3(1-c)}{10}$ and $\beta-c=\frac{1-c}5$, so $G[\beta]=(1-c)-\frac{1-c}4=\frac{3(1-c)}4>0$.
\end{itemize}
Hence the integrand $(\beta-\bstar)G[\beta]$ equals $-\frac9{10}(c^2-\frac c2)\ge0$ on the left half and $-\frac9{40}(1-c)^2\le0$
on the right half.

\emph{Step 2: the Minty integral.} The kink is a single point and does not affect the integral. Since
$\int_0^{1/2}c^2\,dc=\frac1{24}$, $\int_0^{1/2}\frac c2\,dc=\frac1{16}$ and $\int_{1/2}^1(1-c)^2\,dc=\frac1{24}$,
\[
\int_0^1(\beta-\bstar)\,G[\beta]\,dc=-\frac9{10}\Bigl(\frac1{24}-\frac1{16}\Bigr)-\frac9{40}\cdot\frac1{24}
=\frac3{160}-\frac3{320}=\frac3{320}>0 .
\]

\emph{Step 3: the derivative of the $L^2$ distance.} Let $(\beta_t)$ be as in the statement. The $t$-derivative of the
integrand $\frac12(\beta_t-\bstar)^2$ is $(\beta_t-\bstar)\fraction{\partial \beta_t}{\partial t}$. Near $t_0$, $\beta_t$ and $\fraction{\partial \beta_t}{\partial t}$ are bounded by assumption, and $0\le\bstar\le1$, so the derivative of the integrand
is bounded by a constant uniformly in $c$ and in $t$ near $t_0$. Since each $\beta_t$ is measurable, dominated convergence
permits differentiation under the integral sign, and at $t_0$, where $\beta_{t_0}=\beta$ and $\fraction{\partial \beta_t}{\partial t}=G[\beta]$,
\[
\fraction{\mathrm{d}}{\mathrm{d} t}\,\frac12\int_0^1(\beta_t-\bstar)^2\,dc\,\Big|_{t=t_0}=\int_0^1(\beta-\bstar)\,G[\beta]\,dc=\frac3{320}.
\]
\end{proof}

\begin{proof}[Proof of \cref{prop:blowup}]
For this auction $A=1-c$, $B=1$, $\mu(c)=c$, $C=1$ and $\varphi\equiv1$, so \eqref{eq:GF} reads $\fraction{\partial \beta}{\partial t}+H(c,\beta,\fraction{\partial \beta}{\partial c})=0$ with
\[
H(c,z,p)=-(1-c)+\frac{z-c}{p},
\]
and the rest point is $\bstar(c)=(1+c)/2$. Write $\varsigma=400(c-\frac12)$, $a=(\sqrt{40001}-1)/1600$ and $T=1/200$.

\emph{Idea of the proof.} The datum is a smoothed concave corner. Indeed,
\[
b(c)=\frac{1+c}2+a-\frac{\sqrt{1+\varsigma^2}-1}{1600},
\]
and $\sqrt{1+\varsigma^2}\approx|\varsigma|$ outside a window of width about $1/400$ around $c=\frac12$. So the graph of $b$ is close
to a line of slope $\frac34$ left of $\frac12$ and to a line of slope $\frac14$ right of it, and the line of slope $\frac12$ through
$(\frac12,b(\frac12))$ touches $b$ from above. We show four things:
\begin{enumerate}
\item[(1)] the solution stays below three explicit lines that move with the flow: a line $M$ of slope $\frac12$, and lines
$L$ and $R$ that follow the two straight pieces of $b$;
\item[(2)] by time $T$, $\min\{L,R\}$, and hence $\beta$, lies strictly below $M$, although $\beta$ touches $M$ at $t=0$;
\item[(3)] therefore $\beta-M$, slightly tilted in time, attains an interior maximum at a positive time $t_0$ and a type $c_0$;
\item[(4)] $\beta(t_0,\cdot)$ is not differentiable at $c_0$.
\end{enumerate}
Step (4) rests on the following observation. At a point where $\beta$ has slope $\frac12$ and lies below $M$, the
equation lets $\beta$ decrease at most as fast as $M$; but the maximum in (3) forces $\beta$ to decrease faster than $M$.
Only a corner, where no smooth function touches $\beta$ from below, is compatible with this.

\emph{Step 0: the datum.} The radicand is positive, so $b$ is $C^\infty$, and $\fraction{\mathrm{d} b}{\mathrm{d} c}(c)=\frac12-\frac14\,\varsigma/\sqrt{1+\varsigma^2}$, so
$\frac14<\fraction{\mathrm{d} b}{\mathrm{d} c}<\frac34$. At $c=1$ we have $\sqrt{1+\varsigma^2}=\sqrt{40001}$, so $b(1)=1$; with $0<\fraction{\mathrm{d} b}{\mathrm{d} c}<1$ this gives $c\le b(c)\le1$, and $b$
is feasible. Let $\beta$ be the solution from $b$ given by \cref{thm:main}(i).

\emph{Step 1: three upper barriers.} Lines are the natural barriers for this equation. For $U(t,c)=u(t)c+v(t)$ with $u>0$,
\begin{equation}\label{eq:blowup-affine}
\fraction{\partial U}{\partial t}+H(c,U,u)=c\Bigl(\fraction{\mathrm{d} u}{\mathrm{d} t}+2-\frac1u\Bigr)+\Bigl(\fraction{\mathrm{d} v}{\mathrm{d} t}-1+\frac vu\Bigr).
\end{equation}
So $U$ solves the equation exactly if $\fraction{\mathrm{d} u}{\mathrm{d} t}=1/u-2$ and $\fraction{\mathrm{d} v}{\mathrm{d} t}=1-v/u$, and $U$ is a supersolution if the right-hand side is
nonnegative for $0\le c\le1$ (\cref{lem:barriers}(b)).
\begin{itemize}
\item \emph{The line $M$.} The slope $\frac12$ is stationary, and $\fraction{\mathrm{d} v}{\mathrm{d} t}=1-2v$ gives the exact solution
\[
M(t,c)=\frac{1+c}2+a\e^{-2t},
\]
the rest point shifted up by an amount that decays at the damping rate $B/(\fraction{\mathrm{d} \bstar}{\mathrm{d} c})=2$. Since $\sqrt{1+\varsigma^2}\ge1$, we
have $b\le M(0,\cdot)$, with equality at $c=\frac12$.
\item \emph{The lines $L$ and $R$.} Since $\sqrt{1+\varsigma^2}\ge-\varsigma$ and $\sqrt{1+\varsigma^2}\ge\varsigma$, and $\sqrt{40001}\le2001/10$,
\[
b(c)\le\frac34c+\frac{8001}{16000},\qquad b(c)\le\frac14c+\frac{12001}{16000}\qquad(0\le c\le1);
\]
these are the two straight pieces of $b$, raised by $1/16000$. Under the exact affine flow, their slopes move at the
rates $\frac43-2=-\frac23$ and $4-2=2$, that is, toward $\frac12$ from both sides, and their intercepts at the rates
$1-\frac43\cdot\frac{8001}{16000}\approx0.333$ and $1-4\cdot\frac{12001}{16000}\approx-2.000$. We freeze these rates at their
initial values and round the intercept rates up, to $\frac{103}{300}$ and $-\frac{11}6$:
\[
\begin{aligned}
L(t,c)&=\Bigl(\frac34-\frac{2t}{3}\Bigr)c+\frac{8001}{16000}+\frac{103t}{300},\\
R(t,c)&=\Bigl(\frac14+2t\Bigr)c+\frac{12001}{16000}-\frac{11t}{6}.
\end{aligned}
\]
Rounding up makes the residual positive on the short interval $[0,T]$: with $p_L=\fraction{\partial L}{\partial c}$ and $p_R=\fraction{\partial R}{\partial c}$,
\eqref{eq:blowup-affine} gives
\[
\begin{aligned}
p_L\bigl(\fraction{\partial L}{\partial t}+H(c,L,p_L)\bigr)
&=\frac{121}{16000}+t\Bigl(\frac{703}{900}-\frac{8c}{9}\Bigr)>0,\\
p_R\bigl(\fraction{\partial R}{\partial t}+H(c,R,p_R)\bigr)
&=\frac{2003}{48000}+t\Bigl(8c-\frac{15}{2}\Bigr)>0
\end{aligned}
\qquad\text{on }\cB_T.
\]
\end{itemize}
On $\cB_T$ all three functions lie above cost and have positive slopes, so by \cref{lem:barriers}(b) the clipped
functions $\min\{1,L\}$, $\min\{1,R\}$ and $\min\{1,M\}$ are supersolutions. They dominate the datum and equal one at $c=1$,
so comparison (\cref{lem:comparison}) gives
\begin{equation}\label{eq:blowup-upper}
\beta\le L,\ R,\ M\qquad\text{on }\cB_T.
\end{equation}

\emph{Step 2: by time $T$ the corner has dropped below $M$.} By \eqref{eq:blowup-upper}, $\beta(T,\cdot)\le\lambda L(T,\cdot)+(1-\lambda)R(T,\cdot)$ for
every $\lambda\in[0,1]$. We choose $\lambda$ so that this combination is parallel to $M(T,\cdot)$; then comparing the two lines
reduces to comparing their intercepts. The slopes at time $T$ are $\fraction{\partial L}{\partial c}(T)=\frac{56}{75}$ and $\fraction{\partial R}{\partial c}(T)=\frac{13}{50}$, and
$\lambda\frac{56}{75}+(1-\lambda)\frac{13}{50}=\frac12$ gives $\lambda=\frac{36}{73}$. The intercepts at time $T$ are
$\frac{120427}{240000}$ for $L$ and $\frac{35563}{48000}$ for $R$, so the combination has intercept
$\frac{10914527}{17520000}<0.62298$. The line $M(T,\cdot)$ has intercept $\frac12+a\e^{-2T}\ge\frac12+\frac{19701}{160000}>0.62313$, because
$a\ge\frac{199}{1600}$ (from $\sqrt{40001}\ge200$) and $\e^{-2t}\ge1-2t\ge\frac{99}{100}$ for $t\le T$. The intercepts differ by more
than $\frac{3}{20000}$, so
\begin{equation}\label{eq:blowup-gap}
\beta(T,c)\le\frac{36}{73}L(T,c)+\frac{37}{73}R(T,c)<M(T,c)-\frac{3}{20000}\qquad(0\le c\le1).
\end{equation}
Thus $\beta$ touches $M$ at $(0,\frac12)$ but lies uniformly below $M$ at time $T$. We now show that this transition forces
a corner.

\emph{Step 3: an interior maximum at a positive time.} Let
\[
f(t,c)=\beta(t,c)-M(t,c)+\frac t{100}.
\]
The tilt $t/100$ favors later times. It is small enough that $f(T,\cdot)$ stays negative, since $T/100=\frac1{20000}<\frac3{20000}$,
and it is the excess speed that Step 4 turns into a contradiction. On the boundary of $[0,T]\times[0,1]$, other than at
$t=0$:
\begin{itemize}
\item at $t=T$, \eqref{eq:blowup-gap} gives $f(T,c)<-\frac3{20000}+\frac1{20000}=-\frac1{10000}$;
\item at $c=0$, $\beta\le L$ and $M(t,0)\ge\frac12+\frac{19701}{160000}$ give $f(t,0)\le\frac1{16000}+t\bigl(\frac{103}{300}+\frac1{100}\bigr)-\frac{19701}{160000}<-\frac1{10}$;
\item at $c=1$, $\beta=1$ gives $f(t,1)=-a\e^{-2t}+\frac t{100}<-\frac1{10}$.
\end{itemize}
At the initial contact point, $f(0,\frac12)=b(\frac12)-M(0,\frac12)=0$. Since $\beta$ is continuous, there is $\tau\in(0,T)$ with
$f(\tau,\frac12)>-\frac1{40000}$. On the compact set $[\tau,T]\times[0,1]$, $f$ attains its maximum at some $(t_0,c_0)$, and since this
maximum exceeds $-\frac1{40000}$, it is not attained at $t=T$, $c=0$ or $c=1$. Hence $0<\tau\le t_0<T$ and $0<c_0<1$. With
$k=f(t_0,c_0)$,
\begin{equation}\label{eq:blowup-contact}
\beta(s,x)\le M(s,x)+k-\frac{s}{100}\qquad(t_0\le s\le T,\ 0\le x\le1),
\end{equation}
with equality at $(t_0,c_0)$. In words: from time $t_0$ on, $\beta$ lies below a copy of $M$ that sinks at the extra speed
$\frac1{100}$, and it touches this copy at $(t_0,c_0)$.

\emph{Step 4: $\beta(t_0,\cdot)$ is not differentiable at $c_0$.} Suppose it is. Since $c_0$ is an interior maximum of
$f(t_0,\cdot)$, we get $\fraction{\partial \beta}{\partial c}(t_0,c_0)=\fraction{\partial M}{\partial c}=\frac12$.

\emph{Heuristic.} If $\beta$ were smooth near $(t_0,c_0)$, the equation and $\beta\le M$ would give
$\fraction{\partial \beta}{\partial t}=(1-c_0)-2(\beta-c_0)\ge(1-c_0)-2(M-c_0)=\fraction{\partial M}{\partial t}$ there, whereas \eqref{eq:blowup-contact} gives $\fraction{\partial \beta}{\partial t}\le\fraction{\partial M}{\partial t}-\frac1{100}$.
The rigorous argument replaces the derivatives of $\beta$ by those of a smooth function touching $\beta$ from below,
which is what the supersolution property tests; it uses only the differentiability of $\beta(t_0,\cdot)$ at $c_0$.

\emph{Construction of the test function.} Fix $\eta=10^{-5}$; any $\eta<1/802$ works, see the end of this step.
Differentiability gives $\rho>0$ with $[c_0-\rho,c_0+\rho]\subset(0,1)$ and
\begin{equation}\label{eq:blowup-cone}
\beta(t_0,x)\ge\beta(t_0,c_0)+\frac{x-c_0}{2}-\frac\eta2|x-c_0|
\qquad(|x-c_0|\le\rho).
\end{equation}
For small $h>0$ with $t_0+h<T$, put $\delta=250h$ and
\[
W(s,x)=\underbrace{M(s,x)+k-\frac{s}{100}}_{\text{upper contact function}}
+\underbrace{\frac{s-t_0}{200}}_{\text{lift}}
-\underbrace{\eta\sqrt{(x-c_0)^2+\delta^2}}_{\text{smoothed cone}}
\]
on $Q=[t_0,t_0+h]\times[c_0-\rho,c_0+\rho]$. The first part is the function of \eqref{eq:blowup-contact}. The lift raises it at
half the extra speed, so that by time $t_0+h$ it rises above $\beta$ at $c_0$. The cone bends it down away from $c_0$, so
that it stays below $\beta$ at time $t_0$ and on the sides of $Q$. The parameter $\delta$ makes $W$ smooth, and $\delta=250h$ keeps
the smoothing cost $\eta\delta=h/400$ at half the lift $h/200$. Precisely:
\begin{itemize}
\item \emph{Bottom of $Q$.} By the equality in \eqref{eq:blowup-contact} and $\fraction{\partial M}{\partial c}=\frac12$, the first part of $W$ at time $t_0$ equals
$\beta(t_0,c_0)+(x-c_0)/2$. With \eqref{eq:blowup-cone}, $\beta(t_0,x)-W(t_0,x)\ge\frac\eta2|x-c_0|\ge0$, and at $x=c_0\pm\rho$ the gap is at
least $\eta\rho/2$.
\item \emph{Sides of $Q$.} By continuity of $\beta$ and $W$, for $h$ small enough, $\beta-W>0$ at $x=c_0\pm\rho$ for all
$s\in[t_0,t_0+h]$.
\item \emph{Top center of $Q$.} By \eqref{eq:blowup-contact}, $\beta(t_0+h,c_0)-W(t_0+h,c_0)\le-\frac h{200}+\eta\delta=-\frac h{400}<0$.
\end{itemize}
Hence $\beta-W$ attains a negative minimum $m$ on $Q$ at a point $(s_*,x_*)$ with $s_*>t_0$ and $|x_*-c_0|<\rho$, and $\psi=W+m$
touches $\beta$ from below there. This gives an admissible test on $\cB_{t_0+h}$: multiply $\psi$ by a smooth cutoff in $[0,1]$
that equals one near $(s_*,x_*)$ and is supported in $(t_0,\infty)\times(c_0-\rho,c_0+\rho)$. Since $\beta\ge0$ and $\psi\le\beta$ on the
intersection of this set with $\cB_{t_0+h}$, the resulting function lies below $\beta$ on that box and has the same
derivatives at the contact point. This also covers $s_*=t_0+h$, since \cref{def:solution} tests at the terminal time of
every horizon.

\emph{The contradiction.} At the contact point write $p=\fraction{\partial \psi}{\partial c}(s_*,x_*)$ and $r=M(s_*,x_*)-x_*$. Then
\[
\frac12-\eta\le p\le\frac12+\eta,\qquad
\fraction{\partial \psi}{\partial t}=\fraction{\partial M}{\partial t}-\frac1{200},\qquad 0\le r\le1,
\]
and $p>0$, so the supersolution condition requires $\fraction{\partial \psi}{\partial t}+H(x_*,\beta(s_*,x_*),p)\ge0$. Since $H$ increases in $z$ and
$\beta\le M$, and since $\fraction{\partial M}{\partial t}-(1-x_*)=-2r$ because $M$ solves the equation with slope $\frac12$,
\[
\fraction{\partial \psi}{\partial t}+H(x_*,\beta(s_*,x_*),p)
\le-2r-\frac1{200}+\frac rp
=\frac{(1-2p)\,r-p/200}{p}
\le\frac{2\eta-(\frac12-\eta)/200}{p}<0 .
\]
This is the heuristic in quantitative form: $-2r+r/p$ is the difference between the speed the equation allows at slope
$p$ and the speed of $M$, which vanishes at $p=\frac12$, and $-\frac1{200}$ is the excess speed that remains after the lift.
The last inequality holds because $2\eta<(\frac12-\eta)/200$, that is, $\eta<1/802$. This contradicts the supersolution
inequality, so $\beta(t_0,\cdot)$ is not differentiable at $c_0$; it is continuous by \cref{thm:main}(i).

\emph{Step 5: no classical solution.} A feasible $C^1$ classical solution with positive slope on $\cB_T$ is a solution in the
sense of \cref{def:solution} by \cref{lem:barriers}(a), so comparison would make it equal to $\beta$ on $\cB_T$, which
contradicts Step 4.
\end{proof}

%=====================================================================
\section{Proof of Theorem~\ref{thm:main}}\label{app:main}\label{sec:proof1}
%=====================================================================

Throughout, $(A,B,\mu)$ satisfy \cref{ass:main}, $\varphi$ satisfies \cref{ass:metric}, and subsolutions and supersolutions are those
of \cref{def:solution}. The continuum argument combines seven lemmas: an algebraic identity, the formula for the rest point,
the comparison principle, barriers, a restart property, monotonicity in the type, and lower barriers. Existence in part~(i) is supplied by the grid-approximation lemma \cref{lem:scheme-feasible} in \cref{app:grid}.

The lemmas combine as follows.
\begin{itemize}
\item \emph{Rest point.} \Cref{lem:rest} writes the rest point as a weighted average of the anchors of higher types. This
gives its existence, uniqueness, bounds and monotonicity. \Cref{lem:identity} rewrites the payoff field at $\bstar+y$ in terms of
the error $y$ alone; it is the tool for checking that functions built from $\bstar$ are barriers.
\item \emph{Part (i).} Existence comes from \cref{lem:scheme-feasible}. The comparison principle (\cref{lem:comparison}) gives
uniqueness, and \cref{lem:monotone} shows that each time slice of the solution is weakly increasing, hence feasible.
\item \emph{Part (ii).} \Cref{lem:barriers} turns explicit functions that satisfy the differential inequality pointwise into
subsolutions and supersolutions, also after clipping at the cap $C$ or at the anchor $\mu$. With \cref{lem:identity} it shows that
the clipped shifts $\min\{C,\bstar+e(t)\}$ and $\max\{\mu,\bstar-e(t)\}$ are barriers when the scalar envelope $e$ shrinks slowly
enough. Comparison traps the solution between them, so $\cE\le e$. \Cref{lem:shift} lets us restart this argument at any time $t_0$
with $e(t_0)=\cE(t_0)$; this turns the bound $\cE\le e$ into a bound on the Dini derivative of $\cE$.
\item \emph{Part (iii).} \Cref{lem:lowerbarrier} packages \cref{lem:barriers}(a) and \cref{lem:comparison} for explicit
subsolutions; an explicit barrier for a power family then bounds the error from below.
\end{itemize}
Within this appendix, the conditions \condref{G1} and \condref{G2} on the weight enter only through the comparison principle.

\begin{lemma}[Rest-point error identity]\label{lem:identity}
Let $A^{\circ},B^{\circ},\mu^{\circ},z^{\circ},u^{\circ}$ be real numbers with $A^{\circ}u^{\circ}=B^{\circ}(z^{\circ}-\mu^{\circ})$. Then, for all real $y_0,y_1$ with
$u^{\circ}+y_1\ne0$,
\begin{equation}\label{eq:identity}
A^{\circ}-B^{\circ}\,\frac{z^{\circ}+y_0-\mu^{\circ}}{u^{\circ}+y_1}
=
\frac{A^{\circ}\,y_1-B^{\circ}\,y_0}{u^{\circ}+y_1}.
\end{equation}
\end{lemma}

\begin{proof}
The hypothesis says that the value $z^{\circ}$ and the slope $u^{\circ}$ make the field vanish at a single type. Multiply the left-hand side
of \eqref{eq:identity} by $u=u^{\circ}+y_1$ and use the hypothesis to cancel the terms without $y_0$ or $y_1$:
\[
A^{\circ}u-B^{\circ}(z^{\circ}+y_0-\mu^{\circ})
=\bigl(A^{\circ}u^{\circ}-B^{\circ}(z^{\circ}-\mu^{\circ})\bigr)+A^{\circ}y_1-B^{\circ}y_0
=A^{\circ}y_1-B^{\circ}y_0 .
\]
Dividing by $u\ne0$ gives \eqref{eq:identity}.
\end{proof}

With the values $A(c)$, $B(c)$, $\mu(c)$, $\bstar(c)$, $\fraction{\mathrm{d} \bstar}{\mathrm{d} c}(c)$, $y_0=y(c)$ and $y_1=\fraction{\mathrm{d} y}{\mathrm{d} c}(c)$, where $y=\beta-\bstar$, the identity
\eqref{eq:identity} is \eqref{eq:intro-identity}; the hypothesis of the lemma is the rest-point equation \eqref{eq:rest}. In terms of the
Hamiltonian of \cref{sec:solutions}, it reads
\begin{equation}\label{eq:B-hamiltonian-identity}
H\Bigl(c,\bstar+y_0,\fraction{\mathrm{d} \bstar}{\mathrm{d} c}+y_1\Bigr)=\frac{B\,y_0-A\,y_1}{\fraction{\mathrm{d} \bstar}{\mathrm{d} c}+y_1}
\qquad\text{at every }c<1\text{ with }\fraction{\mathrm{d} \bstar}{\mathrm{d} c}(c)+y_1\ne0.
\end{equation}
We use it to verify barriers: with $y_1=0$ for shifts of the rest point in part (ii), and with a general error profile $y$ in part (iii).

\begin{lemma}[Rest points]\label{lem:rest}
Put $\Omega(c)=\exp\bigl(-\int_0^cB/A\bigr)$ for $c<1$
and $\Omega(1)=\lim_{c\uparrow1}\Omega(c)$, so that $\Omega$ is positive and weakly decreasing on $[0,1)$ with
$-\fraction{\mathrm{d} \Omega}{\mathrm{d} c}=(B/A)\,\Omega$.
\begin{enumerate}
\item[(a)] The function
\begin{equation}\label{eq:average}
\bstar(c)=\frac{\Omega(1)\,\mu(1)-\int_c^1\mu(s)\,d\Omega(s)}
{\Omega(1)-\int_c^1d\Omega(s)}
\quad(0\le c<1),\qquad
\bstar(1)=\mu(1),
\end{equation}
is the unique rest point of $(A,B,\mu)$, that is, the unique continuous function on $[0,1]$ with
$\bstar(1)=\mu(1)$ and $A\,\fraction{\mathrm{d} \bstar}{\mathrm{d} c}=B(\bstar-\mu)$ on $[0,1)$.
\item[(b)] The rest point satisfies $\mu\le\bstar\le\mu(1)$ and is weakly increasing.
\end{enumerate}
\end{lemma}

\begin{proof}
\emph{Idea.} The rest-point equation $A\,\fraction{\mathrm{d} \bstar}{\mathrm{d} c}=B(\bstar-\mu)$ is a linear first-order equation with a terminal condition at $c=1$,
and $\Omega$ is its integrating factor: for a solution, $\fraction{\mathrm{d} (\Omega\bstar)}{\mathrm{d} c}=\mu\,\fraction{\mathrm{d} \Omega}{\mathrm{d} c}$. Integrating from $c$ to $1$ gives
\eqref{eq:average}. Standard theory for the equation does not apply at $c=1$, because $B/A$ need not be integrable there (in the
main examples $A(1)=0$). We therefore check the formula directly and prove uniqueness by a monotonicity argument.

On $[0,1)$ the function $\Omega$ is continuously differentiable, because $B/A$ is continuous there by \condref{A1} and \condref{A2}. It is
monotone and bounded, so the limit $\Omega(1)\ge0$ exists, and the integrals in \eqref{eq:average} are Stieltjes integrals with
$d\Omega(s)=-(B/A)(s)\,\Omega(s)\,ds$ on $[0,1)$.

\emph{Step 1: the denominator.} Since $\int_c^1d\Omega=\Omega(1)-\Omega(c)$, the denominator of \eqref{eq:average} equals $\Omega(c)>0$.

\emph{Step 2: bounds, the first claim of (b).} The measure $-d\Omega$ is nonnegative. Hence \eqref{eq:average} expresses $\bstar(c)$ as a
weighted average of the values $\mu(s)$ with $c\le s\le1$: the weight $\Omega(1)$ sits at $s=1$, and the weight $-d\Omega(s)$ is spread over
$[c,1)$. All these values lie in $[\mu(c),\mu(1)]$ by \condref{A3}, so $\mu(c)\le\bstar(c)\le\mu(1)=C$. Equivalently, subtracting $C=\mu(1)$ from
\eqref{eq:average} gives
\begin{equation}\label{eq:B-deficit}
\bstar(c)=C-\frac1{\Omega(c)}\int_c^1\bigl(C-\mu(s)\bigr)\,\frac{B(s)}{A(s)}\,\Omega(s)\,ds\qquad(0\le c<1),
\end{equation}
and the integrand lies between $0$ and $(C-\mu(c))(B/A)\Omega$, whose integral over $[c,1)$ is $\Omega(c)-\Omega(1)\le\Omega(c)$.

\emph{Step 3: the rest-point equation.} For $c<1$, the numerator $\Omega(c)\bstar(c)=\Omega(1)\mu(1)-\int_c^1\mu\,d\Omega$ is differentiable
in $c$ with derivative $\mu(c)\,\fraction{\mathrm{d} \Omega}{\mathrm{d} c}(c)$, since the integrand $\mu\,(B/A)\,\Omega$ is continuous on $[0,1)$. The product rule gives
\[
\Omega\,\fraction{\mathrm{d} \bstar}{\mathrm{d} c}+\bstar\,\fraction{\mathrm{d} \Omega}{\mathrm{d} c}=\mu\,\fraction{\mathrm{d} \Omega}{\mathrm{d} c},
\qquad\text{that is,}\qquad
\Omega\,\fraction{\mathrm{d} \bstar}{\mathrm{d} c}=(\bstar-\mu)\Bigl(-\fraction{\mathrm{d} \Omega}{\mathrm{d} c}\Bigr)=\frac BA\,(\bstar-\mu)\,\Omega .
\]
Dividing by $\Omega>0$ and multiplying by $A$ gives $A\,\fraction{\mathrm{d} \bstar}{\mathrm{d} c}=B(\bstar-\mu)$ on $[0,1)$, with a one-sided derivative at $c=0$.

\emph{Step 4: continuity at the top type.} The function $\bstar$ is continuous on $[0,1)$ by Step 3. By Step 2,
$\mu(c)\le\bstar(c)\le\mu(1)$ for $c<1$, and $\mu$ is continuous, so $\bstar(c)\to\mu(1)=\bstar(1)$ as $c\uparrow1$.

\emph{Step 5: uniqueness.} Let $y$ be the difference of two rest points. Then $y$ is continuous on $[0,1]$, $y(1)=0$, and
$A\,\fraction{\mathrm{d} y}{\mathrm{d} c}=By$ on $[0,1)$. Hence
\[
\fraction{\mathrm{d} (y^2)}{\mathrm{d} c}=2y\,\fraction{\mathrm{d} y}{\mathrm{d} c}=2\,\frac BA\,y^2\ge0\qquad\text{on }[0,1),
\]
so $y^2$ is weakly increasing on $[0,1]$, and $y(c)^2\le y(1)^2=0$.

\emph{Step 6: monotonicity, the second claim of (b).} By Steps 2 and 3, $\fraction{\mathrm{d} \bstar}{\mathrm{d} c}=(B/A)(\bstar-\mu)\ge0$ on $[0,1)$; with continuity
on $[0,1]$, $\bstar$ is weakly increasing.

Steps 1--5 prove (a), and Steps 2 and 6 prove (b).
\end{proof}

\begin{lemma}[Comparison]\label{lem:comparison}
Let $T>0$, let $\sub$ be an upper semicontinuous subsolution with $\sub\le C$ and $\super$ a lower semicontinuous
supersolution with $\mu\le\super\le C$ on $\cB_T$. If $\sub\le\super$ at $t=0$ and at $c=1$, then $\sub\le\super$ on $\cB_T$.
\end{lemma}

\begin{proof}
We double the variables \citep{crandall1983viscosity,crandall1992users}.

\emph{Idea.} Suppose that $\sub>\super$ somewhere. If both functions were smooth, then at a point where the weighted gap
$\e^{-\lambda t}(\sub-\super)$ is maximal they would have the same slope $u$, and subtracting the two differential inequalities would
give a contradiction for two reasons.
\begin{itemize}
\item The zero-order term $-\varphi(c,z)A(c)$ is not monotone in $z$, since $\varphi$ may increase with $z$; by \condref{G1} it can raise the gap at
most at the relative rate $L_\varphi\norm A$. The weight $\e^{-\lambda t}$ with $\lambda>L_\varphi\norm A$ outweighs this.
\item The singular term $B\,\varphi(c,z)(z-\mu)/u$ is, by \condref{G2}, strictly larger at the larger value $z=\sub$ than at $z=\super$. With the
same slope $u>0$ it therefore pushes $\sub$ down harder than $\super$. Only the sign of $u$ matters here, not its size.
\end{itemize}
Since $\sub$ and $\super$ are only semicontinuous, we maximize over pairs of points with a quadratic penalty on their distance. The penalty
supplies smooth test functions that touch $\sub$ from above and $\super$ from below, with the same slope. One more point needs care: the
subsolution condition yields a positive test slope only above the anchor, where the state constraint is inactive. We place the contact
point of $\sub$ above the anchor by using the lower bound $\super\ge\mu$.

Suppose that $\sub(t_{\star},c_{\star})>\super(t_{\star},c_{\star})$ for some $(t_{\star},c_{\star})\in\cB_T$. Let $\norm A=\max A$, which bounds $|A|$
because $A\ge0$ on $[0,1]$ by \condref{A1} and \condref{A2}, and put
\[
\lambda=L_\varphi\norm A+1 .
\]

\emph{Step 1: the doubled functional.} For $\iota>0$ let
\[
\Xi(s,x,s',x')
=
\e^{-\lambda s}\bigl(\sub(s,x)-\super(s',x')\bigr)
-\frac{(s-s')^2+(x-x')^2}{2\iota}
\qquad\bigl((s,x),(s',x')\in\cB_T\bigr).
\]
It is upper semicontinuous, because $\sub$ is upper semicontinuous, $\super$ is lower semicontinuous and the weight is continuous and
positive. So it attains its maximum $\Xi_\iota$ on the compact set $\cB_T\times\cB_T$, at a point $(t,c,t',c')$ that depends on $\iota$.
Write $z_1=\sub(t,c)$ and $z_2=\super(t',c')$. Testing at the doubled point $(t_{\star},c_{\star},t_{\star},c_{\star})$ gives
\[
\e^{-\lambda t}(z_1-z_2)-\frac{(t-t')^2+(c-c')^2}{2\iota}=\Xi_\iota\ge\Xi_0=\e^{-\lambda t_{\star}}(\sub-\super)(t_{\star},c_{\star})>0 .
\]
Hence $z_1>z_2$. Since $z_1\le C$ and $z_2\ge\mu(c')\ge\mu(0)$,
\[
\frac{(t-t')^2+(c-c')^2}{2\iota}\le\e^{-\lambda t}(z_1-z_2)\le C-\mu(0)=R,
\]
so the two points are at distance at most $\sqrt{2\iota R}$.

\emph{Step 2: the limit $\iota\to0$.} Take a sequence $\iota\to0$. The points lie in the compact set $\cB_T$ and the values in $[\mu(0),C]$, so
along a subsequence $(t,c)$ and $(t',c')$ converge, to a common point $(t_{\infty},c_{\infty})$ by Step 1, and $z_1\to\bar z_1$,
$z_2\to\bar z_2$. In the limit:
\begin{itemize}
\item $\e^{-\lambda t_{\infty}}(\bar z_1-\bar z_2)\ge\Xi_0>0$, since the penalty is nonnegative;
\item $\bar z_1\le\sub(t_{\infty},c_{\infty})$ and $\bar z_2\ge\super(t_{\infty},c_{\infty})$, by semicontinuity;
\item $\mu(c_{\infty})\le\bar z_2<\bar z_1\le C$, since $z_2\ge\mu(c')$ and $\mu$ is continuous.
\end{itemize}
By the first two items, $(\sub-\super)(t_{\infty},c_{\infty})\ge\bar z_1-\bar z_2>0$. Since $\sub\le\super$ at $t=0$ and at $c=1$, we get
$t_{\infty}>0$ and $c_{\infty}<1$. Consequently, for all small $\iota$ in the subsequence,
\begin{equation}\label{eq:B-interior}
t,t'>0,\qquad c,c'<1,\qquad z_1>\mu(c),
\end{equation}
where the last inequality holds because $z_1\to\bar z_1>\mu(c_{\infty})$ and $\mu(c)\to\mu(c_{\infty})$. This is where the lower bound
$\super\ge\mu$ is used: it keeps the contact point of $\sub$ above the anchor.

\emph{Step 3: the two test functions.} Fix such an $\iota$. Freezing $(t',c')$ in $\Xi\le\Xi_\iota$ and solving for $\sub$ gives
$\sub\le\psi_1$ on $\cB_T$, and freezing $(t,c)$ gives $\super\ge\psi_2$ on $\cB_T$, where
\[
\begin{aligned}
\psi_1(s,x)&=z_2+\e^{\lambda s}\Bigl(\Xi_\iota+\frac{(s-t')^2+(x-c')^2}{2\iota}\Bigr),\\
\psi_2(s,x)&=z_1-\e^{\lambda t}\Bigl(\Xi_\iota+\frac{(t-s)^2+(c-x)^2}{2\iota}\Bigr).
\end{aligned}
\]
Equality holds at $(t,c)$ and $(t',c')$, respectively. So $\sub-\psi_1$ attains its maximum over $\cB_T$ at $(t,c)$, and $\super-\psi_2$
attains its minimum over $\cB_T$ at $(t',c')$. Both test functions are smooth. With
\[
u_\iota=\e^{\lambda t}\,\frac{c-c'}{\iota},\qquad \tau_\iota=\e^{\lambda t}\,\frac{t-t'}{\iota},
\]
and using $\e^{\lambda t}\bigl(\Xi_\iota+((t-t')^2+(c-c')^2)/(2\iota)\bigr)=z_1-z_2$, their derivatives at the contact points are
\[
\fraction{\partial \psi_1}{\partial t}(t,c)=\lambda(z_1-z_2)+\tau_\iota,\qquad \fraction{\partial \psi_1}{\partial c}(t,c)=u_\iota,\qquad
\fraction{\partial \psi_2}{\partial t}(t',c')=\tau_\iota,\qquad \fraction{\partial \psi_2}{\partial c}(t',c')=u_\iota .
\]
The two slopes coincide, and the time derivatives differ exactly by $\lambda(z_1-z_2)$, which comes from the weight.

By \eqref{eq:B-interior}, both contact points are tested points, and $\sub(t,c)=z_1>\mu(c)$. The subsolution condition above the anchor
therefore gives $u_\iota>0$ and
$\lambda(z_1-z_2)+\tau_\iota+\varphi(c,z_1)H(c,z_1,u_\iota)\le0$. Since $u_\iota>0$, the supersolution condition applies at $(t',c')$ and gives
$\tau_\iota+\varphi(c',z_2)H(c',z_2,u_\iota)\ge0$. Subtracting the second inequality from the first removes $\tau_\iota$; with
$H(c,z,u)=-A(c)+B(c)(z-\mu(c))/u$ we obtain
\begin{equation}\label{eq:B-doubling}
\begin{aligned}
&\underbrace{\lambda(z_1-z_2)-\varphi(c,z_1)A(c)+\varphi(c',z_2)A(c')}_{\text{zero-order part}}\\
&\qquad+\underbrace{\frac{
B(c)\varphi(c,z_1)(z_1-\mu(c))
-
B(c')\varphi(c',z_2)(z_2-\mu(c'))
}{u_\iota}}_{\text{singular part}}
\le0 .
\end{aligned}
\end{equation}

\emph{Step 4: both parts are eventually positive.} We let $\iota\to0$ along the subsequence. The points $(c,z_1)$ and $(c',z_2)$ lie in $\cK$
and converge to $(c_{\infty},\bar z_1)$ and $(c_{\infty},\bar z_2)$, which lie in $\cK$ by Step 2, and $\varphi$ is continuous on $\cK$.
\begin{itemize}
\item \emph{Zero-order part, by \condref{G1}.} It tends to
$\lambda(\bar z_1-\bar z_2)-A(c_{\infty})\bigl(\varphi(c_{\infty},\bar z_1)-\varphi(c_{\infty},\bar z_2)\bigr)$. By \condref{G1}, applied with
$\mu(c_{\infty})\le\bar z_2\le\bar z_1\le C$, and $0\le A\le\norm A$, this is at least
$(\lambda-L_\varphi\norm A)(\bar z_1-\bar z_2)=\bar z_1-\bar z_2>0$.
\item \emph{Numerator of the singular part, by \condref{G2}.} It tends to
\[
B(c_{\infty})
\bigl(
\varphi(c_{\infty},\bar z_1)(\bar z_1-\mu(c_{\infty}))
-
\varphi(c_{\infty},\bar z_2)(\bar z_2-\mu(c_{\infty}))
\bigr)
>0,
\]
by \condref{G2}, since $\bar z_1>\bar z_2$, and $B(c_{\infty})>0$, which holds because $c_{\infty}<1$.
\end{itemize}
For small $\iota$ both parts of \eqref{eq:B-doubling} are therefore positive. Since $u_\iota>0$, the left-hand side of
\eqref{eq:B-doubling} is positive, a contradiction. The size of $u_\iota$, which may tend to $0$ or to $\infty$, plays no role.
\end{proof}

\begin{lemma}[Barriers]\label{lem:barriers}
Let $T>0$, and let $U:\cB_T\to\R$ be continuous with (one-sided at the boundary) partial derivatives $\fraction{\partial U}{\partial t}$, $\fraction{\partial U}{\partial c}$ at every point of $(0,T]\times[0,1)$.
\begin{enumerate}
\item[(a)] If $\mu\le U\le C$, $\fraction{\partial U}{\partial c}>0$ and
$\fraction{\partial U}{\partial t}+\varphi(c,U)H(c,U,\fraction{\partial U}{\partial c})\le0$ (respectively $\ge0$) on $(0,T]\times[0,1)$, then $U$ is a
subsolution (respectively a supersolution).
\item[(b)] If $\mu\le U$ and
$\fraction{\partial U}{\partial t}+\varphi(c,U)H(c,U,\fraction{\partial U}{\partial c})\ge0$ wherever $U<C$ and $\fraction{\partial U}{\partial c}>0$, then
$\min\{C,U\}$ is a supersolution.
\item[(c)] If $U\le C$, and $\fraction{\partial U}{\partial c}>0$ and
$\fraction{\partial U}{\partial t}+\varphi(c,U)H(c,U,\fraction{\partial U}{\partial c})\le0$ wherever $U>\mu$, then
$\max\{\mu,U\}$ is a subsolution.
\end{enumerate}
\end{lemma}

\begin{proof}
\emph{Idea.} At a contact point the derivatives of the test function are compared with those of $U$, and the differential inequality of
$U$ is transferred to the test function. The only property of the equation needed is monotonicity in the slope above the anchor. In (b)
and (c), at points where the clipping is active, the flat cap or anchor forces the test function to have a slope or a time derivative of
the sign that the definition allows.

We use two facts throughout. Let $(t_{\star},c_{\star})\in(0,T]\times[0,1)$, and evaluate all derivatives there.
\begin{itemize}
\item[(F1)] \emph{One-sided derivatives at a contact point.} If $V-\psi$ attains its maximum over $\cB_T$ at $(t_{\star},c_{\star})$ and $V$ has
partial derivatives there, then comparing with earlier times $s<t_{\star}$ and larger types $x>c_{\star}$, which lie in $\cB_T$, gives
\begin{equation}\label{eq:B-onesided}
\fraction{\partial \psi}{\partial t}\le \fraction{\partial V}{\partial t},\qquad \fraction{\partial \psi}{\partial c}\ge \fraction{\partial V}{\partial c} .
\end{equation}
At a minimum both inequalities reverse. Equalities hold at interior points, but only these inequalities are needed; they are what
makes the terminal time $t_{\star}=T$ and the bottom type $c_{\star}=0$ admissible.
\item[(F2)] \emph{Monotonicity in the slope.} If $(c,z)\in\cK$ and $c<1$, then
$u\mapsto\varphi(c,z)H(c,z,u)=\varphi(c,z)\bigl(-A(c)+B(c)(z-\mu(c))/u\bigr)$ is weakly decreasing on $(0,\infty)$, since $\varphi>0$, $B\ge0$
and $z-\mu(c)\ge0$.
\end{itemize}

(a) \emph{Subsolution.} Let $U-\psi$ attain its maximum over $\cB_T$ at $(t_{\star},c_{\star})$. By \eqref{eq:B-onesided},
$\fraction{\partial \psi}{\partial t}\le \fraction{\partial U}{\partial t}$ and $\fraction{\partial \psi}{\partial c}\ge \fraction{\partial U}{\partial c}>0$.
\begin{itemize}
\item If $U>\mu(c_{\star})$, then $\fraction{\partial \psi}{\partial c}>0$, and by (F2)
\[
\fraction{\partial \psi}{\partial t}+\varphi(c_{\star},U)H(c_{\star},U,\fraction{\partial \psi}{\partial c})\le \fraction{\partial U}{\partial t}+\varphi(c_{\star},U)H(c_{\star},U,\fraction{\partial U}{\partial c})\le0 .
\]
\item If $U\le\mu(c_{\star})$, then $U=\mu(c_{\star})$, and $H(c_{\star},\mu(c_{\star}),\fraction{\partial U}{\partial c})=-A(c_{\star})$. The differential inequality reads
$\fraction{\partial U}{\partial t}\le\varphi(c_{\star},\mu(c_{\star}))A(c_{\star})$, and $\fraction{\partial \psi}{\partial t}\le \fraction{\partial U}{\partial t}$ gives the anchor clause of \cref{def:solution}.
\end{itemize}
\emph{Supersolution.} Let $U-\psi$ attain its minimum at $(t_{\star},c_{\star})$ with $\fraction{\partial \psi}{\partial c}>0$. By \eqref{eq:B-onesided} reversed,
$\fraction{\partial \psi}{\partial t}\ge \fraction{\partial U}{\partial t}$ and $0<\fraction{\partial \psi}{\partial c}\le \fraction{\partial U}{\partial c}$. By (F2),
$\fraction{\partial \psi}{\partial t}+\varphi(c_{\star},U)H(c_{\star},U,\fraction{\partial \psi}{\partial c})\ge \fraction{\partial U}{\partial t}+\varphi(c_{\star},U)H(c_{\star},U,\fraction{\partial U}{\partial c})\ge0$.

(b) Let $V=\min\{C,U\}$, and let $V-\psi$ attain its minimum over $\cB_T$ at $(t_{\star},c_{\star})$ with $\fraction{\partial \psi}{\partial c}>0$.
\begin{itemize}
\item If $U(t_{\star},c_{\star})<C$, then $V\le U$ everywhere with equality at $(t_{\star},c_{\star})$, so $U-\psi$ also attains its minimum there.
By \eqref{eq:B-onesided} reversed, $\fraction{\partial \psi}{\partial t}\ge \fraction{\partial U}{\partial t}$ and $0<\fraction{\partial \psi}{\partial c}\le \fraction{\partial U}{\partial c}$. So $U<C$ and $\fraction{\partial U}{\partial c}>0$ at this point, the hypothesis applies, and
(F2) gives the supersolution inequality as in (a), with $V=U$ at the point.
\item If $U(t_{\star},c_{\star})\ge C$, then $V=C$ at $(t_{\star},c_{\star})$ and $V\le C$ everywhere. For every $q\in\cB_T$,
$C-\psi(t_{\star},c_{\star})\le V(q)-\psi(q)\le C-\psi(q)$, so $\psi$ attains its maximum over $\cB_T$ at $(t_{\star},c_{\star})$. Comparing with
larger types gives $\fraction{\partial \psi}{\partial c}\le0$, which contradicts $\fraction{\partial \psi}{\partial c}>0$. So this case does not occur: a test function with positive slope
cannot touch the flat cap from below.
\end{itemize}

(c) Let $V=\max\{\mu,U\}$, and let $V-\psi$ attain its maximum over $\cB_T$ at $(t_{\star},c_{\star})$.
\begin{itemize}
\item If $U(t_{\star},c_{\star})>\mu(c_{\star})$, then $V\ge U$ everywhere with equality at $(t_{\star},c_{\star})$, so $U-\psi$ attains its maximum
there. By \eqref{eq:B-onesided}, $\fraction{\partial \psi}{\partial t}\le \fraction{\partial U}{\partial t}$ and $\fraction{\partial \psi}{\partial c}\ge \fraction{\partial U}{\partial c}>0$, and (F2) gives the condition above the anchor as in (a).
\item If $U(t_{\star},c_{\star})\le\mu(c_{\star})$, then $V=\mu(c_{\star})$ at $(t_{\star},c_{\star})$, so only the anchor clause has to be checked.
Since $V\ge\mu$ everywhere and $\mu$ does not depend on time, for every $s\in[0,T]$
\[
\mu(c_{\star})-\psi(s,c_{\star})\le V(s,c_{\star})-\psi(s,c_{\star})\le\mu(c_{\star})-\psi(t_{\star},c_{\star}).
\]
So $\psi(\cdot,c_{\star})$ attains its minimum over $[0,T]$ at $t_{\star}>0$. Comparing with earlier times gives
$\fraction{\partial \psi}{\partial t}\le0\le\varphi(c_{\star},\mu(c_{\star}))A(c_{\star})$.\qedhere
\end{itemize}
\end{proof}

\begin{lemma}[Restart]\label{lem:shift}
Let $0\le t_0\le T$. If $\sub$ is a bounded subsolution on $\cB_T$, then
$\sub(t_0+\cdot,\cdot)$ is a subsolution on $\cB_{T-t_0}$; if $\super$ is a bounded supersolution on $\cB_T$, then
$\super(t_0+\cdot,\cdot)$ is a supersolution on $\cB_{T-t_0}$.
\end{lemma}

\begin{proof}
\emph{Idea.} A test function for the restarted function only has to stay above it at times after $t_0$; before $t_0$ it may lie below
$\sub$. We add a term that vanishes to high order at the contact time and is large before $t_0$. This makes the test function admissible
on all of $\cB_T$ and leaves its derivatives at the contact point unchanged. Boundedness of $\sub$ tells us how large the term must be.

Let $\gamma=\sub(t_0+\cdot,\cdot)$, and let $\gamma-\psi$ attain its maximum over $\cB_{T-t_0}$ at $(t_{\star},c_{\star})$ with $t_{\star}>0$ and
$c_{\star}<1$. Choose $K\ge|\sub|$ on $\cB_T$ and $M\ge|\psi(\sigma-t_0,x)|$ for $(\sigma,x)\in\cB_T$; $M$ exists since $\psi$ is continuous. Put
\[
\tilde\psi(\sigma,x)
=
\psi(\sigma-t_0,x)+\Lambda\,(\sigma-t_0-t_{\star})^4,
\qquad
\Lambda=\frac{2K+2M+1}{t_{\star}^4}.
\]
We claim that $\sub-\tilde\psi$ attains its maximum over $\cB_T$ at $(t_0+t_{\star},c_{\star})$, where it equals
$v_{\star}=\gamma(t_{\star},c_{\star})-\psi(t_{\star},c_{\star})\ge-K-M$.
\begin{itemize}
\item At times $\sigma\ge t_0$, the point $(\sigma-t_0,x)$ lies in $\cB_{T-t_0}$, so $\sub(\sigma,x)-\psi(\sigma-t_0,x)\le v_{\star}$, and the quartic term is
nonnegative.
\item At times $\sigma<t_0$, we have $(\sigma-t_0-t_{\star})^4\ge t_{\star}^4$, so
$\sub(\sigma,x)-\tilde\psi(\sigma,x)\le K+M-(2K+2M+1)<-K-M\le v_{\star}$.
\end{itemize}
The function $\tilde\psi$ is continuously differentiable, and the quartic term has zero derivative at $\sigma=t_0+t_{\star}$. So the value of
$\sub$ and the derivatives of $\tilde\psi$ at $(t_0+t_{\star},c_{\star})\in(0,T]\times[0,1)$ are those of $\gamma$ and $\psi$ at $(t_{\star},c_{\star})$. Both
clauses of the subsolution condition involve only these quantities, so the condition for $\sub$ at $(t_0+t_{\star},c_{\star})$ is the
required condition for $\gamma$ at $(t_{\star},c_{\star})$. For the supersolution, subtract the quartic term instead of adding it, and
replace maxima by minima.
\end{proof}

\begin{lemma}[Monotonicity]\label{lem:monotone}
Let $\sub$ be a continuous subsolution on $\cB_T$ with $\sub\ge\mu$. Then $\sub(t,\cdot)$ is weakly increasing for every
$t\in(0,T]$.
\end{lemma}

\begin{proof}
\emph{Idea.} Above the anchor, the subsolution condition forbids contact from above by a test function with nonpositive slope. If
$\sub(t,\cdot)$ decreased somewhere, a plane that falls in $c$, slowly enough, would lie above the graph of $\sub$ near the decrease
and touch it at a point above the anchor. Quadratic penalties localize the contact in time and keep it to the left of the decrease.

Suppose that $\gamma=\sub(t,c_1)-\sub(t,c_2)>0$ for some $t\in(0,T]$ and $c_1<c_2$. The constants are chosen as follows.
\begin{itemize}
\item The slope $a=\gamma/(2(c_2-c_1))$: the falling plane $-a(x-c_1)$ drops only by $\gamma/2$ between $c_1$ and $c_2$, while $\sub(t,\cdot)$
drops by $\gamma$.
\item A bound $M\ge|\sub|$ on $\cB_T$, and $\rho_0>0$ such that $|\sub-\sub(t,c_2)|<\gamma/8$ at the points of $\cB_T$ within distance $\rho_0$ of
$(t,c_2)$ in each coordinate; $\rho_0$ exists by continuity.
\item The localization radius $\rho=\min\{\rho_0/2,\,t/2,\,\gamma/(8a)\}$, and the penalty $K=(2M+a+1)/\rho^2$, so that a penalty of $K\rho^2$ outweighs
every possible gain.
\end{itemize}
Let
\[
\psi(s,x)
=
-a(x-c_1)
+K(s-t)^2
+K\bigl((x-c_2)^+\bigr)^2 ,
\]
which is continuously differentiable. The function $\sub-\psi$ is continuous, so it attains its maximum over $\cB_T$ at some point $(s,x)$.
Since $\psi(t,c_1)=0$, the maximum is at least $\sub(t,c_1)$:
\begin{equation}\label{eq:B-mono-max}
\sub(t,c_1)\le\sub(s,x)+a(x-c_1)-K(s-t)^2-K\bigl((x-c_2)^+\bigr)^2 .
\end{equation}
We locate $(s,x)$, using $|\sub|\le M$ and $a(x-c_1)\le a$.
\begin{itemize}
\item \emph{$|s-t|<\rho$.} Otherwise $K(s-t)^2\ge2M+a+1$, and the right-hand side of \eqref{eq:B-mono-max} is at most $-M-1<\sub(t,c_1)$.
\item \emph{$x<c_2$.} If $x\ge c_2+\rho$, the penalty in $x$ gives the same contradiction. If $c_2\le x<c_2+\rho$, the point $(s,x)$ lies within
$\rho<\rho_0$ of $(t,c_2)$, so $\sub(s,x)<\sub(t,c_2)+\gamma/8$; moreover
$a(x-c_1)=a(c_2-c_1)+a(x-c_2)\le\gamma/2+a\rho\le\gamma/2+\gamma/8$. The right-hand side of \eqref{eq:B-mono-max} is then less than
$\sub(t,c_2)+3\gamma/4=\sub(t,c_1)-\gamma/4$, a contradiction.
\end{itemize}
Thus $s>t-\rho\ge t/2>0$ and $x<c_2\le1$, so $(s,x)$ is a tested point. It lies above the anchor: by \eqref{eq:B-mono-max},
$\sub(s,x)\ge\sub(t,c_1)-a(x-c_1)$, and
\begin{itemize}
\item if $x\ge c_1$, then $\sub(s,x)\ge\sub(t,c_1)-\gamma/2>\sub(t,c_2)\ge\mu(c_2)\ge\mu(x)$;
\item if $x<c_1$, then $\sub(s,x)>\sub(t,c_1)\ge\mu(c_1)\ge\mu(x)$.
\end{itemize}
Here we used $\sub\ge\mu$ and \condref{A3}. The subsolution condition above the anchor requires $\fraction{\partial \psi}{\partial c}(s,x)>0$. But $x<c_2$, so the penalty in $x$
has zero derivative there and $\fraction{\partial \psi}{\partial c}(s,x)=-a<0$, a contradiction.
\end{proof}

\begin{lemma}[Lower barriers]\label{lem:lowerbarrier}
Let $\beta$ be a solution from a feasible $b$, and let
$\beta_{\mathrm{low}}:\R^2\to\R$ be continuous on $[0,\infty)\times[0,1]$, have partial derivatives at
every point of $(0,\infty)\times[0,1)$, and satisfy
$\beta_{\mathrm{low}}(0,\cdot)\le b$ and $\beta_{\mathrm{low}}(t,1)\le C$.
If $\beta_{\mathrm{low}}\ge\mu$, $\fraction{\partial \beta_{\mathrm{low}}}{\partial c}>0$ and
\[
\fraction{\partial \beta_{\mathrm{low}}}{\partial t}
\le
\varphi(c,\beta_{\mathrm{low}})
\left(
A-B\frac{\beta_{\mathrm{low}}-\mu}{\fraction{\partial \beta_{\mathrm{low}}}{\partial c}}
\right)
\]
on $(0,\infty)\times[0,1)$, then $\beta_{\mathrm{low}}\le\beta$.
\end{lemma}

\begin{proof}
The plan is to apply \cref{lem:barriers}(a) and then \cref{lem:comparison} on every $\cB_T$. The only hypothesis not stated directly is the
cap $\beta_{\mathrm{low}}\le C$. At $t=0$ it holds because $\beta_{\mathrm{low}}(0,\cdot)\le b\le C$. For $t>0$, the function $\beta_{\mathrm{low}}(t,\cdot)$
is continuous on $[0,1]$ with positive derivative on $[0,1)$, hence increasing, so $\beta_{\mathrm{low}}(t,c)\le\beta_{\mathrm{low}}(t,1)\le C$.
Thus $\mu\le\beta_{\mathrm{low}}\le C$. The differential inequality is $\fraction{\partial \beta_{\mathrm{low}}}{\partial t}+\varphi(c,\beta_{\mathrm{low}})H(c,\beta_{\mathrm{low}},\fraction{\partial \beta_{\mathrm{low}}}{\partial c})\le0$,
so \cref{lem:barriers}(a) makes $\beta_{\mathrm{low}}$ a subsolution on every $\cB_T$. The solution $\beta$ is a continuous supersolution with
$\mu\le\beta\le C$, and $\beta_{\mathrm{low}}\le\beta$ at $t=0$ (since $\beta(0,\cdot)=b$) and at $c=1$ (since $\beta(t,1)=C$). \Cref{lem:comparison}
gives $\beta_{\mathrm{low}}\le\beta$ on every $\cB_T$.
\end{proof}

\begin{proof}[Proof of \cref{thm:main}]
\emph{Rest point.}
By \cref{lem:rest}, there is exactly one rest point $\bstar$, and
$\mu\le\bstar\le C$ with $\bstar$ weakly increasing. Formula \eqref{eq:average} involves only $A$, $B$ and $\mu$, so $\bstar$ does not depend on
$\varphi$.

\emph{(i)} \Cref{lem:scheme-feasible} gives a solution $\beta$ from $b$. If $\beta_1$ and $\beta_2$ are two solutions from $b$, then on every
$\cB_T$ the function $\beta_1$ is a continuous subsolution with $\beta_1\le C$, the function $\beta_2$ is a continuous supersolution with
$\mu\le\beta_2\le C$, and both equal $b$ at $t=0$ and $C$ at $c=1$. \Cref{lem:comparison} gives $\beta_1\le\beta_2$, and exchanging the
roles gives $\beta_1=\beta_2$. By \cref{def:solution}, $\beta$ is continuous on every $\cB_T$, hence on $[0,\infty)\times[0,1]$, and satisfies
$\mu\le\beta\le C$ and $\beta(t,1)=C$. For $t>0$, \cref{lem:monotone} shows that $\beta(t,\cdot)$ is weakly increasing, and $\beta(0,\cdot)=b$ is
weakly increasing. Hence $\beta(t,\cdot)$ is feasible for every $t\ge0$.

\emph{(ii)} \emph{Idea.} We bound $\beta$ above and below by the rest point shifted by a scalar envelope $e(t)$ and clipped to the feasible
interval $[\mu,C]$. By \eqref{eq:B-hamiltonian-identity}, the field at the shift $\bstar\pm e$ pushes it back toward $\bstar$, at type $c$ at
the relative rate $\varphi A(c)/(\bstar(c)-\mu(c))$, which is at least $\varphi_-A(c)/w(c)$. So the shifts are barriers as long as $e$
shrinks no faster than this rate at every type that matters. Comparison then gives $\cE\le e$. Restarting at an arbitrary time
$t_0$ with $e(t_0)=\cE(t_0)$ bounds the Dini derivative of $\cE$ by $\fraction{\mathrm{d} e}{\mathrm{d} t}(t_0)$. Each claim of (ii) follows from a suitable choice of $e$.

\emph{Step 1: clipped shifts of the rest point are barriers.} Let $e:[0,\infty)\to(0,\infty)$ be differentiable and satisfy the
\emph{envelope criterion}
\begin{equation}\label{eq:B-envelope}
-\fraction{\mathrm{d} e}{\mathrm{d} s}(s)\,w(c)\le\varphi_-A(c)\,e(s)
\qquad\text{whenever }s\ge0,\ 0\le c<1\text{ and }e(s)\le w(c).
\end{equation}
Types with $w(c)<e(s)$ impose no condition, because there the clipping is active: $\bstar+e>\mu+w=C$ and $\bstar-e<C-w=\mu$. We show that, for every $T>0$, the clipped shifts
\[
U_+(s,c)=\min\{C,\bstar(c)+e(s)\},\qquad
U_-(s,c)=\max\{\mu(c),\bstar(c)-e(s)\}
\]
are a supersolution and a subsolution on $\cB_T$, respectively.
At every type with $\fraction{\mathrm{d} \bstar}{\mathrm{d} c}>0$, \eqref{eq:B-hamiltonian-identity} with $y_0=\pm e$ and $y_1=0$, followed by the rest-point equation
$B/(\fraction{\mathrm{d} \bstar}{\mathrm{d} c})=A/(\bstar-\mu)$, gives
\begin{equation}\label{eq:B-shift}
\varphi(c,\bstar\pm e)\,H\Bigl(c,\bstar\pm e,\fraction{\mathrm{d} \bstar}{\mathrm{d} c}\Bigr)
=\pm\varphi(c,\bstar\pm e)\,\frac{A\,e}{\bstar-\mu}.
\end{equation}
\begin{itemize}
\item \emph{Upper barrier.} Apply \cref{lem:barriers}(b) to $U=\bstar+e(s)$, which satisfies $U\ge\mu$ and has $\fraction{\partial U}{\partial s}=\fraction{\mathrm{d} e}{\mathrm{d} s}$ and $\fraction{\partial U}{\partial c}=\fraction{\mathrm{d} \bstar}{\mathrm{d} c}$. At a
point with $U<C$ and $\fraction{\partial U}{\partial c}>0$ we have $\bstar>\mu$, because $\fraction{\mathrm{d} \bstar}{\mathrm{d} c}=B(\bstar-\mu)/A$, and $e<C-\bstar\le w$, so \eqref{eq:B-envelope}
applies. With \eqref{eq:B-shift}, $0<\bstar-\mu\le w$, and $\varphi\ge\varphi_-$ on $\cK$,
\[
\fraction{\partial U}{\partial s}+\varphi(c,U)H(c,U,\fraction{\partial U}{\partial c})=\fraction{\mathrm{d} e}{\mathrm{d} s}+\varphi(c,U)\frac{A\,e}{\bstar-\mu}\ge\fraction{\mathrm{d} e}{\mathrm{d} s}+\varphi_-\frac{A\,e}{w}\ge0 .
\]
\item \emph{Lower barrier.} Apply \cref{lem:barriers}(c) to $U=\bstar-e(s)\le C$. At a point with $U>\mu$ we have $\bstar-\mu>e>0$, so
$\fraction{\partial U}{\partial c}=\fraction{\mathrm{d} \bstar}{\mathrm{d} c}>0$, and $e<\bstar-\mu\le w$, so \eqref{eq:B-envelope} applies. With \eqref{eq:B-shift},
\[
\fraction{\partial U}{\partial s}+\varphi(c,U)H(c,U,\fraction{\partial U}{\partial c})=-\fraction{\mathrm{d} e}{\mathrm{d} s}-\varphi(c,U)\frac{A\,e}{\bstar-\mu}\le-\fraction{\mathrm{d} e}{\mathrm{d} s}-\varphi_-\frac{A\,e}{w}\le0 .
\]
\end{itemize}

\emph{Step 2: the envelope bound from any time.} Let $t_0\ge0$, and let $e$ be as in Step 1 with $e(0)\ge\cE(t_0)$. We show
\begin{equation}\label{eq:B-envbound}
\cE(t_0+s)\le e(s)\qquad(s\ge0).
\end{equation}
Let $\gamma(s,c)=\beta(t_0+s,c)$. The solution is bounded, since $\mu(0)\le\beta\le C$, so by \cref{lem:shift} on $\cB_{t_0+T}$, $\gamma$ is a subsolution and a supersolution on $\cB_T$ for every $T>0$; it is
continuous, satisfies $\mu\le\gamma\le C$ and $\gamma(s,1)=C$, and so it is a solution from $\beta(t_0,\cdot)$. We compare $\gamma$ with the barriers
of Step 1 by \cref{lem:comparison}.
\begin{itemize}
\item \emph{$\gamma\le U_+$.} At $s=0$: $\gamma(0,\cdot)=\beta(t_0,\cdot)\le C$ and $\beta(t_0,\cdot)\le\bstar+\cE(t_0)\le\bstar+e(0)$. At $c=1$: both equal $C$,
since $\bstar(1)=C$. Moreover $\mu\le U_+\le C$, because $\bstar\ge\mu$.
\item \emph{$U_-\le\gamma$.} At $s=0$: $\beta(t_0,\cdot)\ge\mu$ and $\beta(t_0,\cdot)\ge\bstar-e(0)$. At $c=1$: both equal $C$. Moreover $U_-\le C$.
\end{itemize}
Hence $\bstar-e(s)\le U_-\le\gamma(s,\cdot)\le U_+\le\bstar+e(s)$, which is \eqref{eq:B-envbound}.

\emph{Step 3: the rate $\Psi$.} First, $\Psi$ is positive and weakly increasing on $(0,R]$. Let $0<E\le R$. The set $\{c\in[0,1]:w(c)\ge E\}$ is
compact, because $w$ is continuous; it contains $c=0$, because $w(0)=R$; and it excludes $c=1$, because $w(1)=0$. On it, $A/w$ is continuous
and positive by \condref{A2}, so the minimum $\Psi(E)$ is attained and positive. A larger $E$ gives a smaller set, hence a weakly larger minimum.

Second, the condition $-\fraction{\mathrm{d} e}{\mathrm{d} s}\le\varphi_-\Psi(e)\,e$ implies the envelope criterion. Let $e>0$ be differentiable with
$-\fraction{\mathrm{d} e}{\mathrm{d} s}\le\varphi_-\Psi(e)\,e$. If $e(s)\le w(c)$ with $c<1$, then $c$ belongs to the set defining $\Psi(e(s))$, so $\Psi(e(s))\le A(c)/w(c)$,
and multiplying by $w(c)>0$ gives \eqref{eq:B-envelope}. By Step 2,
\begin{equation}\label{eq:B-psi-bound}
-\fraction{\mathrm{d} e}{\mathrm{d} s}\le\varphi_-\Psi(e)\,e\ \text{ and }\ e(0)\ge\cE(t_0)
\quad\Longrightarrow\quad
\cE(t_0+s)\le e(s)\quad(s\ge0).
\end{equation}

\emph{Step 4: elementary properties of $\cE$.} Since $\beta(t,\cdot)$ and $\bstar$ take values in $[\mu,C]$, we have $\cE(t)\le\norm w=R$, and
$\cE(t)=0$ exactly when $\beta(t,\cdot)=\bstar$. For every $\delta>0$ the constant envelope $e\equiv\cE(t_0)+\delta$ satisfies
\eqref{eq:B-envelope}, because its left-hand side vanishes. So $\cE(t)\le\cE(t_0)+\delta$ for $t\ge t_0$, and $\cE$ is weakly decreasing.

\emph{Step 5: the strict Dini bound.} Fix $t_0$ with $V=\cE(t_0)>0$, and put $r=\varphi_-\Psi(V/2)$, which is positive by Step 3 since
$0<V/2\le R$. We want an envelope that starts at $V$ and decreases as fast as possible at $s=0$, subject to
$-\fraction{\mathrm{d} e}{\mathrm{d} s}\le\varphi_-\Psi(e)\,e$. The rate $\Psi(e)$ can be small when $e$ is small, so we keep $e\ge V/2$, where $\Psi(e)\ge\Psi(V/2)$, and let $e$ approach
the floor $V/2$ at the constant rate $r$:
\[
e(s)=\frac V2\bigl(1+\e^{-rs}\bigr).
\]
Then $V/2\le e\le V\le R$, and $-\fraction{\mathrm{d} e}{\mathrm{d} s}=r(e-V/2)\le re=\varphi_-\Psi(V/2)\,e\le\varphi_-\Psi(e)\,e$ by monotonicity of $\Psi$. By
\eqref{eq:B-psi-bound} and $e(0)=V=\cE(t_0)$,
\[
\frac{\cE(t_0+s)-\cE(t_0)}{s}\le\frac{e(s)-e(0)}{s}\to\fraction{\mathrm{d} e}{\mathrm{d} s}(0)=-\frac r2V\qquad(s\downarrow0).
\]
Hence $D^+\cE(t_0)\le-\frac12\varphi_-\Psi(\cE(t_0)/2)\cE(t_0)<0$. (A floor $\theta V$ with $0<\theta<1$ gives
$-(1-\theta)\varphi_-\Psi(\theta V)V$ in the same way; the theorem uses $\theta=\frac12$.) With Step 4, $\cE$ vanishes exactly at the rest point and
has a negative Dini derivative wherever it is positive, so it is a Lyapunov functional.

\emph{Step 6: convergence.} If $R=0$, then $\mu\equiv C$, so $\beta=\bstar=C$. Let $R>0$ and $\varepsilon>0$. The envelope is built from a cutoff
type. Since $w$ is continuous with $w(1)=0$, there is $d\in[0,1)$ with $w(d)\le\varepsilon/2$, and $a=\min_{[0,d]}A$ is positive by \condref{A2}.
Types $c\le d$ are damped at the rate $\varphi_-A(c)/w(c)\ge\varphi_-a/R$ or faster. Types $c>d$ have cost gap at most $w(d)$, and their error
is at most $w(d)$ anyway. This suggests an envelope that decays at the rate $\varphi_-a/R$ down to the floor $w(d)$:
\[
e(s)=w(d)+R\,\e^{-\varphi_-as/R}.
\]
It satisfies $e(0)\ge R\ge\cE(0)$. To check \eqref{eq:B-envelope}, let $e(s)\le w(c)$ with $c<1$, and write $\zeta=R\,\e^{-\varphi_-as/R}>0$. Then
$w(c)\ge w(d)+\zeta>w(d)$, so $\mu(c)<\mu(d)$; by \condref{A3}, $c<d$, and hence $A(c)\ge a$. Using $w(c)\le R$,
\[
-\fraction{\mathrm{d} e}{\mathrm{d} s}(s)\,w(c)=\frac{\varphi_-a}{R}\,\zeta\,w(c)\le\varphi_-a\,\zeta\le\varphi_-A(c)\,\zeta\le\varphi_-A(c)\,e(s).
\]
By Step 2 with $t_0=0$, $\cE(s)\le\varepsilon/2+R\,\e^{-\varphi_-as/R}$, which is at most $\varepsilon$ for large $s$. Since $\varepsilon$ was arbitrary, $\cE(t)\to0$.

% Coauthor note: This is the central object for the convergence proof as it constructs the ``vanishing barrier envelope.'' Actually, this method seems novel and we should discuss if the method should become the central contribution.

\emph{Step 7: power rates.} Let $A\ge aw^p$ on $[0,1)$ with $a>0$ and $p\ge1$, and fix $t_0\ge0$. If $\cE(t_0)=0$, then $\cE=0$ on $[t_0,\infty)$ by
Step 4, and both claims hold. Otherwise let $e$ solve
\[
\fraction{\mathrm{d} e}{\mathrm{d} s}=-\varphi_-a\,e^p,\qquad e(0)=\cE(t_0);
\]
explicitly, $e(t-t_0)$ is the right-hand side of \eqref{eq:power}. It is positive and differentiable. It satisfies \eqref{eq:B-envelope}: if
$e\le w(c)$, then $e^{p-1}\le w(c)^{p-1}$ because $p\ge1$, and
\[
-\fraction{\mathrm{d} e}{\mathrm{d} s}\,w(c)=\varphi_-a\,e\,\bigl(e^{p-1}w(c)\bigr)\le\varphi_-a\,w(c)^p\,e\le\varphi_-A(c)\,e .
\]
Step 2 gives \eqref{eq:power}. Since $\cE(t)-\cE(t_0)\le e(t-t_0)-e(0)$ for $t\ge t_0$ and $\fraction{\mathrm{d} e}{\mathrm{d} s}(0)=-\varphi_-a\,\cE(t_0)^p$, the upper right Dini
derivative of $\cE$ at $t_0$ is at most $-\varphi_-a\,\cE(t_0)^p$.

\emph{(iii)} \emph{Idea.} The upper bound comes from (ii). For the lower bound we construct an explicit subsolution
$\beta_{\mathrm{low}}=\bstar+y_-$ whose error $y_-$ decays at each type at a fixed multiple of the local damping rate of that type, and
apply \cref{lem:lowerbarrier}. For $p>1$ the damping rate vanishes at the top type. The types near the top are therefore slow, and
evaluating $y_-$ at the types whose damping time is of order $t$ gives the power lower bound.

\emph{Step 1: the family and its rest point.} Let $p\ge1$, $\varphi\equiv1$ (which satisfies \cref{ass:metric} with $L_\varphi=0$ and
$\varphi_-=1$), and
\[
A(c)=1-c,\qquad B\equiv1,\qquad \mu(c)=1-(1-c)^{1/p}.
\]
These coefficients satisfy \cref{ass:main}, with $C=1$, $R=1$, $w=(1-c)^{1/p}$ and $A=1-c=w^p$. So the growth condition of (ii) holds with
equality and $a=\varphi_-=1$. For $p=1$ this is the uniform duopoly. The rest point is an affine function of the cost gap: for
$\bstar=1-\kappa w$ we have $A\,\fraction{\mathrm{d} \bstar}{\mathrm{d} c}=(\kappa/p)\,w$ and $\bstar-\mu=(1-\kappa)\,w$, so $\kappa=p/(p+1)$ solves the rest-point equation,
and $\bstar(1)=1$. By \cref{lem:rest} it is the rest point:
\begin{equation}\label{eq:B-family-rest}
\bstar=1-\frac p{p+1}\,w,\qquad \fraction{\mathrm{d} \bstar}{\mathrm{d} c}=\frac1{p+1}(1-c)^{1/p-1}.
\end{equation}

\emph{Step 2: the initial strategy and the upper bound.}
Put $\Theta=p/(2(p+1))$, $\Gamma=2(p+1)^2/p$, and $b=1-\Theta w$. The strategy $b$ lies halfway between the rest point and the cap:
$b-\bstar=\Theta w$ and $C-\bstar=2\Theta w$. It is continuous and weakly increasing with $\mu=1-w\le b\le1$, so it is feasible, and
$\cE(0)=\Theta\,w(0)=\Theta$. By (ii) with $a=\varphi_-=1$ and $t_0=0$, the error satisfies $\cE(t)\le\frac14\e^{-t}$ for $p=1$, and for $p>1$
\[
\cE(t)\le\Theta\bigl(1+(p-1)\Theta^{p-1}t\bigr)^{-1/(p-1)}\le\Theta\,m^{-1/(p-1)}(1+t)^{-1/(p-1)},\qquad m=\min\{1,(p-1)\Theta^{p-1}\}.
\]

\emph{Step 3: the lower barrier.} We look for an error profile proportional to the initial error, $y_-=\Theta wZ$, with a decay factor
$0<Z\le1$. For $B=1$, \eqref{eq:B-hamiltonian-identity} shows that $\beta_{\mathrm{low}}=\bstar+y_-$ satisfies the inequality of
\cref{lem:lowerbarrier} exactly when
\begin{equation}\label{eq:B-lower-target}
\fraction{\partial y_-}{\partial t}\cdot\fraction{\partial \beta_{\mathrm{low}}}{\partial c}\le A\,\fraction{\partial y_-}{\partial c}-y_-
\qquad\text{wherever }\fraction{\partial \beta_{\mathrm{low}}}{\partial c}>0 .
\end{equation}
The local damping rate at the rest point is $B/(\fraction{\mathrm{d} \bstar}{\mathrm{d} c})=(p+1)(1-c)^{1-1/p}$ by \eqref{eq:B-family-rest}. We let each type decay at the
constant multiple $\Gamma/(p+1)$ of its own rate:
\[
Z(t,c)=
\begin{cases}
\e^{-\Gamma t}, & p=1,\\
\e^{-\Gamma t(1-c)^{1-1/p}}, & p>1,
\end{cases}
\qquad
y_-=\Theta wZ,
\qquad
\beta_{\mathrm{low}}=\bstar+y_-.
\]
The multiple $\Gamma/(p+1)=2(1+1/p)$ is the product of two factors lost in the estimates below: the slope of $\beta_{\mathrm{low}}$ is
only bounded below by half the slope of $\bstar$, which is where the choice of $b$ halfway to the cap enters, and the transport term
$A\,\fraction{\partial y_-}{\partial c}$ contributes a factor $1+1/p$. The value of $\Gamma$ is the smallest that closes these estimates, as the following computation shows. We have
\[
\fraction{\partial y_-}{\partial t}=-\Gamma\Theta(1-c)Z,\qquad
\fraction{\partial y_-}{\partial c}=-\frac\Theta p(1-c)^{1/p-1}Z+\Theta\Gamma\Bigl(1-\frac1p\Bigr)tZ\ge-\frac\Theta p(1-c)^{1/p-1}Z .
\]
With $Z\le1$ and $\frac p{p+1}-\Theta=\Theta$, this gives three bounds on $(0,\infty)\times[0,1)$:
\begin{itemize}
\item $\fraction{\partial \beta_{\mathrm{low}}}{\partial c}\ge\bigl(\frac p{p+1}-\Theta\bigr)\frac1p(1-c)^{1/p-1}=\frac\Theta p(1-c)^{1/p-1}>0$;
\item $A\,\fraction{\partial y_-}{\partial c}-y_-\ge-\frac\Theta p(1-c)^{1/p}Z-\Theta wZ=-\bigl(1+\frac1p\bigr)\Theta wZ$;
\item $\fraction{\partial y_-}{\partial t}\cdot\fraction{\partial \beta_{\mathrm{low}}}{\partial c}\le-\frac{\Gamma\Theta^2}p\,wZ$, since $\fraction{\partial y_-}{\partial t}\le0$.
\end{itemize}
So \eqref{eq:B-lower-target} holds as soon as $\Gamma\Theta^2/p\ge(1+1/p)\Theta$, that is, $\Gamma\ge(p+1)/\Theta$. Our $\Gamma$ satisfies
$\Gamma\Theta=p+1$ with equality. (Any datum $b=1-\theta w$ with $0<\theta<\frac p{p+1}$ works in the same way, with the error profile $(\frac p{p+1}-\theta)wZ$ and
$\Gamma=(p+1)/\theta$.) The remaining
hypotheses of \cref{lem:lowerbarrier} are immediate: $\beta_{\mathrm{low}}$ is continuous on $[0,\infty)\times[0,1]$ with partial derivatives on
$(0,\infty)\times[0,1)$, $\beta_{\mathrm{low}}(0,\cdot)=\bstar+\Theta w=b$, $\beta_{\mathrm{low}}(t,1)=1=C$, and $\beta_{\mathrm{low}}\ge\bstar\ge\mu$.
\Cref{lem:identity,lem:lowerbarrier} give
\[
\beta-\bstar\ge y_-=\Theta\,w\,Z .
\]

\emph{Step 4: evaluation.}
If $p=1$, then $\Theta=\frac14$ and $\Gamma=8$, and evaluating at $c=0$ gives
\[
\frac14\e^{-8t}
\le
\cE(t)
\le
\frac14\e^{-t},
\]
so exponential decay is optimal in order. For $p=1$ the family is the two-bidder first-price auction with uniform costs,
and this lower bound is the case $f\equiv1$ of \cref{cor:auction}(d): there $u_-=\frac12$, $\Gamma=8$, and the datum
$\bstar+\frac12u_-w$ is the present $b$. If $p>1$, then $(1-c)^{1-1/p}=w^{p-1}$, so $y_-(t,\cdot)=\Theta w\,\e^{-\Gamma tw^{p-1}}$. For large $t$, as a function of $w$, this is largest at
$w^{p-1}=1/((p-1)\Gamma t)$: at the types whose damping time is of order $t$, so that the damping has had just enough time to act. We use
the nearby choice
$1-c=(1+t)^{-p/(p-1)}$. Then $w=(1+t)^{-1/(p-1)}$ and $tw^{p-1}=t/(1+t)\le1$, so $Z\ge\e^{-\Gamma}$, and
\[
\cE(t)
\ge
\Theta\e^{-\Gamma}(1+t)^{-1/(p-1)}.
\]
With Step 2, $\cE(t)\asymp(1+t)^{-1/(p-1)}$. In particular no bound $M(1+t)^{-r}$ with $r>1/(p-1)$ holds, so the exponent $1/(p-1)$ in
\eqref{eq:power} is optimal. Finally, choosing
$p=1+2/\nu$ shows that for every $\nu>0$ some admissible coefficients have
$\cE(t)\not=O((1+t)^{-\nu})$: then $1/(p-1)=\nu/2$, and $(1+t)^{-\nu/2}/(1+t)^{-\nu}=(1+t)^{\nu/2}\to\infty$.
\end{proof}

\subsection{Proof of Example~\ref{ex:pooled}}\label{app:example}

\begin{proof}[Proof of \cref{ex:pooled}]
\emph{Idea.} The datum $b\equiv1$ and the cap condition $\beta(t,1)=1$ suggest looking for solutions among the lines through $(1,1)$,
$\beta=1-u(t)(1-c)$ with slope $u(t)$. Along such lines \eqref{eq:GF} reduces to an ordinary differential equation for the slope, and
$\mathcal T$ is the time at which this equation reaches a given slope. For this game $\bstar(c)=(1+c)/2$, by \cref{prop:minty} or directly
from \eqref{eq:rest}.

\emph{Step 1: the slope equation.} For $\beta=1-u(1-c)$ we have $\fraction{\partial \beta}{\partial c}=u$, $\beta-c=(1-c)(1-u)$ and $\fraction{\partial \beta}{\partial t}=-(1-c)\,\fraction{\mathrm{d} u}{\mathrm{d} t}$, while the
right-hand side of \eqref{eq:GF} is
\[
A-B\,\frac{\beta-\mu}{\fraction{\partial \beta}{\partial c}}
=(1-c)-\frac{(1-c)(1-u)}u
=-\frac{1-2u}{u}\,(1-c).
\]
So $\beta$ solves \eqref{eq:GF} exactly when $\fraction{\mathrm{d} u}{\mathrm{d} t}=(1-2u)/u$. Separating variables with $u(0)=0$ gives $t=\int_0^uv/(1-2v)\,dv=\mathcal T(u)$,
since $v/(1-2v)=-\frac12+\frac12(1-2v)^{-1}$.

\emph{Step 2: the inverse $u$.} The function $\mathcal T$ is continuous on $[0,\frac12)$ with $\mathcal T(0)=0$,
$\fraction{\mathrm{d} \mathcal T}{\mathrm{d} u}(u)=u/(1-2u)>0$ for $u>0$, and $\mathcal T(u)\to\infty$ as $u\to\frac12$. So it is an increasing bijection of $[0,\frac12)$
onto $[0,\infty)$, and its inverse $u$ is continuous and increasing with $u(0)=0$ and $u(t)>0$ for $t>0$. For $t>0$ the inverse is
differentiable with $\fraction{\mathrm{d} u}{\mathrm{d} t}=\frac{1}{\fraction{\mathrm{d} \mathcal T}{\mathrm{d} u}(u)}=(1-2u)/u$.

\emph{Step 3: $\beta$ is the solution.} By Steps 1 and 2, for $t>0$, $\beta=1-u(t)(1-c)$ is a classical solution with positive slope $u(t)$.
Since $0\le u<\frac12$, it satisfies $c\le\beta\le1$. By \cref{lem:barriers}(a), it is a subsolution and a supersolution on every $\cB_T$.
It is continuous on $[0,\infty)\times[0,1]$, with $\beta(0,\cdot)=1$ and $\beta(t,1)=1$, so it is a solution from $b$, and by
\cref{thm:main}(i) it is the solution.

\emph{Step 4: bounds on the slope.}
\begin{itemize}
\item \emph{Upper bound.} The function $\mathcal T(u)-u^2/2$ vanishes at $0$ and has derivative $u/(1-2u)-u=2u^2/(1-2u)\ge0$, so
$\mathcal T(u)\ge u^2/2$. With $u=u(t)$ this gives $t\ge u(t)^2/2$, that is, $u(t)\le\sqrt{2t}$.
\item \emph{Lower bound.} If $u(t)\ge\frac14$ there is nothing to show. If $u=u(t)\le\frac14$, apply
$-\log(1-y)\le y/(1-y)\le y+2y^2$, valid for $0\le y\le\frac12$, with $y=2u$. The first inequality is $\log x\le x-1$ at
$x=1/(1-y)$, and the second is equivalent to $y^2(1-2y)\ge0$. Hence
$t=\mathcal T(u)\le-\frac u2+\frac14(2u+8u^2)=2u^2$, that is, $u(t)\ge\sqrt{t/2}$.
\end{itemize}
Together, $u(t)\ge\min\{\sqrt{t/2},\frac14\}$, and $u(t)$ is of order $\sqrt t$ for small $t$.

\emph{Step 5: the error.} We have $\beta-\bstar=(1-c)(\frac12-u)\ge0$, which is largest at $c=0$, so $\cE(t)=\frac12-u(t)$. Finally
$\mathcal T(u)=t$ is
$\log(1-2u)=-4t-2u$, that is,
\[
\frac12-u
=
\frac12\e^{-4t-2u}.
\]
Since $0\le u<\frac12$, this gives $\frac12\e^{-1-4t}\le\cE(t)\le\frac12\e^{-4t}$.
\end{proof}

%=====================================================================
\section{Proofs for Section~\ref{sec:main}}\label{app:equilibrium}
%=====================================================================

\subsection{Proof of Theorem~\ref{thm:selection}}\label{sec:proof2}

In an allocation game, a tie-breaking rule lets the action $z$ win with some
probability $P(z)$, so that the payoff is $\kappa(z-m(c))P(z)$; here $Q(x^+(z))\le P(z)\le Q(x^-(z))$, and $P(z)<Q(x^-(z))$ whenever
$Q(x^+(z))<Q(x^-(z))$ and a tied player does not surely beat every tied rival. Let $x^*$ be the implied type of $\bstar$, that is,
$x^*(z)=(\bstar)^{-1}(z)$ for $\bstar(0)<z<\bstar(1)$, $x^*(z)=0$ for $z\le\bstar(0)$ and $x^*(z)=1$ for $z\ge\bstar(1)$. Where $\bstar$ is
continuous and strictly increasing, as in (b) and (e), $x^*$ is continuous and weakly increasing on $\R$.

\begin{proof}[Proof of \cref{thm:selection}]
\emph{(a) Regular equilibria.} \emph{Idea.} At a regular equilibrium each interior type chooses an interior action, so the
first-order condition $G[\beta](c)=0$ holds, and by \eqref{eq:field} this is the rest-point equation. The rest-point
equation has only one solution with the boundary value $C$ at $c=1$.

\emph{Step 1: the first-order condition.} Let $\beta$ be a regular symmetric equilibrium and $c\in(0,1)$, and put
$u=\fraction{\mathrm{d} \beta}{\mathrm{d} c}(c)>0$. Since $\beta$ is feasible, $\beta(1)=C$, and since it is strictly increasing, $\beta(c)<C$. By
\cref{def:mobius}, $\pi_c(\cdot;\beta)$ is differentiable at $\beta(c)$ with derivative $G[\beta](c)=A(c)-B(c)(\beta(c)-\mu(c))/u$.
Every action $z\in(\beta(c),C]$ is feasible and earns at most $\pi_c(\beta(c);\beta)$, so the right difference quotients at
$\beta(c)$ are nonpositive, and their limit gives $G[\beta](c)\le0$, that is,
\[
B(c)\bigl(\beta(c)-\mu(c)\bigr)\ge A(c)\,u>0
\]
by \condref{A2}. Hence $\beta(c)>\mu(c)$: the action of type $c$ is interior to $[\mu(c),C]$. At an interior maximum the derivative
vanishes, so $A(c)\fraction{\mathrm{d} \beta}{\mathrm{d} c}(c)=B(c)(\beta(c)-\mu(c))$. Thus $\beta$ solves \eqref{eq:rest} on $(0,1)$.

\emph{Step 2: uniqueness.} Let $y=\beta-\bstar$. Both functions are continuous on $[0,1]$, solve $A\,\fraction{\mathrm{d} \beta}{\mathrm{d} c}=B(\beta-\mu)$ on
$(0,1)$ and equal $C$ at $c=1$. Subtracting the two equations gives $\fraction{\mathrm{d} y}{\mathrm{d} c}=(B/A)y$, so $\fraction{\mathrm{d} (y^2)}{\mathrm{d} c}=2(B/A)y^2\ge0$ on $(0,1)$. Hence
$y^2$ is weakly increasing on $[0,1]$, and $y(c)^2\le y(1)^2=0$. Thus $\beta=\bstar$.

\emph{(b) Allocation games.} \emph{Idea.} The rest point is an average of the costs $m$ of higher types. Hence the payoff of type
$c$ from posting the action of type $x$ has the integral form $\kappa\int_x^1(m(s)-m(c))q(s)\,ds$, which is largest at $x=c$
because $m$ is increasing (Steps 1 and 2). Conversely, the equilibrium inequalities against mimicking determine the equilibrium payoff up to a
constant (Step 3), and pooling is broken by a slight undercut (Step 4). For approximate equilibria we compare the winning
probabilities under $\beta$ and $\bstar$ action by action (Step 5).

\emph{Step 1: the rest point and the mimicking payoff.} Here $B/A=q/Q=-\fraction{\mathrm{d} Q}{\mathrm{d} c}/Q$, so $\Omega=Q/Q(0)$ in \cref{lem:rest}, with
$\Omega(1)=0$ and $d\Omega=-q\,ds/Q(0)$, and \eqref{eq:average} becomes
\[
\bstar(c)=\frac1{Q(c)}\int_c^1q(s)\,m(s)\,ds\qquad(0\le c<1),\qquad \bstar(1)=m(1)=C .
\]
From here on, Steps 1 and 2 use only $Q(c)=\int_c^1q$ with an integrable $q\ge0$ whose integral over every interval is
positive, with $\bstar$ defined by this formula. Then
$\bstar(x)$ is an average of $m$ over $[x,1]$ with weights $q/Q(x)$ that charge every subinterval. Since $m$ is strictly
increasing, $m<\bstar<m(1)$ on $[0,1)$. Moreover, for $x<y<1$, writing $Q(x)=Q(y)+\int_x^yq$,
\[
Q(x)\bigl(\bstar(y)-\bstar(x)\bigr)=\int_x^yq(s)\bigl(\bstar(y)-m(s)\bigr)\,ds>0,
\]
because $m(s)\le m(y)<\bstar(y)$ for $s\le y$. So $\bstar$ is strictly increasing; it is continuous on $[0,1)$ as a quotient of
continuous functions and at $1$ because $m\le\bstar\le m(1)$. Its range is $[\bstar(0),C]$. For a type $c$ let
$\Pi_c(x)=\pi_c(\bstar(x);\bstar)$ be its payoff when it posts the action of type $x$. Since $x^*(\bstar(x))=x$,
$Q(x)=\int_x^1q$ and $Q(x)\bstar(x)=\int_x^1qm$,
\[
\Pi_c(x)=\kappa\bigl(\bstar(x)-m(c)\bigr)Q(x)=\kappa\int_x^1\bigl(m(s)-m(c)\bigr)q(s)\,ds .
\]

\emph{Step 2: $\bstar$ is an equilibrium against every action.} By the integral form,
\[
\Pi_c(c)-\Pi_c(x)=
\begin{cases}
\kappa\int_c^x\bigl(m(s)-m(c)\bigr)q(s)\,ds\ge0, & x\ge c,\\[1mm]
\kappa\int_x^c\bigl(m(c)-m(s)\bigr)q(s)\,ds\ge0, & x<c,
\end{cases}
\]
because $m$ is increasing: mimicking a higher type gives up the profit on the types in between, and mimicking a lower
type adds only losses. Now fix a type $c$ and an action $z\in\R$.
\begin{itemize}
\item If $z<\bstar(0)$, then $x^*(z)=0$ and $\pi_c(z;\bstar)=\kappa(z-m(c))Q(0)<\Pi_c(0)\le\Pi_c(c)$.
\item If $z\ge C$, then $x^*(z)=1$ and $\pi_c(z;\bstar)=0=\Pi_c(1)\le\Pi_c(c)$.
\item Otherwise $z=\bstar(x)$ with $x=x^*(z)\in[0,1]$, and $\pi_c(z;\bstar)=\Pi_c(x)\le\Pi_c(c)$.
\end{itemize}
So $\pi_c(z;\bstar)\le\pi_c(\bstar(c);\bstar)$ for every $z$, and the equilibrium inequality holds globally.

\emph{Step 3: uniqueness among strictly increasing equilibria.} We use revealed preference, as in the integral form of
equilibrium payoffs \citep{myerson1981optimal,milgrom2002envelope}. Let $\beta$ be a strictly increasing feasible equilibrium,
and let $\pi^\beta(c)=\kappa(\beta(c)-m(c))Q(c)$ be its equilibrium payoff. A type $c$ that posts $\beta(c')$ faces no ties, has
implied type $c'$ and earns
\[
\kappa\bigl(\beta(c')-m(c)\bigr)Q(c')=\pi^\beta(c')+\kappa\bigl(m(c')-m(c)\bigr)Q(c').
\]
This is at most $\pi^\beta(c)$: by the equilibrium property if $\beta(c')\ge m(c)$, so that $\beta(c')$ is feasible for type $c$,
and otherwise because it is nonpositive while $\pi^\beta(c)\ge0$. Hence
\begin{equation}\label{eq:C-revealed}
\pi(c')+\kappa\bigl(m(c')-m(c)\bigr)Q(c')\le\pi(c)\qquad(c,c'\in[0,1])
\end{equation}
holds for $\pi=\pi^\beta$, and by Step~2 also for $\pi=\pi^*$, the same expression for $\bstar$. For $c\le c'$, \eqref{eq:C-revealed}
for the pairs $(c,c')$ and $(c',c)$ gives
\[
\kappa\bigl(m(c')-m(c)\bigr)Q(c')\le\pi(c)-\pi(c')\le\kappa\bigl(m(c')-m(c)\bigr)Q(c).
\]
Both $\pi^\beta(c)-\pi^\beta(c')$ and $\pi^*(c)-\pi^*(c')$ lie in this interval, so $D=\pi^\beta-\pi^*$ satisfies
\[
|D(c)-D(c')|\le\kappa\bigl(m(c')-m(c)\bigr)\bigl(Q(c)-Q(c')\bigr)\qquad(c\le c').
\]
The right side is a product of two small increments, which is what makes $D$ constant. Let $\varepsilon>0$, and
partition $[c,1]$ into points $c=x_0<\dots<x_N=1$ with $m(x_{k+1})-m(x_k)\le\varepsilon$, which is possible by uniform continuity
of $m$. Summing the bound over the partition gives
$|D(c)-D(1)|\le\kappa\varepsilon\sum_k(Q(x_k)-Q(x_{k+1}))\le\kappa\varepsilon Q(0)$. Since $\varepsilon$ is arbitrary, $D(c)=D(1)$, and
$D(1)=0$ because $Q(1)=0$. Hence $\kappa(\beta(c)-\bstar(c))Q(c)=D(c)=0$, so $\beta=\bstar$ where $Q>0$, that is on $[0,1)$, and
$\beta(1)=m(1)=\bstar(1)$ by feasibility.

\emph{Step 4: no pooling.} Let $\beta$ be weakly increasing and feasible, and an equilibrium under a tie-breaking rule of
(b). Since $q=B/\kappa>0$ on $[0,1)$ by \condref{A2}, $Q$ is strictly decreasing. Suppose that $c<c'$ pool at the action
$p=\beta(c)=\beta(c')$. Then $x^-(p)\le c$ and $x^+(p)\ge c'$, so
$Q(x^+(p))\le Q(c')<Q(c)\le Q(x^-(p))=:\ell$, and the tie-breaking rule gives $P(p)<\ell$. Every lower action $z<p$ has
$x^+(z)\le x^-(p)$, because every type above $x^-(p)$ posts at least $p>z$, and therefore wins with probability $P(z)\ge Q(x^+(z))\ge\ell$. So undercutting $p$ raises the winning
probability from $P(p)$ to at least $\ell$, at the cost of a smaller margin. Since $p\ge m(c')>m(c)$, the margin $p-m(c)$ is
positive, and the undercut
\[
z=m(c)+\bigl(p-m(c)\bigr)\frac{\ell+P(p)}{2\ell}
\]
is feasible for type $c$ and lies below $p$, because $\frac{\ell+P(p)}{2\ell}<1$. It is chosen so that the margin shrinks by the
factor $\frac{\ell+P(p)}{2\ell}$, the midpoint between $\frac{P(p)}\ell$ and $1$; at the factor $\frac{P(p)}\ell$ the undercut would
just break even. It earns
\[
\kappa\bigl(z-m(c)\bigr)P(z)\ge\kappa\bigl(z-m(c)\bigr)\ell=\kappa\bigl(p-m(c)\bigr)\frac{\ell+P(p)}2>\kappa\bigl(p-m(c)\bigr)P(p),
\]
which is more than type $c$ earns at $p$, a contradiction. Hence $\beta$ is strictly increasing, $x^-=x^+$ at its actions, the
tie-breaking rule plays no role, and Step~3 gives $\beta=\bstar$. Conversely, $\bstar$ is strictly increasing, so a
tie-breaking rule never applies to it, and by Step~2 it is an equilibrium under every tie-breaking rule.

\emph{Step 5: approximate equilibria.} Let $\beta$ be feasible, $\delta=\norm{\beta-\bstar}$, and fix a tie-breaking rule with
winning probability $P$. The idea is that $\beta$ and $\bstar$ differ by at most $\delta$ in action, so the implied types of an
action under $\beta$ are those of nearby actions under $\bstar$.
\begin{itemize}
\item \emph{Sandwich.} For every action $z$,
\begin{equation}\label{eq:C-sandwich}
x^*(z-\delta)\le x^-(z),\qquad x^+(z)\le x^*(z+\delta).
\end{equation}
Indeed, a type $s<x^*(z-\delta)$ has $\bstar(s)<z-\delta$, so $\beta(s)\le\bstar(s)+\delta<z$ and $s\le x^-(z)$; a type $s>x^*(z+\delta)$ has
$\bstar(s)>z+\delta$, so $\beta(s)\ge\bstar(s)-\delta>z$ and $s\ge x^+(z)$. Since $Q$ is decreasing,
$Q(x^*(z+\delta))\le P(z)\le Q(x^*(z-\delta))$.
\item \emph{Deviation payoffs.} For $z\ge m(c)$, the upper bound on $P(z)$, Step~2 at the action $z-\delta$, and
$Q\le Q(0)$ give
\[
\kappa\bigl(z-m(c)\bigr)P(z)\le\kappa\bigl(z-\delta-m(c)\bigr)Q\bigl(x^*(z-\delta)\bigr)+\kappa\delta Q\bigl(x^*(z-\delta)\bigr)\le\Pi_c(c)+\kappa\delta Q(0).
\]
For $z<m(c)$ the payoff is at most $0\le\Pi_c(c)$.
\item \emph{Own payoff.} Let $z_0=\beta(c)\ge m(c)$, $z_1=z_0+\delta\ge\bstar(c)$ and $x_1=x^*(z_1)\ge c$. By \eqref{eq:C-sandwich}, $P(z_0)\ge Q(x_1)$.
Since $\bstar(x_1)\le z_1\le\bstar(c)+2\delta$, the slope hypothesis gives $u_-(x_1-c)\le2\delta$, and with $\bar q=\bar B/\kappa$, so
that $Q(c')-Q(c'')\le\bar q(c''-c')$ for $c'\le c''$, we get $Q(x_1)\ge Q(c)-2\bar q\delta/u_-$. Using $z_0-m(c)\ge\bstar(c)-m(c)-\delta$,
\[
\kappa\bigl(z_0-m(c)\bigr)P(z_0)\ge\kappa\bigl(\bstar(c)-m(c)-\delta\bigr)Q(x_1)
\ge\Pi_c(c)-\frac{2\kappa\bar q\delta}{u_-}\bigl(\bstar(c)-m(c)\bigr)-\kappa\delta Q(0),
\]
where $\Pi_c(c)=\kappa(\bstar(c)-m(c))Q(c)$.
\end{itemize}
Subtracting the two bounds, every action $z$ gains at most
\[
\kappa\bigl(z-m(c)\bigr)P(z)-\kappa\bigl(z_0-m(c)\bigr)P(z_0)\le2\delta\Bigl(\kappa Q(0)+\frac{\kappa\bar q\,(\bstar(c)-m(c))}{u_-}\Bigr)
\]
over the own action. With $\kappa Q(0)=A(0)$, $\kappa\bar q=\bar B$ and $\bstar(c)-m(c)\le m(1)-m(0)=R$, this is the bound
$2\norm{\beta-\bstar}(A(0)+R\bar B/u_-)$.

\emph{(c) Interdependent and affiliated costs.} \emph{Idea.} As in (b) we show that the payoff $\Pi_c(x)=\pi_c(\bstar(x);\bstar)$
from mimicking type $x$ is single-peaked at $x=c$. Here we use its derivative in $x$, which the
rest-point identity turns into an expression whose sign is $c-x$ under the stated monotonicity conditions. This needs
$\bstar$ strictly increasing, so that every action in the range of $\bstar$ is the action of exactly one type.

\emph{Interdependent costs, Step 1: the rest point is strictly increasing.} Here $\pi_c(z;\beta)=\int_{x(z)}^1(z-\bar m(c,s))g(s)\,ds$ with
the density $g$ of the lowest rival type, $A(c)=\int_c^1g$, $B=g$ and $\mu(c)=\bar m(c,c)$. Suppose that $\bstar(c)=\mu(c)$ for some
$c<1$, and fix $d\in(c,1)$. On $[c,d]$ the function $y=\bstar-\mu(c)$ satisfies $y(c)=0$ and
\[
0\le\fraction{\mathrm{d} y}{\mathrm{d} x}=\frac BA\bigl(\bstar-\mu\bigr)\le\frac BA\,y\le K\,y,
\]
with $K=\max_{[c,d]}B/A$, because $\mu(c)\le\mu\le\bstar$ there. By Gr\"onwall's inequality $y=0$ on $[c,d]$. As $d<1$ is arbitrary,
$\bstar=\mu(c)$ on $[c,1)$, and continuity at $1$ gives $\mu(1)=\bstar(1)=\mu(c)$, which contradicts $\mu(c)<\mu(1)$. Hence $\bstar>\mu$
on $[0,1)$, and $\fraction{\mathrm{d} \bstar}{\mathrm{d} c}=B(\bstar-\mu)/A>0$ there: the rest point is strictly increasing, and no types tie.

\emph{Interdependent costs, Step 2: the mimicking payoff.} For $x\in[0,1]$,
$\Pi_c(x)=\bstar(x)\int_x^1g-\int_x^1\bar m(c,s)g(s)\,ds$. For $x\in(0,1)$ its derivative is
$\fraction{\mathrm{d} \bstar}{\mathrm{d} x}(x)A(x)-(\bstar(x)-\bar m(c,x))g(x)$, and the rest-point identity $\fraction{\mathrm{d} \bstar}{\mathrm{d} x}(x)A(x)=g(x)(\bstar(x)-\bar m(x,x))$ at $x$
turns it into
\[
\fraction{\mathrm{d} \Pi_c}{\mathrm{d} x}(x)=g(x)\bigl(\bar m(c,x)-\bar m(x,x)\bigr).
\]
The markup terms cancel, and only the difference between the expected costs of type $c$ and of type $x$, given the
lowest rival type $x$, remains. If $\bar m(\cdot,x)$ is
weakly increasing, this is nonnegative for $x<c$ and nonpositive for $x>c$, so $\Pi_c\le\Pi_c(c)$ on $[0,1]$.

\emph{Interdependent costs, Step 3: actions outside the range.} Actions below $\bstar(0)$ win against every rival, and the payoff
$\int_0^1(z-\bar m(c,s))g(s)\,ds$ increases in $z$, so it is at most $\Pi_c(0)$. Actions of at least $C$ earn $0=\Pi_c(1)$. Hence $\bstar$
is a symmetric Bayes--Nash equilibrium against every action.

\emph{Affiliated costs, Step 1: the rest point is strictly increasing.} Here $\pi_c(z;\beta)=(z-c)\Sigma_c(x(z))$, so $A(c)=\Sigma_c(c)$,
$B(c)=g_c(c)$ and $\mu(c)=c$. If $\bstar(c)=c$ for some $c<1$, the markup $y=\bstar-c$ satisfies $y(c)=0$ and
$\fraction{\mathrm{d} y}{\mathrm{d} c}(c)=(B/A)(\bstar(c)-c)-1=-1$, so $y<0$ just to the right of $c$, which contradicts $\bstar\ge\mu$. Hence $\bstar>\mu$ on $[0,1)$,
and $\bstar$ is strictly increasing as before.

\emph{Affiliated costs, Step 2: the sign of the derivative.} The payoff from mimicking $x$ is $\Pi_c(x)=(\bstar(x)-c)\Sigma_c(x)$,
with derivative $\fraction{\mathrm{d} \bstar}{\mathrm{d} x}(x)\Sigma_c(x)-(\bstar(x)-c)g_c(x)$. We multiply by $\Sigma_x(x)>0$ and use the rest-point identity
$\Sigma_x(x)\fraction{\mathrm{d} \bstar}{\mathrm{d} x}(x)=g_x(x)(\bstar(x)-x)$:
\[
\Sigma_x(x)\,\fraction{\mathrm{d} \Pi_c}{\mathrm{d} x}(x)=a\bigl(\bstar(x)-x\bigr)-b\bigl(\bstar(x)-c\bigr),\qquad a=g_x(x)\Sigma_c(x),\quad b=g_c(x)\Sigma_x(x).
\]
Both $a$ and $b$ are nonnegative, and $b\le a$ exactly when $\mathfrak{h}_c(x)\le\mathfrak{h}_x(x)$, where $\mathfrak{h}_c(x)=g_c(x)/\Sigma_c(x)$
is the hazard rate. Since $\mathfrak{h}_c(x)$ is weakly decreasing in $c$, we have $b\le a$ for $x<c$ and $a\le b$ for $x>c$. With
$\bstar(x)\ge x$:
\begin{itemize}
\item for $x<c$ and $\bstar(x)\ge c$, the right side is $a(c-x)+(a-b)(\bstar(x)-c)\ge0$;
\item for $x<c$ and $\bstar(x)<c$, both $a(\bstar(x)-x)\ge0$ and $-b(\bstar(x)-c)\ge0$;
\item for $x>c$, the right side is $(a-b)(\bstar(x)-x)+b(c-x)\le0$.
\end{itemize}
So $\fraction{\mathrm{d} \Pi_c}{\mathrm{d} x}$ has the sign of $c-x$, and $\Pi_c\le\Pi_c(c)$ on $[0,1]$.

\emph{Affiliated costs, Step 3: actions outside the range.} An action $z<\bstar(0)$ earns $(z-c)\Sigma_c(0)\le\Pi_c(0)$, and an action
$z\ge C$ earns $0=\Pi_c(1)$. Hence $\bstar$ is a symmetric Bayes--Nash equilibrium against every action.

\emph{(d) The nonlinear family.} This part is proved separately below.

\emph{(e) Near the rest point.} \emph{Idea.} Payoffs depend on the rivals' strategy only through the implied types
$x^\pm(z)$, and by the sandwich \eqref{eq:C-sandwich} these are within a small distance of $x^*(z)$ when $\norm{\beta-\bstar}$ is
small. The deviation gains against $\beta$ are therefore uniformly close to those against $\bstar$, for every tie-breaking
rule. The two claims then follow by choosing the tolerance.

\emph{Step 1: payoffs as functions of the implied type.} Write
$I_c(z,x)=\int_0^xV_c(z,s)g_c(s)\,ds+\int_x^1W_c(z,s)g_c(s)\,ds$, so that \eqref{eq:payoff} is $I_c(z,x(z))$, and let $\bar W$ be the
maximum of $(|W_c|+|V_c|)g_c$ over $[0,1]\times[\mu(0),C]\times[0,1]$. Since $\fraction{\partial I_c}{\partial x}=(V_c-W_c)g_c$,
\[
|I_c(z,x)-I_c(z,x')|\le\bar W|x-x'| .
\]
Under a tie-breaking rule, the payoff at a feasible action $z$ differs from $I_c(z,x^-(z))$ only through the integrand on
$(x^-(z),x^+(z))$, where $W_c$ is replaced by a convex combination of $W_c$ and $V_c$; so it lies within
$\bar W(x^+(z)-x^-(z))$ of $I_c(z,x^-(z))$.

\emph{Step 2: deviation payoffs.} Let $\delta>0$ and let $\beta$ be feasible with $\norm{\beta-\bstar}\le\delta$. By \eqref{eq:C-sandwich},
$x^-(z)$, $x^+(z)$ and $x^*(z)$ all lie in $[x^*(z-\delta),x^*(z+\delta)]$. With Step~1,
\[
\bigl|\pi_c(z;\beta)-I_c\bigl(z,x^*(z)\bigr)\bigr|\le2\bar W\,\omega(\delta),\qquad \omega(\delta)=\sup_{z\in\R}\bigl(x^*(z+\delta)-x^*(z-\delta)\bigr).
\]
Since $x^*$ is continuous and constant outside $[\bstar(0),C]$, it is uniformly continuous, and $\omega(\delta)\to0$ as $\delta\to0$.

\emph{Step 3: the own action.} At $z=\beta(c)$, Step~2 gives
$|\pi_c(\beta(c);\beta)-I_c(\beta(c),x^*(\beta(c)))|\le2\bar W\omega(\delta)$. The function $(c,z)\mapsto I_c(z,x^*(z))$ is continuous, because
$W_cg_c$ and $V_cg_c$ are jointly continuous and $x^*$ is continuous, and hence uniformly continuous on
$[0,1]\times[\mu(0),C]$. Since $|\beta(c)-\bstar(c)|\le\delta$ and $x^*(\bstar(c))=c$, the difference
$|I_c(\beta(c),x^*(\beta(c)))-I_c(\bstar(c),c)|$ tends to $0$ as $\delta\to0$, uniformly in $c$.

\emph{Step 4: deviation gains.} Against $\bstar$, which is strictly increasing and creates no ties,
$\pi_c(z;\bstar)=I_c(z,x^*(z))$ and $\pi_c(\bstar(c);\bstar)=I_c(\bstar(c),c)$. By Steps 2 and 3,
$\pi_c(z;\beta)-\pi_c(\beta(c);\beta)$ differs from $\pi_c(z;\bstar)-\pi_c(\bstar(c);\bstar)$ by an amount that tends to $0$ as $\delta\to0$,
uniformly over types, feasible actions, tie-breaking rules and feasible strategies $\beta$ with $\norm{\beta-\bstar}\le\delta$.

\emph{Step 5: the two claims.} If $\bstar$ is an equilibrium, every gain against $\bstar$ is at most $0$; a $\delta$ for which the
amount in Step~4 is at most $\varepsilon$ makes every such $\beta$ an $\varepsilon$-equilibrium. If instead type $c$ gains
$\varepsilon_0>0$ at $z$ against $\bstar$, a $\delta$ for which the amount is at most $\varepsilon_0/2$ leaves a gain of at least $\varepsilon_0/2$.
\end{proof}

\begin{proof}[Proof of \cref{thm:selection}(d)]
\emph{Idea.} The dynamics sees only $(A,B,\mu)$, which do not depend on $\theta$, so neither does the rest point. Whether
the rest point is an equilibrium depends on the whole payoff. We compute the payoff $\Pi_c(x)=\pi_c(\bstar(x);\bstar)$ of type
$c$ when it posts the action of type $x$. The gap $\Pi_c(c)-\Pi_c(x)$ factors as $(x-c)^2$ times a bracket, and the sign of
the bracket is decided by the maximum of a cubic.

\emph{Step 1: the rest point.} The coefficients are $A=1-c$, $B=1$ and $\mu(c)=c$ for every $\theta$ (\cref{prop:examples}), and
$\bstar(c)=(1+c)/2$ satisfies $A\,\fraction{\mathrm{d} \bstar}{\mathrm{d} c}=(1-c)/2=\bstar-c$ and $\bstar(1)=1$. By uniqueness of the rest point (\cref{lem:rest}), it is the rest point
for every $\theta\ge0$.

\emph{Step 2: the mimicking payoff.} Since $\fraction{\mathrm{d}}{\mathrm{d} s}\bigl(-(s-c)^2(1-s)\bigr)=(s-c)(3s-c-2)=\zeta_c(s)$, we have
$\int_x^1\zeta_c(s)\,ds=(x-c)^2(1-x)$ for $x\in[0,1]$. With $g\equiv1$, an action $z$ with implied type $x=x(z)$ therefore earns
\[
\pi_c(z;\beta)=(1-x)(z-c)+\theta z^2(x-c)^2(1-x).
\]
Against $\bstar$ the action of type $x$ is $z=\frac{1+x}2$, so
\begin{gather*}
\Pi_c(x)=(1-x)\Bigl(\frac{1+x}2-c\Bigr)+\theta\,\frac{(1+x)^2}4\,(x-c)^2(1-x),\\
\Pi_c(c)-\Pi_c(x)=(x-c)^2\Bigl(\frac12-\frac\theta4(1-x)(1+x)^2\Bigr).
\end{gather*}
For the gap, $\Pi_c(c)=\frac12(1-c)^2$, and the $\theta$-free part is $\frac12(1-c)^2-(1-x)(\frac{1+x}2-c)=\frac12(x-c)^2$.

\emph{Step 3: the cubic.} The identity
\[
\frac{32}{27}-(1-x)(1+x)^2=\Bigl(x-\frac13\Bigr)^2\Bigl(x+\frac53\Bigr)
\]
shows that the maximum of $(1-x)(1+x)^2$ over $[0,1]$ is $\frac{32}{27}$, attained only at $x=\frac13$. Hence the bracket in Step~2 is
nonnegative for all $x\in[0,1]$ if and only if $\frac\theta4\cdot\frac{32}{27}\le\frac12$, that is, $\theta\le\frac{27}{16}$.

\emph{Step 4: sufficiency for $\theta\le\frac{27}{16}$.} Let $z\in[c,1]$ be a feasible action of type $c$. If $z\ge\bstar(0)=\frac12$,
then $z=\bstar(x)$ with $x=2z-1\in[0,1]$, and $\pi_c(z;\bstar)=\Pi_c(x)\le\Pi_c(c)$ by Steps 2 and 3. If $z<\frac12$, the action wins
surely and earns $z-c+\theta c^2z^2$. This is increasing in $z\ge0$, so it is at most its value at $z=\frac12$, the action of type
$0$, which is $\Pi_c(0)\le\Pi_c(c)$.

\emph{Step 5: the deviation for $\theta>\frac{27}{16}$.} At $c=0$ and $x=\frac13$ the gap is
$\Pi_0(0)-\Pi_0(\frac13)=\frac19(\frac12-\frac{8\theta}{27})<0$. So type $0$ gains by posting the feasible action $\bstar(\frac13)=\frac23$.

\emph{Step 6: unrestricted actions.} The feasible action set matters, as noted after \cref{thm:selection}. For $\theta>0$ and
$c\in(0,1]$, the action $z=-3-3/(\theta c^2)$ wins surely, and $|z|\ge3$ and $\theta c^2|z|\ge3$ give $\theta c^2z^2\ge3|z|$. So it earns at
least $-|z|-c+3|z|=2|z|-c>\frac12\ge\frac12(1-c)^2=\Pi_c(c)$.
\end{proof}

\begin{proof}[Proof of \cref{cor:auction}]
\emph{Idea.} In procurement coordinates the auction is the allocation game of the first row of \cref{tab:members}. So
(a) and (c) are \cref{thm:selection}(b), and (b) is \cref{thm:main}, with explicit constants. The one new estimate is a
lower bound $u_-$ on the slope of the rest point, which enters (c) and (d). For (d) we build an explicit lower barrier.

\emph{Step 1: coefficients, rest point and slope.} By \cref{prop:examples}, the procurement auction is the allocation game with
$\kappa=1$, $m(c)=c$ and $Q=S^{n-1}$, so $A=S^{n-1}$, $B=(n-1)S^{n-2}f$, $\mu(c)=c$ and $C=1$, and these coefficients satisfy
\cref{ass:main}. Since $B/A=-\fraction{\mathrm{d} A}{\mathrm{d} c}/A$ and $A(0)=1$, we have $\Omega=A$ in \cref{lem:rest}, with $\Omega(1)=0$. Integrating by parts in
\eqref{eq:average},
\[
\bstar(c)=\frac{-\int_c^1s\,dA(s)}{A(c)}=c+\frac1{A(c)}\int_c^1A .
\]
Then $\fraction{\mathrm{d} \bstar}{\mathrm{d} c}=B(\bstar-c)/A=(n-1)fS^{-n}\int_c^1S^{n-1}$. Since $f\le f_+$ and $S^{n-1}f=-\frac1n\,\fraction{\mathrm{d} (S^n)}{\mathrm{d} s}$,
\[
\int_c^1S^{n-1}\ge\frac1{f_+}\int_c^1S^{n-1}f=\frac{S(c)^n}{nf_+},
\]
hence $\fraction{\mathrm{d} \bstar}{\mathrm{d} c}\ge(n-1)f/(nf_+)\ge u_-$ with $u_-=(n-1)f_-/(nf_+)$ as in (d). By the mean value theorem,
$\bstar(c')-\bstar(c)\ge u_-(c'-c)$ for $c\le c'$.

\emph{Reflection.} The coordinates $c=1-v$ and $\beta(c)=1-\alpha(1-c)$ turn a bid $\alpha(v)$ into the action $z=1-\alpha(v)$, and
$v-\alpha(v)=z-c$. The highest bid wins exactly when the lowest action wins, and ties correspond to ties. So payoffs,
tie-breaking rules, feasible strategies, weak monotonicity and the uniform distance are all preserved, and each claim
may be proved in procurement coordinates.

\emph{(a)} Here $F_V(v)=S(1-v)$, so $A(1-v)=F_V(v)^{n-1}$, and the substitution $s\mapsto1-s$ turns $\int_{1-v}^1S^{n-1}$ into
$\int_0^vF_V^{n-1}$. So the reflection of $\bstar$ is $1-\bstar(1-v)=\alpha^*(v)$, and \cref{thm:selection}(b) gives the claims.

\emph{(b)} Feasible bid functions reflect to feasible strategies, so \cref{thm:main}(i) gives the solution, and
\cref{thm:main}(ii) shows that $\cE$ is a strict Lyapunov functional. For the rates we apply \cref{thm:main}(ii) with
$\varphi\equiv1$, so $\varphi_-=1$, and $t_0=0$. Since $S(c)\ge f_-(1-c)$ and $w=1-c$, we have $A\ge f_-^{\,n-1}w^{n-1}$, so $p=n-1$ and
$a=f_-^{\,n-1}$. Both $\beta(0,\cdot)$ and $\bstar$ take values in $[c,1]$, so $\cE(0)\le1=R$. For $n=2$, \eqref{eq:power} gives
$\cE(t)\le\cE(0)\e^{-f_-t}\le\e^{-f_-t}$. For $n\ge3$ it gives
$\cE(t)\le\bigl(\cE(0)^{-(n-2)}+(n-2)f_-^{\,n-1}t\bigr)^{-1/(n-2)}$, which increases in $\cE(0)$, so $\cE(0)\le1$ gives the claim.

\emph{(c)} Here $A(0)=1$, $R=1$, $B\le\bar B=(n-1)f_+$ because $S\le1$, and $\bstar(c')-\bstar(c)\ge u_-(c'-c)$ by Step~1. Since $\beta(t,\cdot)$ is
feasible, \cref{thm:selection}(b) applies with the constant
\[
2\Bigl(A(0)+\frac{R\bar B}{u_-}\Bigr)=2\Bigl(1+\frac{(n-1)f_+}{u_-}\Bigr)=2\Bigl(1+\frac{nf_+^2}{f_-}\Bigr),
\]
and (b) bounds $\cE(t)=\norm{\beta(t,\cdot)-\bstar}$.

\emph{(d)} \emph{Idea.} The datum is the rest point plus a bump $\frac12u_-w$ that vanishes at the top type. We show that
the error of type $c$ decays no faster than $\e^{-\Gamma tw^{n-2}}$, with $w=1-c$. For $n\ge3$ this rate degenerates at the top,
where the gradient is weak, and the slowest type gives the bound. The amplitude $\frac12u_-$ is half the slope bound of
Step~1, so that the barrier stays strictly increasing, and $\Gamma$ is chosen so that the barrier inequality closes.

\emph{Step 2: the datum.} The datum $b=\bstar+\frac12u_-w$ is the reflection of $\alpha_0$. It is increasing, since
$b(c')-b(c)\ge(u_--\frac12u_-)(c'-c)\ge0$ for $c\le c'$; it satisfies $b\ge\bstar\ge c$; and $b\le1$ because $1-\bstar(c)\ge u_-w$ by Step~1.
So $b$ is feasible.

\emph{Step 3: the barrier.} Let $y_-=\frac12u_-w\,\e^{-\Gamma tw^{n-2}}$ and $\beta_{\mathrm{low}}=\bstar+y_-$. Then
\[
\fraction{\partial y_-}{\partial t}=-\Gamma w^{n-2}y_-\le0,\qquad
\fraction{\partial y_-}{\partial c}=\tfrac12u_-\bigl(-1+(n-2)\Gamma tw^{n-2}\bigr)\e^{-\Gamma tw^{n-2}}\ge-\tfrac12u_-\e^{-\Gamma tw^{n-2}},
\]
so $\fraction{\partial \beta_{\mathrm{low}}}{\partial c}\ge u_--\frac12u_-=\frac12u_-$. Since $S\le f_+w$ and $f\le f_+$, we have $A\le f_+^{\,n-1}w^{n-1}$ and
$Bw\le(n-1)f_+^{\,n-1}w^{n-1}$, so
\[
A+Bw\le nf_+^{\,n-1}w^{n-1}=\tfrac12u_-\Gamma w^{n-1};
\]
this is the reason for the choice $\Gamma=2nf_+^{\,n-1}/u_-$. Therefore
\[
A\fraction{\partial y_-}{\partial c}-By_-\ge-\tfrac12u_-\e^{-\Gamma tw^{n-2}}(A+Bw)\ge-\tfrac12u_-\cdot\tfrac12u_-\Gamma w^{n-1}\e^{-\Gamma tw^{n-2}}
=\tfrac12u_-\,\fraction{\partial y_-}{\partial t}\ge\fraction{\partial \beta_{\mathrm{low}}}{\partial c}\,\fraction{\partial y_-}{\partial t} ,
\]
where the last step uses $\fraction{\partial y_-}{\partial t}\le0$ and $\fraction{\partial \beta_{\mathrm{low}}}{\partial c}\ge\frac12u_-$. By \cref{lem:identity}, with $y_0=y_-$ and $y_1=\fraction{\partial y_-}{\partial c}$, the
left side equals $\fraction{\partial \beta_{\mathrm{low}}}{\partial c}\bigl(A-B(\beta_{\mathrm{low}}-\mu)/\fraction{\partial \beta_{\mathrm{low}}}{\partial c}\bigr)$, so dividing by $\fraction{\partial \beta_{\mathrm{low}}}{\partial c}>0$ gives
$\fraction{\partial \beta_{\mathrm{low}}}{\partial t}\le A-B(\beta_{\mathrm{low}}-\mu)/\fraction{\partial \beta_{\mathrm{low}}}{\partial c}$.

\emph{Step 4: the bound.} We have $\beta_{\mathrm{low}}(0,\cdot)=b$, $\beta_{\mathrm{low}}(t,1)=\bstar(1)=1$ and $\beta_{\mathrm{low}}\ge\bstar\ge\mu$, so
\cref{lem:lowerbarrier} gives $\beta-\bstar\ge y_-$, and $\cE(t)\ge\sup_cy_-(t,c)$. For $n=2$, evaluate at $c=0$, where $y_-(t,0)=\frac12u_-\e^{-\Gamma t}$.
For $n\ge3$ maximize $w\e^{-\Gamma tw^{n-2}}$ over $w\in(0,1]$: the derivative vanishes at $w^{n-2}=1/((n-2)\Gamma t)$, which lies in $(0,1]$ if
$(n-2)\Gamma t\ge1$, and the value there is $(\e\,(n-2)\Gamma t)^{-1/(n-2)}$.
\end{proof}

%=====================================================================
\section{Proofs for Section~\ref{sec:grid}}\label{app:grid}
%=====================================================================

Write $\mathrm{pr}_i(z)=\min\{C,\max\{\mu_i,z\}\}$ for the projection of type $i<N$ onto its feasible actions.

\paragraph{Roadmap.} All results of this appendix rest on one property of the update, proved in \cref{lem:monotone-update}:
under the CFL condition \eqref{eq:cfl}, raising the action of a type, or the action of the next type, never lowers the updated
action. This is a discrete comparison principle, and it is used in two ways. In \cref{thm:grid}(b) the iterates are trapped
between two explicit barriers built from the grid rest point, so the error $E_k$ obeys a scalar recursion. In
\cref{lem:scheme-feasible} monotonicity, together with consistency of the update and the comparison principle
\cref{lem:comparison}, gives convergence to the solution as the grid is refined; this lemma also provides existence in
\cref{thm:main}(i). \Cref{prop:safeguards} shows that the slope floor and the positivity of the weight cannot be dropped, and
\cref{cor:learn} combines the three parts of \cref{thm:grid}.

\begin{lemma}[Monotone update]\label{lem:monotone-update}
Let the step $h$ satisfy the CFL condition \eqref{eq:cfl} for $\eta\ge0$, let $i<N$ and $|\xi|\le\eta$, and let $z_1,z_2$ be actions of type
$i$ and $z_1^+,z_2^+$ actions of type $i+1$ with $\mu_i\le z_1\le z_2\le C$, $z_1^+\le C$ and $z_1^+\le z_2^+$. Then
\[
z_1+h\varphi_i(z_1)\Bigl(A_i-\frac{\Delta_i(z_1-\mu_i)}{\max\{z_1^+-z_1,\ell\Delta_i\}}+\xi\Bigr)\le
z_2+h\varphi_i(z_2)\Bigl(A_i-\frac{\Delta_i(z_2-\mu_i)}{\max\{z_2^+-z_2,\ell\Delta_i\}}+\xi\Bigr).
\]
In particular, the update of \cref{alg:grid} with a fixed error vector is monotone on feasible strategies.
\end{lemma}

\begin{proof}
\emph{Idea.} The updated action is the old action $z$ plus $h$ times the weighted field. The field decreases in $z$, so the
update is increasing in $z$ as long as $h$ times the rate at which the weighted field decreases is at most one. This rate has
two sources: the weight $\varphi_i(z)$ varies with $z$, which \condref{G1} controls, and the damping term
$\Delta_i\varphi_i(z)(z-\mu_i)/\max\{z^+-z,\ell\Delta_i\}$ increases in $z$. The floor bounds its denominator below by
$\ell\Delta_i$. The two inequalities of \eqref{eq:cfl} are exactly the resulting bounds when the floor is active and when it is
inactive.

\emph{Step 1: a common neighbour.} The right-hand side is weakly increasing in $z_2^+$: the damping term has the nonnegative
numerator $\Delta_i\varphi_i(z_2)(z_2-\mu_i)$, and its denominator weakly increases with $z_2^+$. So it suffices to prove the
claim with $z_2^+$ replaced by $z_1^+$. Write $z^+=z_1^+$. Since $z^+\le C$, we have $z^+-\mu_i\le C-\mu(0)=R$.

\emph{Step 2: the constants.} Put
\[
\mathcal D(z)=\varphi_i(z)(z-\mu_i),\qquad
\mathcal U(z)=z+h\varphi_i(z)(A_i+\xi)-\frac{h\Delta_i\mathcal D(z)}{\max\{z^+-z,\ell\Delta_i\}},
\]
so that the claim reads $\mathcal U(z_1)\le\mathcal U(z_2)$. Let $\mu_i\le u\le v\le C$. Since $0\le A_i\le\norm A$, we have
$|A_i+\xi|\le\norm A+\eta$, and \condref{G1} gives $(A_i+\xi)(\varphi_i(v)-\varphi_i(u))\ge-L_\varphi(\norm A+\eta)(v-u)$. The restoring term
$\mathcal D$ is Lipschitz with constant $\hat\varphi=\varphi_++L_\varphi R$: by \condref{G1} and $u-\mu_i\le R$,
\begin{equation}\label{eq:D-mono-damping}
\mathcal D(v)-\mathcal D(u)=\varphi_i(v)(v-u)+\bigl(\varphi_i(v)-\varphi_i(u)\bigr)(u-\mu_i)\le\hat\varphi\,(v-u),
\qquad
0\le\mathcal D(u)\le\hat\varphi\,(u-\mu_i).
\end{equation}

\emph{Step 3: the floor is active at $u$,} that is, $z^+-u\le\ell\Delta_i$; then it is also active at $v$. Both denominators equal
$\ell\Delta_i$, so by Step~2
\[
\begin{aligned}
\mathcal U(v)-\mathcal U(u)&=(v-u)+h(A_i+\xi)\bigl(\varphi_i(v)-\varphi_i(u)\bigr)-\frac h\ell\bigl(\mathcal D(v)-\mathcal D(u)\bigr)\\
&\ge(v-u)\Bigl(1-hL_\varphi(\norm A+\eta)-\frac{h\hat\varphi}\ell\Bigr),
\end{aligned}
\]
which is nonnegative by the first inequality of \eqref{eq:cfl}.

\emph{Step 4: the floor is inactive at $v$,} that is, $z^+-v\ge\ell\Delta_i$; then it is also inactive at $u$. Now
\[
\frac{\mathcal D(v)}{z^+-v}-\frac{\mathcal D(u)}{z^+-u}=\frac{(\mathcal D(v)-\mathcal D(u))(z^+-u)+\mathcal D(u)(v-u)}{(z^+-u)(z^+-v)}.
\]
By \eqref{eq:D-mono-damping} the numerator is at most $\hat\varphi(v-u)(z^+-u)+\hat\varphi(u-\mu_i)(v-u)=\hat\varphi(v-u)(z^+-\mu_i)\le\hat\varphi R\,(v-u)$,
and the denominator is at least $(\ell\Delta_i)^2$. Multiplying by $h\Delta_i$,
\[
\mathcal U(v)-\mathcal U(u)\ge(v-u)\Bigl(1-hL_\varphi(\norm A+\eta)-\frac{h\hat\varphi R}{\ell^2\Delta_i}\Bigr),
\]
which is nonnegative by the second inequality of \eqref{eq:cfl}.

\emph{Step 5: gluing.} If the floor is active at $z_1$, Step~3 with $[u,v]=[z_1,z_2]$ applies; if it is inactive at $z_2$, Step~4 applies.
In the remaining case $z^+-z_2<\ell\Delta_i<z^+-z_1$, the kink $z^+-\ell\Delta_i$ lies between $z_1$ and $z_2$, the floor is just
inactive there, and Step~4 on $[z_1,z^+-\ell\Delta_i]$ followed by Step~3 on $[z^+-\ell\Delta_i,z_2]$ gives the claim.

\emph{Step 6: the update.} For feasible strategies $\beta\le\beta'$ and type $i<N$, apply the inequality with $z_1=\beta_i$,
$z_2=\beta'_i$, $z_1^+=\beta_{i+1}$ and $z_2^+=\beta'_{i+1}$; then apply the weakly increasing map $\mathrm{pr}_i$. The top type
keeps the action $C$.
\end{proof}

\begin{lemma}[Grid approximation from feasible data]\label{lem:scheme-feasible}
Let $b$ be feasible. For $N\ge1$ let $0=c^N_0<\dots<c^N_N=1$ be grids whose mesh tends to $0$, let
$\ell_N\to0$ and $\eta_N\to0$, let $|\xi_i^k|\le\eta_N$, and let $h_N$ satisfy \eqref{eq:cfl} for $\eta_N$. Then there is a solution $\beta$ from $b$, and, for every $T>0$,
\[
\max\bigl\{
|\beta^{N,k}_i-\beta(kh_N,c_i^N)|:
0\le kh_N\le T,\ 0\le i\le N
\bigr\}\to0.
\]
\end{lemma}

\begin{proof}
\emph{Idea.} We follow \citet{barles1991convergence}: a scheme that is monotone, stable and consistent converges to the
solution of the limit equation whenever that equation has a comparison principle. Here monotonicity is \cref{lem:monotone-update},
stability is the feasibility of the iterates, and the comparison principle is \cref{lem:comparison}. We form the upper and lower
half-relaxed limits $\sub\ge\super$ of the iterates \citep{barles1987discontinuous} and show three things: both start from
$b$ (Step~1 and Step~2, by explicit barriers); $\sub$ is a subsolution (Step~3); and $\super$ is a supersolution (Step~4). Steps~3
and~4 pass to the limit in a single update at a grid point where the iterates touch a test function. \Cref{lem:comparison} then
gives $\sub\le\super$, so the two limits agree, and their common value is the solution; equality of the two limits is the same as
uniform convergence (Step~5). No equicontinuity of the iterates is needed. The singular damping term needs care twice. At
$t=0$ it can be as large as $R/\ell_N$, which is unbounded, so the lower barrier of Step~1 is built on a profile whose slope is
bounded below. In the limit of Step~3 the floored denominator tends to the slope of the test function, which is why a
nonpositive test slope above the anchor leads to a contradiction.

\emph{Step 0: preliminaries.} Write $c_i=c_i^N$ and fix $T>0$. Because of the projection, the iterates are feasible:
$\mu(c_i)\le\beta^{N,k}_i\le C$ and $\beta^{N,k}_N=C$; moreover $\beta^{N,0}_i=b(c_i)$. The first inequality of
\eqref{eq:cfl} gives $h_N\hat\varphi\le\ell_N$, so $h_N\to0$.

\emph{Step 1: barriers at $t=0$.} \emph{Upper barrier.} At a feasible strategy the damping term is nonnegative, so
$G_i\le A_i\le\norm A$. With $\varphi_i\le\varphi_+$ and $\xi^k_i\le\eta_N$, each update raises an action by at most
$h_N\varphi_+(\norm A+\eta_N)$, and the projection does not raise it further, because the old action is at least $\mu_i$.
By induction over $k$,
\begin{equation}\label{eq:D-upper-initial}
\beta^{N,k}_i\le b(c_i)+kh_N\varphi_+(\norm A+\eta_N).
\end{equation}
\emph{Lower barrier.} An update can lower an action by as much as $h_N\varphi_+(R/\ell_N+\eta_N)$, and this rate per unit of
time is unbounded as $\ell_N\to0$. Fix $\varepsilon>0$. We compare the iterates with a profile whose slope in the loss coordinate
is at least $\varepsilon/X(1)$; on such a profile the damping term is at most $R/(\varepsilon/X(1))$, uniformly in $N$. To obtain it,
tilt $b$ down by
\[
\underline b_i=b(c_i)-\varepsilon-\frac{\varepsilon}{X(1)}\bigl(X(1)-X(c_i)\bigr).
\]
Since $b$ is weakly increasing, $\underline b_{i+1}-\underline b_i\ge(\varepsilon/X(1))\Delta_i$, and the tilt costs at most $2\varepsilon$.
The bound on the damping term fixes the speed at which the barrier sinks,
\[
\Theta_N=\varphi_+\Bigl(\frac{RX(1)}\varepsilon+\eta_N\Bigr),
\]
and we let $\underline b^k_i=\max\{\mu_i,\underline b_i-\Theta_Nkh_N\}$ for $i<N$ and $\underline b^k_N=C$. These profiles are
feasible, since $\underline b_i\le b(c_i)\le C$. Let $N$ be so large that $\ell_N\le\varepsilon/X(1)$. We claim that the update of
$\underline b^k$ at type $i<N$ is at least $\underline b^{k+1}_i$.
\begin{itemize}
\item If $\underline b^{k+1}_i=\mu_i$, this holds because the projection keeps every update above $\mu_i$.
\item Otherwise $\underline b^k_i=\underline b_i-\Theta_Nkh_N>\mu_i$, and $\underline b^k_{i+1}\ge\underline b_{i+1}-\Theta_Nkh_N$ (also for $i+1=N$,
because $\underline b_N\le C$). Hence $\underline b^k_{i+1}-\underline b^k_i\ge(\varepsilon/X(1))\Delta_i\ge\ell_N\Delta_i$: the floor is
inactive, and since $A_i\ge0$ and $\underline b^k_i-\mu_i\le R$, we get $G_i(\underline b^k)\ge-RX(1)/\varepsilon$. So the weighted
field plus error is at least $-\Theta_N$, and the update is at least
$\min\{C,\underline b^k_i-\Theta_Nh_N\}=\underline b^{k+1}_i$.
\end{itemize}
Since $\underline b^0\le\beta^{N,0}$, \cref{lem:monotone-update} and induction over $k$ give $\underline b^k\le\beta^{N,k}$: the update of
$\underline b^k$ lies above $\underline b^{k+1}$ and below the update of $\beta^{N,k}$. As $(\varepsilon/X(1))(X(1)-X(c_i))\le\varepsilon$,
\begin{equation}\label{eq:D-lower-initial}
\beta^{N,k}_i\ge\underline b^k_i
\ge b(c_i)-2\varepsilon-\Theta_Nkh_N.
\end{equation}

\emph{Step 2: half-relaxed limits} \citep{barles1987discontinuous}.
Let $\sub(t,c)$ and $\super(t,c)$ be the limit superior and inferior of
$\beta^{N,k}_i$ as $N\to\infty$ and $(kh_N,c_i)\to(t,c)$ within $\cB_T$.
Then $\sub$ is upper and $\super$ lower semicontinuous, and
$\mu\le\super\le\sub\le C$, where $\mu\le\super$ uses the continuity of $\mu$. At the top type,
$\sub(t,1)\le C=\mu(1)\le\super(t,1)$, so $\sub(t,1)=\super(t,1)=C$. Along grid points with $kh_N\to0$ and $c_i\to c$, the right-hand
side of \eqref{eq:D-upper-initial} tends to $b(c)$ and that of \eqref{eq:D-lower-initial} to $b(c)-2\varepsilon$, by continuity of $b$
and because $\eta_N$, and hence $\Theta_N$, stays bounded. Since $\varepsilon>0$ is arbitrary,
\[
\sub(0,\cdot)\le b\le\super(0,\cdot).
\]

\emph{Step 3: $\sub$ is a subsolution.}
Let $\psi$ be continuously differentiable and let $\sub-\psi$ attain its maximum over $\cB_T$ at $(t_{\star},c_{\star})\in(0,T]\times[0,1)$.
Adding $(t-t_{\star})^2+(c-c_{\star})^2$ to $\psi$ makes the maximum strict and does not change the derivatives of $\psi$ at
$(t_{\star},c_{\star})$, so we may assume it is strict. Then, along a subsequence of $N$, maximizers $(k,i)$ of
$\beta^{N,k}_i-\psi(kh_N,c_i)$ over the grid points of $\cB_T$ satisfy $(kh_N,c_i)\to(t_{\star},c_{\star})$ and
$\beta^{N,k}_i\to\sub(t_{\star},c_{\star})$; since $t_{\star}>0$ and $c_{\star}<1$, eventually $k\ge1$ and $i<N$. Write
$U=\sub(t_{\star},c_{\star})$, $\Xi_N=\beta^{N,k}_i-\psi(kh_N,c_i)$ for the maximal value and
$[\psi]_i=\psi((k-1)h_N,c_{i+1})-\psi((k-1)h_N,c_i)$. Maximality at the two grid points that enter the update gives
\[
\beta^{N,k-1}_{i'}
\le
\psi((k-1)h_N,c_{i'})+\Xi_N
\qquad(i'=i,i+1).
\]
\emph{The discrete inequality.} We compare the actual update at $(k-1,i)$ with the update of the test function, shifted by $\Xi_N$.
To apply \cref{lem:monotone-update}, the shifted test value must be capped at $C$, and its neighbour is raised so that the floored
denominator becomes that of the test function:
\[
z_1=\beta^{N,k-1}_i,\qquad
z_2=\min\{C,\psi((k-1)h_N,c_i)+\Xi_N\},
\]
\[
z_1^+=\beta^{N,k-1}_{i+1},\qquad
z_2^+=z_2+\max\{z_1^+-z_2,[\psi]_i\}.
\]
The hypotheses of the lemma hold: $\mu_i\le z_1\le z_2\le C$ by maximality, $z_1^+\le C$, and $z_1^+\le z_2^+$. Moreover
$z_1^+-z_2\le\max\{[\psi]_i,\ell_N\Delta_i\}$: if $z_2<C$, then $z_1^+-z_2\le[\psi]_i$ by maximality at $i+1$; if $z_2=C$, then
$z_1^+-z_2\le0<\ell_N\Delta_i$. Hence
\[
\max\{z_2^+-z_2,\ell_N\Delta_i\}=\max\{z_1^+-z_2,[\psi]_i,\ell_N\Delta_i\}=\max\{[\psi]_i,\ell_N\Delta_i\}.
\]
Suppose the lower projection is inactive at $(k,i)$, that is, $\beta^{N,k}_i>\mu_i$. Then $\beta^{N,k}_i$ is at most the
unprojected update from $(z_1,z_1^+)$, which by \cref{lem:monotone-update} is at most the unprojected update from $(z_2,z_2^+)$.
Since $\beta^{N,k}_i-\psi(kh_N,c_i)=\Xi_N$ and $z_2\le\psi((k-1)h_N,c_i)+\Xi_N$, the time difference of $\psi$ is at most
$\beta^{N,k}_i-z_2$, and with $\xi^{k-1}_i\le\eta_N$ we obtain
\begin{equation}\label{eq:D-sub-step}
\frac{\psi(kh_N,c_i)-\psi((k-1)h_N,c_i)}{h_N}
\le
\varphi_i(z_2)
\left(
A_i-\Delta_i\frac{z_2-\mu_i}{\max\{[\psi]_i,\ell_N\Delta_i\}}
+\eta_N
\right).
\end{equation}
This holds also when $z_2=C$, because we only used $z_2\le\psi((k-1)h_N,c_i)+\Xi_N$. If instead the lower projection is active,
$\beta^{N,k}_i=\mu_i$, then maximality at $(k-1,i)$ gives $\psi(kh_N,c_i)-\psi((k-1)h_N,c_i)\le\mu_i-\beta^{N,k-1}_i\le0$. In both
cases, since the damping term in \eqref{eq:D-sub-step} is nonnegative,
\begin{equation}\label{eq:D-sub-anchor}
\frac{\psi(kh_N,c_i)-\psi((k-1)h_N,c_i)}{h_N}\le\varphi_i(z_2)(A_i+\eta_N).
\end{equation}
\emph{The limit.} The time quotient of $\psi$ tends to $\fraction{\partial \psi}{\partial t}(t_{\star},c_{\star})$. By the mean value theorem, applied to $\psi$ and
to $X$ with $\fraction{\mathrm{d} X}{\mathrm{d} c}=B$, the quotient $[\psi]_i/\Delta_i$ tends to $\fraction{\partial \psi}{\partial c}(t_{\star},c_{\star})/B(c_{\star})$. Further
$\psi((k-1)h_N,c_i)+\Xi_N\to U\le C$, so $z_2\to U$; and $\mu_i\to\mu(c_{\star})$, $A_i\to A(c_{\star})$ and, by continuity of $\varphi$ on
$\cK$, $\varphi_i(z_2)\to\varphi(c_{\star},U)>0$.
\begin{itemize}
\item \emph{Above the anchor, $U>\mu(c_{\star})$.} Since $\beta^{N,k}_i\to U$, eventually $\beta^{N,k}_i>\mu_i$, so
\eqref{eq:D-sub-step} holds. Divide its denominator by $\Delta_i$. If $\fraction{\partial \psi}{\partial c}(t_{\star},c_{\star})\le0$, then
$\max\{[\psi]_i/\Delta_i,\ell_N\}$ is positive and tends to $0$, because $\ell_N\to0$, while $z_2-\mu_i\to U-\mu(c_{\star})>0$; the
right-hand side tends to $-\infty$ and the left-hand side to $\fraction{\partial \psi}{\partial t}(t_{\star},c_{\star})$, a contradiction. Hence
$\fraction{\partial \psi}{\partial c}(t_{\star},c_{\star})>0$, and the limit of \eqref{eq:D-sub-step} is
$\fraction{\partial \psi}{\partial t}+\varphi(c_{\star},U)H(c_{\star},U,\fraction{\partial \psi}{\partial c})\le0$ at $(t_{\star},c_{\star})$.
\item \emph{At the anchor, $U\le\mu(c_{\star})$,} so $U=\mu(c_{\star})$. The limit of \eqref{eq:D-sub-anchor} is the anchor clause
$\fraction{\partial \psi}{\partial t}(t_{\star},c_{\star})\le\varphi(c_{\star},\mu(c_{\star}))A(c_{\star})$.
\end{itemize}

\emph{Step 4: $\super$ is a supersolution.}
Let $\super-\psi$ attain a strict minimum over $\cB_T$ at $(t_{\star},c_{\star})\in(0,T]\times[0,1)$ with $\fraction{\partial \psi}{\partial c}(t_{\star},c_{\star})>0$,
let $U=\super(t_{\star},c_{\star})$, and choose minimizers $(k,i)$, $\Xi_N$ and $[\psi]_i$ as in Step~3. Minimality gives
$\beta^{N,k-1}_{i'}\ge\psi((k-1)h_N,c_{i'})+\Xi_N$ for $i'=i,i+1$, and also $\beta^{N,k}_{i+1}\ge\psi(kh_N,c_{i+1})+\Xi_N$ in the same round.
\emph{The cap is inactive at $(k,i)$.} Because $\fraction{\partial \psi}{\partial c}(t_{\star},c_{\star})>0$, eventually $\psi(kh_N,c_{i+1})>\psi(kh_N,c_i)$, and
then $\beta^{N,k}_i=\psi(kh_N,c_i)+\Xi_N<\beta^{N,k}_{i+1}\le C$. So $\beta^{N,k}_i$ is at least the unprojected update from
$(\beta^{N,k-1}_i,\beta^{N,k-1}_{i+1})$. \emph{The discrete inequality.} Now the shifted test value is raised to the anchor, so that it
is feasible, and lies below the iterate:
\[
z_1=\max\{\mu_i,\psi((k-1)h_N,c_i)+\Xi_N\},\qquad
z_1^+=\psi((k-1)h_N,c_{i+1})+\Xi_N,
\]
with $z_2=\beta^{N,k-1}_i\ge z_1$ and $z_2^+=\beta^{N,k-1}_{i+1}\ge z_1^+$, and $z_1^+\le C$. If $z_1>\mu_i$, then $z_1^+-z_1=[\psi]_i$; if
$z_1=\mu_i$, the damping term at $z_1$ vanishes. So in both cases the unprojected update from $(z_1,z_1^+)$ has the floored
denominator $\max\{[\psi]_i,\ell_N\Delta_i\}$ of the test function. By \cref{lem:monotone-update} it is at most the unprojected update from
$(z_2,z_2^+)$, which is at most $\beta^{N,k}_i$. Since $z_1\ge\psi((k-1)h_N,c_i)+\Xi_N$, the time difference of $\psi$ is at least
$\beta^{N,k}_i-z_1$, and with $\xi^{k-1}_i\ge-\eta_N$ we obtain
\begin{equation}\label{eq:D-super-step}
\frac{\psi(kh_N,c_i)-\psi((k-1)h_N,c_i)}{h_N}
\ge
\varphi_i(z_1)
\left(
A_i-\Delta_i\frac{z_1-\mu_i}{\max\{[\psi]_i,\ell_N\Delta_i\}}
-\eta_N
\right).
\end{equation}
\emph{The limit.} As in Step~3, $z_1\to\max\{\mu(c_{\star}),U\}=U$ and $[\psi]_i/\Delta_i\to\fraction{\partial \psi}{\partial c}(t_{\star},c_{\star})/B(c_{\star})>0$, so
the floored denominator divided by $\Delta_i$ has a positive limit, and the limit of \eqref{eq:D-super-step} is
$\fraction{\partial \psi}{\partial t}+\varphi(c_{\star},U)H(c_{\star},U,\fraction{\partial \psi}{\partial c})\ge0$ at $(t_{\star},c_{\star})$.

\emph{Step 5: conclusion.}
By Steps~2--4 and \cref{lem:comparison}, $\sub\le\super$ on $\cB_T$. Hence $\sub=\super$; this common value $\beta_T$ is continuous,
because it is both upper and lower semicontinuous, takes values in $[\mu,C]$, equals $b$ at $t=0$ and $C$ at $c=1$, and is a
subsolution and a supersolution on $\cB_T$. For $T\le T'$, every sequence of grid points admissible for the horizon $T$ is
admissible for $T'$, so on $\cB_T$ the upper limit for $T$ is at most that for $T'$ and the lower limit for $T'$ is at most that
for $T$; this matters only at $t=T$. Hence $\beta_T\le\beta_{T'}\le\beta_T$ on $\cB_T$, and the functions $\beta_T$ define a single
function $\beta$, which is a solution from $b$.

If the convergence claim failed, there would be $\varepsilon>0$ and,
for infinitely many $N$, grid points $(kh_N,c_i^N)\in\cB_T$ with
$|\beta^{N,k}_i-\beta(kh_N,c_i^N)|\ge\varepsilon$. At a limit point $(t,c)$ of these points, the continuity of $\beta$ would give
$\sub(t,c)-\beta(t,c)\ge\varepsilon$ or $\beta(t,c)-\super(t,c)\ge\varepsilon$, a contradiction.
\end{proof}

\begin{proof}[Proof of \cref{thm:grid}]
Write
\[
\mathcal U_i(\beta,\xi)=\mathrm{pr}_i\bigl(\beta_i+h\varphi_i(\beta_i)(G_i(\beta)+\xi)\bigr)
\]
for the update of type $i<N$ with error $\xi$. For $i<N$ we have $c_i<1$ and hence $A_i>0$. By \cref{lem:monotone-update},
$\mathcal U_i(\beta,\xi)\le\mathcal U_i(\beta',\xi)$ whenever $\beta\le\beta'$ are feasible and $|\xi|\le\eta$. The CFL condition gives
$h\varphi_+\le h\hat\varphi\le\ell$. Part~(a) establishes two properties of the grid rest point, \eqref{eq:D-rest-below} and
\eqref{eq:D-rest-cap}; they are the input for the barriers of part~(b).

\emph{(a)} \emph{Step 1: fixed points do not depend on the step or the weight.} For $z\in[\mu_i,C]$, $s>0$ and real $G$,
\[
\mathrm{pr}_i(z+sG)=z
\quad\Longleftrightarrow\quad
G=0,\ \text{or }z=\mu_i\text{ and }G\le0,\ \text{or }z=C\text{ and }G\ge0 .
\]
Indeed, if $G>0$ and $z<C$, the projection lies above $z$; if $G<0$ and $z>\mu_i$, it lies below $z$; in the three listed cases it
returns $z$. With $s=h\varphi_i(z)>0$, the condition does not involve $h\varphi_i(z)$. So the feasible fixed points of the update
without errors do not depend on the weight or on the step.

\emph{Step 2: two properties of the grid rest point.} Downward induction shows that $\hat\beta$ is feasible: the second term of
\eqref{eq:gridrest} is at least $\mu_i$, and the first term is a convex combination of $\hat\beta_{i+1}\le C$ and $\mu_i\le C$. Write
$\rho_i=(A_i\hat\beta_{i+1}+\Delta_i\mu_i)/(A_i+\Delta_i)$ for the first term, so that $A_i(\hat\beta_{i+1}-\rho_i)=\Delta_i(\rho_i-\mu_i)$, and
$\hat d_i=\max\{\hat\beta_{i+1}-\hat\beta_i,\ell\Delta_i\}$ for the floored difference at the rest point. For $i<N$:
\begin{align}
\hat\beta_i<C\ &\Longrightarrow\ A_i\hat d_i=\Delta_i(\hat\beta_i-\mu_i)\ \text{ and }\ \hat\beta_i-\mu_i\ge\ell A_i,
\label{eq:D-rest-below}\\
\hat\beta_i=C\ &\Longrightarrow\ w_i\le\ell A_i .
\label{eq:D-rest-cap}
\end{align}
For \eqref{eq:D-rest-below}: if $\hat\beta_i<C$, the second term of \eqref{eq:gridrest} is $\mu_i+\ell A_i$, since otherwise it equals
$C\le\hat\beta_i$; so $\hat\beta_i-\mu_i\ge\ell A_i$. If $\hat\beta_i=\rho_i$, then $A_i(\hat\beta_{i+1}-\hat\beta_i)=\Delta_i(\hat\beta_i-\mu_i)\ge
A_i\ell\Delta_i$. If $\hat\beta_i=\mu_i+\ell A_i>\rho_i$, then $A_i(\hat\beta_{i+1}-\hat\beta_i)\le\Delta_i(\rho_i-\mu_i)\le A_i\ell\Delta_i=
\Delta_i(\hat\beta_i-\mu_i)$. Since $A_i\hat d_i=\max\{A_i(\hat\beta_{i+1}-\hat\beta_i),A_i\ell\Delta_i\}$, in both cases
$A_i\hat d_i=\Delta_i(\hat\beta_i-\mu_i)$. For \eqref{eq:D-rest-cap}: if $w_i>\ell A_i$, the
second term is $\mu_i+\ell A_i<C$, so $\hat\beta_i=C$ would force $\rho_i\ge C$, which is impossible because $\hat\beta_{i+1}\le C$ and
$\mu_i<C$. By \eqref{eq:D-rest-below} we have $G_i(\hat\beta)=0$ below the cap. At the cap the floor is active, because
$\hat\beta_{i+1}\le C=\hat\beta_i$, so $G_i(\hat\beta)=A_i-w_i/\ell\ge0$ by \eqref{eq:D-rest-cap}. By Step~1, $\hat\beta$ is a fixed point
of the update without errors.

\emph{Step 3: uniqueness.} Let $\beta$ be a feasible fixed point of the update without errors. We show $\beta_i=\hat\beta_i$ by
downward induction; $\beta_N=C=\hat\beta_N$. Let $\beta_{i+1}=\hat\beta_{i+1}$, and view $G_i$ as a function of $z=\beta_i$:
$F(z)=A_i-\Delta_i(z-\mu_i)/\max\{\hat\beta_{i+1}-z,\ell\Delta_i\}$. On $[\mu_i,\infty)$ the numerator of the fraction is nonnegative and
strictly increasing and the denominator is positive and weakly decreasing, so $F$ is strictly decreasing. By Step~2,
$F(\hat\beta_i)\ge0$, with equality if $\hat\beta_i<C$. If $\mu_i=C$, there is nothing to show; let $\mu_i<C$.
\begin{itemize}
\item If $\beta_i<\hat\beta_i$, then $\beta_i<C$ and $F(\beta_i)>0$. By Step~1 a fixed point with $\beta_i<C$ needs $F(\beta_i)=0$, or
$\beta_i=\mu_i$ and $F(\beta_i)\le0$; neither holds.
\item If $\beta_i>\hat\beta_i$, then $\hat\beta_i<C$, so $F(\hat\beta_i)=0$ and $F(\beta_i)<0$, while $\beta_i>\mu_i$. By Step~1 a fixed point
with $\beta_i>\mu_i$ needs $F(\beta_i)=0$, or $\beta_i=C$ and $F(\beta_i)\ge0$; neither holds.
\end{itemize}
Hence $\hat\beta$ is the only feasible fixed point.

\emph{Step 4: the floor bias.} For $\ell=0$ the second term of \eqref{eq:gridrest} is $\mu_i$, which is at most the first term, a convex
combination of $\check\beta_{i+1}\ge\mu_i$ and $\mu_i$. So $\check\beta_i=(A_i\check\beta_{i+1}+\Delta_i\mu_i)/(A_i+\Delta_i)\ge\mu_i$. The map
$z\mapsto(A_iz+\Delta_i\mu_i)/(A_i+\Delta_i)$ is weakly increasing with slope $A_i/(A_i+\Delta_i)<1$, and $\min\{\mu_i+\ell A_i,C\}-\check\beta_i\le\ell A_i$.
Comparing the two terms of \eqref{eq:gridrest} with $\check\beta_i$ gives
\[
0\le\hat\beta_i-\check\beta_i\le\max\{\hat\beta_{i+1}-\check\beta_{i+1},\ell A_i\},
\]
and downward induction from $\hat\beta_N-\check\beta_N=0$ gives $0\le\hat\beta-\check\beta\le\ell\norm A$.

\emph{Step 5: no floor bias under the side condition.} Neither \cref{alg:grid} nor \eqref{eq:gridrest} uses $A_N$, so we may set
$A_N=0$. Let $\mu_{i+1}-\mu_i+\ell(A_{i+1}-A_i)\ge\ell\Delta_i$ for all $i<N$. We show by downward induction that
$\hat\beta_i-\mu_i\ge\ell A_i$ and $\hat\beta_i=\rho_i$ for $i<N$; at $i=N$ the first claim reads $0\ge0$. If
$\hat\beta_{i+1}-\mu_{i+1}\ge\ell A_{i+1}$, the side condition gives
\[
\hat\beta_{i+1}-\mu_i=(\hat\beta_{i+1}-\mu_{i+1})+(\mu_{i+1}-\mu_i)\ge\ell A_{i+1}+\mu_{i+1}-\mu_i\ge\ell(A_i+\Delta_i),
\]
so $\rho_i-\mu_i=A_i(\hat\beta_{i+1}-\mu_i)/(A_i+\Delta_i)\ge\ell A_i$. Hence the first term of \eqref{eq:gridrest} dominates,
$\hat\beta_i=\rho_i$, and $\hat\beta_i-\mu_i\ge\ell A_i$. Thus $\hat\beta$ satisfies the recursion of $\check\beta$, and $\hat\beta=\check\beta$.
Moreover $\hat\beta_{i+1}-\hat\beta_i=\Delta_i(\hat\beta_{i+1}-\mu_i)/(A_i+\Delta_i)\ge\ell\Delta_i$, so the floor is inactive at the rest point.

\emph{Step 6: plain gradient play.} Now let $\varphi\equiv1$ and $\eta=0$, so that \eqref{eq:cfl} reads $h\le\ell$ and $hR\le\ell^2\Delta_i$. We show
that one update maps a feasible $\beta$ with $\beta_{i+1}-\beta_i\ge\ell\Delta_i$ for $i<N$ to a strategy with the same two properties,
and that neither the floor nor the projection acts; induction over $k$ then completes (a). Let $\beta^\sharp$ be the update of $\beta$
without floor and projection; the floor is inactive at $\beta$ by assumption. For $i<N$ put $u=(\beta_{i+1}-\beta_i)/\Delta_i\ge\ell$,
$\Theta=(\beta_i-\mu_i)/\Delta_i$ and $\Theta_A=(A_i-A_{i+1})/\Delta_i$; if $i+1<N$, put $\Theta'=(\beta_{i+1}-\mu_{i+1})/\Delta_i$ and
$u'=(\beta_{i+2}-\beta_{i+1})/\Delta_{i+1}\ge\ell$, and if $i+1=N$, read $\Theta'/u'$ as $0$. Since $\beta^\sharp_i=\beta_i+h(A_i-\Theta/u)$, and
likewise at $i+1$ (at $i+1=N$ with $A_N=0$),
\[
\frac{\beta^\sharp_{i+1}-\beta^\sharp_i}{\Delta_i}=u-h\Theta_A+h\Bigl(\frac\Theta u-\frac{\Theta'}{u'}\Bigr).
\]
Since $u'\ge\ell$, the term $-h\Theta'/u'$ is at least $-h\Theta'/\ell$. The side condition bounds $\Theta'$, because
$\Theta+u-\Theta'=(\mu_{i+1}-\mu_i)/\Delta_i\ge\ell(1+\Theta_A)$; and the CFL condition gives $h(\Theta+u)=h(\beta_{i+1}-\mu_i)/\Delta_i\le hR/\Delta_i\le\ell^2$. Hence
\[
\frac{\beta^\sharp_{i+1}-\beta^\sharp_i}{\Delta_i}-\ell\ge u-h\Theta_A+\frac{h\Theta}u-\frac{h(\Theta+u-\ell-\ell\Theta_A)}{\ell}-\ell
=(u-\ell)\Bigl(1-\frac{h(\Theta+u)}{u\ell}\Bigr)\ge0 ,
\]
because $u\ge\ell$ and $h(\Theta+u)\le\ell^2\le u\ell$. Moreover $\beta^\sharp_i-\mu_i\ge(\beta_i-\mu_i)(1-h/\ell)\ge0$, because $A_i\ge0$ and
$\Delta_i/(\beta_{i+1}-\beta_i)\le1/\ell$; and $\beta^\sharp_i<\beta^\sharp_{i+1}<\dots<\beta^\sharp_N=C$. So the projection does not act,
and $\beta^\sharp$ is feasible and respects the floor. The iterates are those of plain gradient play.

\emph{(b)} \emph{Idea.} We trap the iterates between the clipped shifts $U^\pm$ of the rest point by a distance $E$ and show that one
update moves both barriers toward $\hat\beta$, to distance $\Upsilon(E)$. Monotonicity then carries the iterates along, so $E_k$
is bounded by any sequence $e_k$ that obeys the scalar recursion $e_{k+1}\ge\Upsilon(e_k)$. The function $\gamma(E)$ below is a lower bound
on the restoring field at distance $E$; the two rates follow from two choices of $\gamma$.

\emph{Step 0: one round makes the iterate feasible.} The projection maps every action of type $i<N$ into $[\mu_i,C]$, and
$\beta^1_N=C$.

\emph{Step 1: the envelope recursion.} Let $\gamma:[0,\infty)\to[0,\infty)$ satisfy, for $E\in[0,R]$, the two conditions $\ell\gamma(E)\le E$
and $w_i\gamma(E)\le A_iE$ whenever $i<N$ and $0\le E\le w_i$, and let $\Upsilon(E)=E-h\varphi_-\gamma(E)+h\varphi_+\eta$ be weakly increasing.
The second condition is a discrete form of the damping rate $A/w$ of \cref{sec:solutions}; the first says that $\gamma(E)$ does not exceed
$E/\ell$, the restoring field that the floor guarantees to a barrier at the cap. We show that
$E_k\le e_k$ for all $k$ whenever $e_0\ge R$ and $e_{k+1}\ge\min\{R,\Upsilon(\min\{e_k,R\})\}$ for all $k$. We first note that $\gamma(0)=0$,
by the first condition, so $\Upsilon\ge\Upsilon(0)=h\varphi_+\eta\ge0$ on $[0,R]$.

\emph{Step 2: the weight.} Since $\gamma\ge0$, $\eta\ge0$, $|\xi|\le\eta$ and $\varphi_-\le\varphi_i\le\varphi_+$ on feasible actions,
\begin{equation}\label{eq:D-weight}
\varphi_i(z)(G+\xi)\le-\varphi_-\gamma(E)+\varphi_+\eta\ \text{ if }G\le-\gamma(E),\qquad
\varphi_i(z)(G+\xi)\ge\varphi_-\gamma(E)-\varphi_+\eta\ \text{ if }G\ge\gamma(E).
\end{equation}
So it suffices to bound the field $G_i$ of the barriers by $\mp\gamma$.

\emph{Step 3: the barriers.} Let $E\in[0,R]$, $U^+_i=\min\{C,\hat\beta_i+E\}$ and $U^-_i=\max\{\mu_i,\hat\beta_i-E\}$; both are feasible. We show
\begin{equation}\label{eq:D-barrier-step}
\mathcal U_i(U^+,\xi)\le\hat\beta_i+\Upsilon(E),\qquad
\mathcal U_i(U^-,\xi)\ge\hat\beta_i-\Upsilon(E)
\qquad(i<N,\ |\xi|\le\eta).
\end{equation}
Below the cap the key tool is the rest identity $A_i\hat d_i=\Delta_i(\hat\beta_i-\mu_i)$ of \eqref{eq:D-rest-below}. Applied as
\eqref{eq:identity} with $A^{\circ}=A_i$, $B^{\circ}=\Delta_i$, $\mu^{\circ}=\mu_i$, $z^{\circ}=\hat\beta_i$, $u^{\circ}=\hat d_i$, $y_1=0$ and
$y_0=\pm E$, it is the discrete form of \cref{lem:identity}:
\begin{equation}\label{eq:D-discrete-identity}
A_i-\frac{\Delta_i(\hat\beta_i\pm E-\mu_i)}{\hat d_i}=\mp\frac{\Delta_iE}{\hat d_i},
\qquad
\frac{\Delta_iE}{\hat d_i}\ge\gamma(E)\ \text{ if }E\le w_i .
\end{equation}
The inequality holds because $w_i>0$ below the cap and $\gamma(E)w_i\hat d_i\le EA_i\hat d_i=E\Delta_i(\hat\beta_i-\mu_i)\le E\Delta_iw_i$ by the
second condition on $\gamma$. At the cap the floor takes over, through \eqref{eq:D-rest-cap} and the first condition on $\gamma$. The
case analysis is as follows.

\emph{Upper barrier.}
\begin{itemize}
\item \emph{$\hat\beta_i+E<C$.} Then $U^+_i=\hat\beta_i+E$ and $U^+_{i+1}\le\hat\beta_{i+1}+E$, so $U^+_{i+1}-U^+_i\le\hat\beta_{i+1}-\hat\beta_i$, and
$E<w_i$. A smaller floored difference makes the field smaller, so by \eqref{eq:D-discrete-identity}
\[
G_i(U^+)\le A_i-\frac{\Delta_i(\hat\beta_i+E-\mu_i)}{\hat d_i}=-\frac{\Delta_iE}{\hat d_i}\le-\gamma(E),
\]
and by \eqref{eq:D-weight}, $\mathcal U_i(U^+,\xi)\le\max\{\mu_i,\hat\beta_i+\Upsilon(E)\}=\hat\beta_i+\Upsilon(E)$.
\item \emph{$\hat\beta_i+E\ge C>\hat\beta_i$.} Put $E'=C-\hat\beta_i\le E$. The floor is active at $U^+$, because $U^+_{i+1}\le C=U^+_i$; so
$G_i(U^+)=A_i-w_i/\ell$. By \eqref{eq:D-rest-below}, $w_i-E'=\hat\beta_i-\mu_i\ge\ell A_i$, hence $G_i(U^+)\le-E'/\ell\le-\gamma(E')$ by the first
condition on $\gamma$, and $\mathcal U_i(U^+,\xi)\le\hat\beta_i+\Upsilon(E')\le\hat\beta_i+\Upsilon(E)$, because $\Upsilon$ is weakly increasing.
\item \emph{$\hat\beta_i=C$.} Then $\mathcal U_i(U^+,\xi)\le C=\hat\beta_i$, and $\Upsilon(E)\ge0$.
\end{itemize}

\emph{Lower barrier.}
\begin{itemize}
\item \emph{$\hat\beta_i-E>\mu_i$ and $\hat\beta_i<C$.} Then $U^-_i=\hat\beta_i-E$, $U^-_{i+1}\ge\hat\beta_{i+1}-E$ and $E<w_i$, and as above
$G_i(U^-)\ge\Delta_iE/\hat d_i\ge\gamma(E)$.
\item \emph{$\hat\beta_i-E>\mu_i$ and $\hat\beta_i=C$.} The floor bounds the denominator of $G_i(U^-)$ below by $\ell\Delta_i$, so, by
\eqref{eq:D-rest-cap}, $G_i(U^-)\ge A_i-(U^-_i-\mu_i)/\ell=A_i-(w_i-E)/\ell\ge E/\ell\ge\gamma(E)$.
\end{itemize}
In both cases $\mathcal U_i(U^-,\xi)\ge\min\{C,\hat\beta_i-E+h\varphi_-\gamma(E)-h\varphi_+\eta\}=\hat\beta_i-\Upsilon(E)$.
\begin{itemize}
\item \emph{$\hat\beta_i-E\le\mu_i$.} Put $E'=\hat\beta_i-\mu_i\le E$. Then $U^-_i=\mu_i$ and $G_i(U^-)=A_i\ge\gamma(E')$: if $w_i>0$, because
$w_i\gamma(E')\le A_iE'\le A_iw_i$, and if $w_i=0$, because $E'=0$ and $\gamma(0)=0$. Hence
$\mathcal U_i(U^-,\xi)\ge\min\{C,\mu_i+h\varphi_-\gamma(E')-h\varphi_+\eta\}=\hat\beta_i-\Upsilon(E')\ge\hat\beta_i-\Upsilon(E)$.
\end{itemize}

\emph{Step 4: induction.} Feasible strategies differ from $\hat\beta$ by at most $R$, so $E_0\le R\le e_0$. If $E_k\le e_k$, then
$U^-\le\beta^k\le U^+$ for $E=\min\{e_k,R\}$; for $e_k\ge R$ this holds because then $U^+=C$ and $U^-_i=\mu_i$. Monotonicity, applied to the feasible
strategies $U^-$, $\beta^k$ and $U^+$, and \eqref{eq:D-barrier-step} give $E_{k+1}\le\min\{R,\Upsilon(E)\}\le e_{k+1}$.

\emph{Step 5: the first bound.} Here the damping rate $A_i/w_i$ is bounded below only on the types with $w_i>\bar w$, where $A_i\ge a$ and
$w_i\le R$ give $A_i/w_i\ge a/R$. A deviation $E\le\bar w$ therefore earns no guaranteed restoring field, and a larger deviation earns
the rate $a/R$ on its excess. So take $\gamma(E)=(a/R)\max\{E-\bar w,0\}$ for $E\ge0$. Then $\ell\gamma(E)\le E$, because $\ell a\le R$. If
$\bar w<E\le w_i$, then $A_i\ge a$ and $w_i\gamma(E)\le aE\le A_iE$; if $E\le\bar w$, then $\gamma(E)=0$. Moreover $\Upsilon$ is weakly increasing
on $[0,\infty)$ because $h\varphi_-a\le\ell a\le R$. For $e\ge\bar w$ the recursion is affine,
$\Upsilon(e)-\bar w=(1-h\varphi_-a/R)(e-\bar w)+h\varphi_+\eta$, with fixed point $\bar w+R\varphi_+\eta/(\varphi_-a)$. Its solution from $e_0=\bar w+R+R\varphi_+\eta/(\varphi_-a)$ is
\[
e_k=\bar w+(1-h\varphi_-a/R)^kR+R\varphi_+\eta/(\varphi_-a)\ge\bar w ,
\]
which satisfies $e_0\ge R$ and $\Upsilon(\min\{e_k,R\})\le\Upsilon(e_k)=e_{k+1}$.

\emph{Step 6: the second bound.} If $A_i\ge aw_i^p$, then $A_i/w_i\ge aw_i^{p-1}\ge aE^{p-1}$ for $E\le w_i$, so we take $\gamma(E)=aE^p$
for $E\ge0$. Then $\ell\gamma(E)\le\ell aR^{p-1}E\le E$ on $[0,R]$; moreover $w_i\gamma(E)\le aw_i^pE\le A_iE$ for
$E\le w_i$; and $\Upsilon$ is weakly increasing on $[0,R]$ because its derivative is at least $1-h\varphi_-apR^{p-1}\ge0$. For $p=1$ the function $\Upsilon$ is
weakly increasing on $[0,\infty)$, and $e_k=(1-h\varphi_-a)^kR+\varphi_+\eta/(\varphi_-a)$ satisfies $\Upsilon(\min\{e_k,R\})\le\Upsilon(e_k)=e_{k+1}$.
For $p>1$ the recursion is not affine, and we compare it with an ordinary differential equation. We split the restoring term
$h\varphi_-\gamma(e)$ into two halves: one half absorbs the error term $h\varphi_+\eta$ as long as $e$ is above the level
$e_\eta=(2\varphi_+\eta/(\varphi_-a))^{1/p}$, where $\varphi_-\gamma(e)\ge2\varphi_+\eta$; the other half drives the decay. This explains the factor
$\frac12$. Let
\[
\bar e(t)=R\Bigl(1+(p-1)\frac{\varphi_-a}2R^{p-1}t\Bigr)^{-1/(p-1)},
\]
which solves $\fraction{\mathrm{d} \bar e}{\mathrm{d} t}=-\frac{\varphi_-a}2\bar e^p$ with $\bar e(0)=R$ and is convex, and let $e_k=\max\{e_\eta,\bar e(kh)\}$; then $e_0\ge R$.
If $e_\eta\ge R$, then $e_{k+1}\ge R$. Otherwise $\bar e\le R$ gives $e_k\in[e_\eta,R]$, so $\varphi_-\gamma(e_k)\ge2\varphi_+\eta$ and
$\Upsilon(e_k)\le e_k-\frac h2\varphi_-\gamma(e_k)$.
\begin{itemize}
\item If $e_k=\bar e(kh)$, then $\Upsilon(e_k)\le\bar e(kh)+h\,\fraction{\mathrm{d} \bar e}{\mathrm{d} t}(kh)\le\bar e((k+1)h)$, by convexity.
\item If $e_k=e_\eta$, then $\Upsilon(e_k)=e_\eta-2h\varphi_+\eta+h\varphi_+\eta=e_\eta-h\varphi_+\eta\le e_\eta$.
\end{itemize}
In both cases $\Upsilon(e_k)\le e_{k+1}$.

\emph{(c)} \emph{Idea.} With $\ell=0$, \eqref{eq:gridrest} is an upwind discretization of the rest-point equation $A\,\fraction{\mathrm{d} \bstar}{\mathrm{d} c}=B(\bstar-\mu)$, solved
backward from $c=1$. Its errors propagate through a convex combination, so they do not grow, and the local error vanishes with the
mesh. Near $c=1$ we do not need the local error, because there both $\check\beta_i$ and $\bstar(c_i)$ lie in the short interval $[\mu_i,C]$.

Let $\varepsilon>0$ and $y_i=\check\beta_i-\bstar(c_i)$. For $i<N$, Cauchy's mean value theorem for $\bstar$ and $X$ on
$[c_i,c_{i+1}]$, with $\fraction{\mathrm{d} X}{\mathrm{d} c}=B>0$ and $A\,\fraction{\mathrm{d} \bstar}{\mathrm{d} c}=B(\bstar-\mu)$ on $(c_i,c_{i+1})$, gives $s_i\in(c_i,c_{i+1})$ with
$\bstar(c_{i+1})-\bstar(c_i)=\Delta_i(\bstar-\mu)(s_i)/A(s_i)$. Subtracting this from the recursion
$(A_i+\Delta_i)\check\beta_i=A_i\check\beta_{i+1}+\Delta_i\mu_i$ of Step~4 of (a) gives
\begin{gather*}
(A_i+\Delta_i)\,y_i=A_i\,y_{i+1}-\tau_i,\\
\tau_i=\Delta_i(\bstar-\mu)(c_i)-A_i\bigl(\bstar(c_{i+1})-\bstar(c_i)\bigr)=\Delta_i\Bigl((\bstar-\mu)(c_i)-\frac{A(c_i)}{A(s_i)}(\bstar-\mu)(s_i)\Bigr),
\end{gather*}
and $|y_i|\le\max\{|y_{i+1}|,|\tau_i|/\Delta_i\}$. Choose $\bar c<1$ with $w(\bar c)\le\varepsilon$, let $a>0$ be the minimum of $A$ on
$[0,(1+\bar c)/2]$, and let the mesh be at most $(1-\bar c)/2$, and so small that $|(\bstar-\mu)(c)-(\bstar-\mu)(s)|\le\varepsilon/2$ and
$|A(c)-A(s)|\le a\varepsilon/(2(R+1))$ whenever $|c-s|$ is at most the mesh; this is possible because $\bstar-\mu$ and $A$ are uniformly
continuous on $[0,1]$. We use downward induction from $y_N=0$.
\begin{itemize}
\item If $c_i\ge\bar c$, then $\check\beta_i$ and $\bstar(c_i)$ lie in $[\mu_i,C]$, so $|y_i|\le w_i\le w(\bar c)\le\varepsilon$.
\item If $c_i<\bar c$, then $c_i,s_i\in[0,(1+\bar c)/2]$ and, since $0\le\bstar-\mu\le R$ and $A(s_i)\ge a$,
\[
|\tau_i|/\Delta_i\le|(\bstar-\mu)(c_i)-(\bstar-\mu)(s_i)|+R\,|A(s_i)-A(c_i)|/a\le\varepsilon,
\]
so $|y_i|\le\max\{|y_{i+1}|,\varepsilon\}\le\varepsilon$.
\end{itemize}
Hence $|y_i|\le\varepsilon$ for all $i$.
\end{proof}

\begin{proposition}[Safeguards]\label{prop:safeguards}\leavevmode
\begin{enumerate}
\item[(a)] Consider the uniform duopoly $A=1-c$, $B=1$, $\mu=c$ on the grid $c_i=i/N$ and \cref{alg:grid} with $\varphi\equiv1$ and
without floor and projection, so that $G_i=A_i-(\beta^k_i-\mu_i)/(N(\beta^k_{i+1}-\beta^k_i))$. Let $u_k=N(\beta^k_N-\beta^k_{N-1})$ and $\eta>0$. If
$u_0>0$, $\bar u>0$ and $N\eta>(2\bar u-1)/\bar u$, the errors $\xi^k_{N-1}=-\eta$ give $u_k>\bar u$ for some $k$; if $\underline u>0$ and
$N\eta>(1-2\underline u)/\underline u$, the errors $\xi^k_{N-1}=\eta$ give $u_k<\underline u$ for some $k$.
\item[(b)] Let $\varphi\ge0$. If $\varphi(c_i,C)=0$ for all $i<N$, the pooled strategy $\beta_i=C$ is a fixed point of \cref{alg:grid} for
every error sequence, and $\hat\beta_i<C$ whenever $\mu_i+\ell A_i<C$. If $\varphi(c_i,\mu_i)=0$ for all $i<N$, the strategy $\beta_i=\mu_i$
($i<N$), $\beta_N=C$ is a fixed point, and $\hat\beta_i>\mu_i$ whenever $\mu_i<C$.
\end{enumerate}
\end{proposition}

\begin{proof}[Proof of \cref{prop:safeguards}]
(a) \emph{Idea.} The top type is held at $C=1$, so the update of type $N-1$ involves only its own action and the constant $C$. Hence the top slope
$u_k$ follows a scalar recursion in which the error enters with the factor $N$. An error of fixed sign larger than the restoring
reaction pushes $u_k$ out of every window.

Here $u_k=N(1-\beta^k_{N-1})$. Since $A_{N-1}=1/N$ and $\beta^k_{N-1}-\mu_{N-1}=(1-u_k)/N$, the field of
type $N-1$ is $(2u_k-1)/(Nu_k)$ whenever $u_k\ne0$, and
\[
u_{k+1}=u_k-h\Bigl(\frac{2u_k-1}{u_k}+N\xi^k_{N-1}\Bigr).
\]
The reaction $(2u-1)/u=2-1/u$ is increasing in $u>0$.
\begin{itemize}
\item With $\xi^k_{N-1}=-\eta$: as long as $u_0\le u_k\le\bar u$, the increment $u_{k+1}-u_k$ is at least $h(N\eta-(2\bar u-1)/\bar u)>0$. A
sequence that grows by a fixed positive amount while it stays in $[u_0,\bar u]$ exceeds $\bar u$ after finitely many steps.
\item With $\xi^k_{N-1}=\eta$: as long as $u_k\ge\underline u$, the decrement $u_k-u_{k+1}$ is at least $h(N\eta-(1-2\underline u)/\underline u)>0$, because
$(1-2u)/u$ is decreasing in $u>0$. So $u_k$ falls below $\underline u$ after finitely many steps.
\end{itemize}

(b) If $\varphi(c_i,C)=0$, the pooled strategy has $\beta_i+h\varphi_i(\beta_i)(G_i+\xi)=C$ for every error, and the projection keeps
it. By \eqref{eq:gridrest}, $\hat\beta_i<C$ whenever $\mu_i+\ell A_i<C$: both terms of the maximum are then below $C$, the first because
$\hat\beta_{i+1}\le C$ and $\mu_i<C$. So the constant iterate $C$ does not converge to $\hat\beta_i$. If $\varphi(c_i,\mu_i)=0$, the
anchor strategy is fixed alike, because then $\beta_i+h\varphi_i(\beta_i)(G_i+\xi)=\mu_i$; and
$\hat\beta_i\ge\min\{\mu_i+\ell A_i,C\}>\mu_i$ whenever $\mu_i<C$, since $A_i>0$.
\end{proof}

\begin{proof}[Proof of \cref{cor:learn}]
\emph{Idea.} By the triangle inequality,
\[
|\beta^k_i-\bstar(c_i)|\le|\beta^k_i-\hat\beta_i|+|\hat\beta_i-\check\beta_i|+|\check\beta_i-\bstar(c_i)|.
\]
The first bound of \cref{thm:grid}(b) splits the first term into three parts, the cutoff $\bar w$, a transient and an error term;
\cref{thm:grid}(a) bounds the second term, the floor bias, and \cref{thm:grid}(c) the third, the consistency error. Each constant is
chosen to make one of these five parts at most $\varepsilon/5$.

Both $\beta^k_i$ and $\bstar(c_i)$ lie in $[\mu_i,C]$, so the claim holds if $R\le\varepsilon$; let $R>\varepsilon$. The constants are chosen as follows.
\begin{itemize}
\item \emph{Cutoff:} choose $\bar c<1$ with $\bar w=w(\bar c)\le\varepsilon/5$, which is possible because $\mu$ is continuous at $1$, and let
$a>0$ be the minimum of $A$ on $[0,\bar c]$. If $w_i>\bar w$, then $\mu_i<\mu(\bar c)$, so $c_i<\bar c$ and $A_i\ge a$. So the first bound
of \cref{thm:grid}(b) applies with these $\bar w$ and $a$.
\item \emph{Floor:} $\ell=\min\{R/a,\varepsilon/(5\norm A)\}$. The first entry gives the side condition $\ell a\le R$ of that bound, and the
second gives the floor bias $|\hat\beta_i-\check\beta_i|\le\ell\norm A\le\varepsilon/5$ by \cref{thm:grid}(a).
\item \emph{Error bound:} $\eta_0=\varphi_-a\varepsilon/(5R\varphi_+)$, so that the error term $R\varphi_+\eta_0/(\varphi_-a)$ equals $\varepsilon/5$.
\item \emph{Time:} $T=(R/(\varphi_-a))\log(5R/\varepsilon)$. Since $0\le h\varphi_-a/R\le1$ by the CFL condition and $\ell a\le R$, we have
$(1-h\varphi_-a/R)^k\le\e^{-kh\varphi_-a/R}\le\varepsilon/(5R)$ for $kh\ge T$, so the transient $(1-h\varphi_-a/R)^kR$ is at most $\varepsilon/5$.
\item \emph{Mesh:} $\delta$ from \cref{thm:grid}(c) for $\varepsilon/5$, so that $|\check\beta_i-\bstar(c_i)|\le\varepsilon/5$.
\end{itemize}
Hence $E_k\le3\varepsilon/5$ for $kh\ge T$, and $|\beta^k_i-\bstar(c_i)|\le3\varepsilon/5+\varepsilon/5+\varepsilon/5=\varepsilon$.
\end{proof}
\subsection{Proof of \cref{thm:scheme}}

\begin{proof}
\emph{Idea.} If $b$ is weakly increasing, the claim is \cref{lem:scheme-feasible}. Otherwise the iterates start from $b$,
while the claimed limit is the solution from the lower envelope $b^\flat\le b$. The lower bound is a comparison: the
iterates from $b^\flat$ lie below those from $b$ and converge to $\beta$. For the upper bound we show that the scheme
removes the excess $b-b^\flat$ in an initial layer of vanishing length. Two estimates for the iterates do this.
\begin{itemize}
\item The \emph{suffix barrier} (Step~3) bounds every action by the initial action of every higher type, up to a drift
of order $kh$, a term of order $\ell$, and a term that decays like $(1-h\varphi_-/\ell)^k$. The mechanism is the floor. A
type whose action is not below that of the next type sees the floored difference $\ell\Delta_i$, so the damping term of
$G_i$ has size $(\beta_i-\mu_i)/\ell$ and pulls the action down at rate $\varphi_-/\ell$, which is large when $\ell$ is
small. After a short time $t_0\gg\ell$ the iterates therefore lie at most a few $\varepsilon$ above $b^\flat$.
\item The \emph{shift estimate} (Step~4) shows that an excess $\delta$ of one initial vector over another grows by at
most the factor $\e^{L_\varphi(\norm A+1)T}$ up to time $T$.
\end{itemize}
Restarting at time about $t_0$ and comparing with the iterates from $b^\flat$, which converge to $\beta$ by
\cref{lem:scheme-feasible}, gives the upper bound; the time lag $t_0$ is absorbed by the continuity of $\beta$.

\emph{Step 1: monotone data and the envelope.} If $b$ is weakly increasing, then $b^\flat=b$ is feasible, and the claim,
including $T_0=0$, is \cref{lem:scheme-feasible}. In general, $b^\flat(c)=\min_{s\in[c,1]}b(s)$ is attained, because $b$ is
continuous on the compact interval $[c,1]$. It is weakly increasing, because the interval shrinks as $c$ increases, and it
is continuous, with the modulus of continuity of $b$. Moreover $b^\flat\le b\le C$, and $b^\flat(c)\ge\min_{s\ge c}\mu(s)=\mu(c)$
because $\mu$ is weakly increasing. So $b^\flat$ is feasible, and the solution $\beta$ from $b^\flat$ exists and is unique by
\cref{thm:main}(i). The grid vectors $b(c^N_i)$ and $b^\flat(c^N_i)$ are feasible grid strategies, since $\mu(1)=C$ forces
$b(1)=b^\flat(1)=C$.

\emph{Step 2: lower bound.} Let $\beta^{\flat,N,k}$ denote the iterates of \cref{alg:grid} from $b^\flat(c^N_i)$ with the
same errors $\xi^k_i$. By \cref{lem:monotone-update}, applied in every round to the feasible vectors
$\beta^{\flat,N,k}\le\beta^{N,k}$, the order $b^\flat(c^N_i)\le b(c^N_i)$ of the initial vectors persists:
$\beta^{\flat,N,k}\le\beta^{N,k}$ for all $k$. By \cref{lem:scheme-feasible}, $\beta^{\flat,N,k}_i$ converges to
$\beta(kh_N,c^N_i)$ uniformly on the grid points of $\cB_T$. Hence, for every $\varepsilon>0$ and all large $N$,
$\beta^{N,k}_i\ge\beta(kh_N,c^N_i)-\varepsilon$ whenever $0\le kh_N\le T$. This lower bound holds down to $t=0$.

For the upper bound we use two estimates, proved in Steps~3 and~4 for an arbitrary grid. There $\beta^k$ and
$\tilde\beta^k$ are iterates of \cref{alg:grid} from feasible grid strategies $\beta^0$ and $\tilde\beta^0$, which need not
be increasing, with the same errors $|\xi^k_i|\le\eta$ and a step that satisfies \eqref{eq:cfl}. Since
$\hat\varphi=\varphi_++L_\varphi R\ge\varphi_+$, the CFL condition gives $h\varphi_-\le h\varphi_+\le h\hat\varphi\le\ell$, so
\[
p=h\varphi_-/\ell\in[0,1].
\]

\emph{Step 3: suffix barrier.} We show that, for $i\le j\le N$ and every round $k$,
\begin{equation}\label{eq:D2-suffix}
\beta^k_i
\le
\bar b^k_j
=
\beta^0_j
+
D_k
+
2\ell(\norm A+\eta)
+
(1-p)^kR,
\qquad
D_k=kh\varphi_+(\norm A+\eta).
\end{equation}
The four terms have the following roles. The initial action $\beta^0_j$ of the higher type $j$ is the target. The drift
$D_k$ bounds how far type $j$ itself can rise in $k$ rounds. The initial excess $\beta^0_i-\beta^0_j$ is at most $R$, and the
floor shrinks it by the factor $1-p$ per round; this gives $(1-p)^kR$. The term $2\ell(\norm A+\eta)$ is the price of the
floor, as the case analysis below shows.

\emph{Drift.} One round raises an action by at most $h\varphi_+(\norm A+\eta)$:
\begin{equation}\label{eq:D2-drift}
\beta^{k+1}_i\le\beta^k_i+h\varphi_+(\norm A+\eta).
\end{equation}
Indeed, the damping term of $G_i$ is nonnegative, so $G_i(\beta^k)+\xi^k_i\le A_i+\eta\le\norm A+\eta$, and the increment
$h\varphi_i(\beta^k_i)(G_i(\beta^k)+\xi^k_i)$ is at most $h\varphi_+(\norm A+\eta)$ (it is nonpositive if the bracket is
negative). The projection keeps the bound, because $\mathrm{pr}_i(x)\le\max\{\mu_i,x\}$ and $\mu_i\le\beta^k_i$. For $i=N$ the
action stays at $C$. Summing \eqref{eq:D2-drift} over the rounds gives $\beta^k_j\le\beta^0_j+D_k$.

\emph{Induction on $k$.} For $k=0$, $\beta^0_i-\beta^0_j\le C-\mu_j\le C-\mu(0)=R$. Suppose \eqref{eq:D2-suffix} holds for $k$
and all $i\le j$. For $i=j$ the bound for $k+1$ is the drift bound $\beta^{k+1}_j\le\beta^0_j+D_{k+1}$. Let $i<j$, so
$i<N$, and put
\[
X=\min\{C,\bar b^k_j\}.
\]
By the induction hypothesis, applied to $i$ and to $i+1\le j$, and by feasibility, $\beta^k_i\le X$ and
$\beta^k_{i+1}\le X$. We compare with the flat pair in which both the own action and the next action equal $X$:
\cref{lem:monotone-update} with $z_1=\beta^k_i$, $z_1^+=\beta^k_{i+1}$ and $z_2=z_2^+=X$ bounds the update of type $i$
before projection by its value at this pair. At a flat pair the difference is $0$, so the floor is active and the
damping term is $(X-\mu_i)/\ell$:
\begin{equation}\label{eq:D2-flat}
\beta^k_i+h\varphi_i(\beta^k_i)\bigl(G_i(\beta^k)+\xi^k_i\bigr)
\le
V,
\qquad
V=X+h\varphi_i(X)\Bigl(A_i+\xi^k_i-\frac{X-\mu_i}{\ell}\Bigr).
\end{equation}
This comparison is the point of the barrier: it replaces the unknown neighbor by the worst case for the floor, which is
also the case in which the floor acts most strongly. Note $X\ge\beta^k_i\ge\mu_i$. Two cases remain.
\begin{itemize}
\item \emph{$X-\mu_i\ge\ell(\norm A+\eta)$: the floor term dominates.} The bracket in $V$ is at most
$\norm A+\eta-(X-\mu_i)/\ell\le0$, and $\varphi_i(X)\ge\varphi_-$, so
\[
V\le X+h\varphi_-\Bigl(\norm A+\eta-\frac{X-\mu_i}\ell\Bigr)=\mu_i+(1-p)(X-\mu_i)+h\varphi_-(\norm A+\eta).
\]
Now $X\le\bar b^k_j$ and $\bar b^k_j-\mu_i=(\beta^0_j-\mu_i)+D_k+2\ell(\norm A+\eta)+(1-p)^kR$, where
$\beta^0_j-\mu_i\ge\mu_j-\mu_i\ge0$. Multiplying by $1-p\in[0,1]$ does not increase the first two terms, and
$(1-p)\,2\ell(\norm A+\eta)=2\ell(\norm A+\eta)-2h\varphi_-(\norm A+\eta)$ because $p\ell=h\varphi_-$. Hence
\[
V\le\beta^0_j+D_k+2\ell(\norm A+\eta)-h\varphi_-(\norm A+\eta)+(1-p)^{k+1}R\le\bar b^{k+1}_j .
\]
\item \emph{$X-\mu_i<\ell(\norm A+\eta)$: the action is already close to the anchor.} The bracket in $V$ is at most
$\norm A+\eta$, so the increment in $V$ is at most $h\varphi_+(\norm A+\eta)\le\ell(\norm A+\eta)$, and
\[
V<\mu_i+2\ell(\norm A+\eta)\le\beta^0_j+2\ell(\norm A+\eta)\le\bar b^{k+1}_j .
\]
\end{itemize}
The second case explains the factor $2$: one $\ell(\norm A+\eta)$ is the threshold below which the floor term need not
dominate the drift, and the other is the largest possible step. The projection keeps the bound, because
$\beta^{k+1}_i\le\max\{\mu_i,V\}$ and $\mu_i\le\beta^0_j\le\bar b^{k+1}_j$. This completes the induction.

\emph{Step 4: shift estimate.} We show that if $\beta^0\le\tilde\beta^0+\delta$ with $\delta\ge0$, then
\begin{equation}\label{eq:D2-shift}
\beta^k\le\tilde\beta^k+\delta\bigl(1+hL_\varphi(\norm A+\eta)\bigr)^k\qquad\text{for all }k.
\end{equation}
Since $\beta^k$ and $\tilde\beta^k$ are again feasible, it suffices to treat one round. The profile
$Y=\min\{C,\tilde\beta^0+\delta\}$ is feasible and $\beta^0\le Y$, so by \cref{lem:monotone-update} the update of $\beta^0$
is at most that of $Y$. It remains to compare the updates of $Y$ and $\tilde\beta^0$ at a type $i<N$ (at $i=N$ both are
$C$). Write $x=\tilde\beta^0_i$ and $x'=Y_i\in[x,x+\delta]$, and write the update before projection as
\[
z+h\varphi_i(z)(A_i+\xi^0_i)-h\Delta_i\,\frac{\varphi_i(z)(z-\mu_i)}{M},
\]
with $M$ the floored difference.
\begin{itemize}
\item \emph{Floored differences.} If $Y_i=x+\delta$, then $Y_{i+1}-Y_i\le\tilde\beta^0_{i+1}-\tilde\beta^0_i$, since
$Y_{i+1}\le\tilde\beta^0_{i+1}+\delta$; if $Y_i=C$, then $Y_{i+1}-Y_i\le0$. In both cases
$M'=\max\{Y_{i+1}-Y_i,\ell\Delta_i\}\le M=\max\{\tilde\beta^0_{i+1}-\tilde\beta^0_i,\ell\Delta_i\}$.
\item \emph{Damping.} By \condref{G2}, $\varphi_i(x')(x'-\mu_i)\ge\varphi_i(x)(x-\mu_i)\ge0$. With $M'\le M$, the damping term of $Y$
is at least that of $\tilde\beta^0$.
\item \emph{Weight.} By \condref{G1}, $(A_i+\xi^0_i)(\varphi_i(x')-\varphi_i(x))\le(\norm A+\eta)L_\varphi\delta$.
\end{itemize}
Hence the update of $Y$ before projection exceeds that of $\tilde\beta^0$ by at most
$\delta+hL_\varphi(\norm A+\eta)\delta$. The projection is weakly increasing and satisfies
$\mathrm{pr}_i(x+t)\le\mathrm{pr}_i(x)+t$ for $t\ge0$, so it keeps this bound.

\emph{Step 5: conclusion.} Fix $0<T_0\le T$ and $\varepsilon>0$, let $N$ be so large that $\eta_N\le1$, and put
$\Lambda=L_\varphi(\norm A+1)$. We choose the length $t_0$ of the initial layer with three properties: $t_0\le T_0$, so
that every round in the claim lies after the layer; $t_0\varphi_+(\norm A+1)\le\varepsilon$, so that the drift in
\eqref{eq:D2-suffix} is small; and
\[
|\beta(t,c)-\beta(t',c)|\le\varepsilon\qquad\text{whenever }t,t'\in[0,T],\ |t-t'|\le t_0,
\]
which is possible because $\beta$ is uniformly continuous on the compact set $\cB_T$. Let $k_N=\lfloor t_0/h_N\rfloor$.
The CFL condition gives $h_N\le\ell_N/\varphi_+\to0$, so $t_0/2\le k_Nh_N\le t_0$ for large $N$.

\emph{Phase 1: after $k_N$ rounds the iterates are within $4\varepsilon$ above $b^\flat$.} Fix $i$ and let $s\in[c^N_i,1]$
attain $b^\flat(c^N_i)=b(s)$. The largest index $j$ with $c^N_j\le s$ satisfies $j\ge i$ and $|c^N_j-s|\le$ mesh, so, by the
uniform continuity of $b$, $b(c^N_j)\le b(s)+\varepsilon$ once the mesh is small. The suffix barrier
\eqref{eq:D2-suffix} for this $j$ and $k=k_N$ gives
\begin{equation}\label{eq:D2-phase1}
\beta^{N,k_N}_i
\le
b^\flat(c^N_i)+\varepsilon
+t_0\varphi_+(\norm A+1)
+2\ell_N(\norm A+1)
+\e^{-k_Nh_N\varphi_-/\ell_N}R
\le
b^\flat(c^N_i)+4\varepsilon
\end{equation}
for large $N$. Here we used $(1-p)^{k_N}\le\e^{-pk_N}$, and the last two terms are at most $\varepsilon$ each for large $N$
because $\ell_N\to0$ and $k_Nh_N\ge t_0/2$. The floor must vanish for this step: the relaxation time $\ell_N/\varphi_-$ has to
be short compared with $t_0$.

\emph{Phase 2: restart.} Let $\tilde\beta^{N,k}$ be the iterates from $b^\flat(c^N_i)$ with the shifted errors
$\xi^{k_N+k}$, which also satisfy $|\xi^{k_N+k}_i|\le\eta_N$. The iterates $\beta^{N,k_N+k}$ are the iterates from
$\beta^{N,k_N}$ with the same shifted errors. By \eqref{eq:D2-phase1} and the shift estimate \eqref{eq:D2-shift} with
$\delta=4\varepsilon$, using $(1+h_NL_\varphi(\norm A+\eta_N))^k\le\e^{\Lambda kh_N}$,
\[
\beta^{N,k_N+k}_i\le\tilde\beta^{N,k}_i+4\varepsilon\e^{\Lambda T}\qquad(kh_N\le T).
\]
\Cref{lem:scheme-feasible}, applied to the feasible datum $b^\flat$ and the shifted errors, gives
$\tilde\beta^{N,k}_i\le\beta(kh_N,c^N_i)+\varepsilon$ for $kh_N\le T$ and large $N$; the limit is $\beta$ because the solution
from $b^\flat$ is unique. Finally, $\beta(kh_N,c^N_i)\le\beta((k_N+k)h_N,c^N_i)+\varepsilon$ when $(k_N+k)h_N\le T$, by the choice
of $t_0$ and $k_Nh_N\le t_0$.

Every round $k'$ with $T_0\le k'h_N\le T$ satisfies $k'\ge k_N$, because $k_Nh_N\le t_0\le T_0$. Writing $k'=k_N+k$ and
combining the three bounds,
\[
\beta^{N,k'}_i
\le
\beta(k'h_N,c^N_i)
+
2\varepsilon
+
4\varepsilon\e^{\Lambda T}
\]
for large $N$. Since $\varepsilon$ is arbitrary, this proves the upper bound. Together with the lower bound of Step~2, it
proves the claim.
\end{proof}

\paragraph{Nonmonotonicity is distinct from pooling.}
Consider the two-bidder first-price procurement auction with uniform costs, $A=1-c$, $B=1$, $\mu=c$ and $C=1$, and the
common profile $b(c)=\max\{c,1-c\}$. It is continuous with $c\le b(c)\le1$, so \cref{thm:scheme} applies. It decreases on
$[0,\frac12)$, but it pools no interval of types: each bid is used by at most two types. For a bid $z\in(\frac12,1)$ the rival
bids above $z$ exactly when its cost lies in $[0,1-z)$ or in $(z,1]$, so the winning probability is $2(1-z)$ and
\[
\pi(z;b,c)=2(z-c)(1-z),\qquad
\fraction{\partial \pi}{\partial z}(z;b,c)=2(1-2z+c).
\]
At $c=\frac14$ and the current bid $z=b(\frac14)=\frac34$, this derivative is $-\frac12$, a finite number. The lower envelope
is $b^\flat(c)=\frac12$ for $c\le\frac12$ and $b^\flat(c)=c$ for $c\ge\frac12$. By \cref{thm:scheme}, the iterates converge to the
solution $\beta$ from $b^\flat$ at every positive time, and $\beta(t,\frac14)\to b^\flat(\frac14)=\frac12$ as $t\downarrow0$.
So in the limit the action of type $\frac14$ drops from $\frac34$ to $\frac12$ instantly, although its payoff derivative is
finite. The drop comes from the numerical extension of \cref{alg:grid} to nonmonotone vectors (the suffix barrier in the
proof of \cref{thm:scheme}), not from a singularity of the payoff as under pooling. The example thus separates the actual
payoff derivative from the numerical extension in \cref{alg:grid} and the lower-envelope initialization in
\cref{thm:scheme}.

\subsection{Proof of \cref{thm:stochastic}}

\begin{proof}
\emph{(a)} \emph{Idea.} The errors are not assumed to be small; only the weighted series converge. So
\cref{thm:grid}(b), whose bounds contain the error bound $\eta$, does not apply. We use instead that the error enters
the update additively, and we absorb it into the tail $Z^k$ of the series: the shifted iterates $\beta^k+Z^k$ obey the
error-free update up to perturbations of order $h_k\|Z^k\|_1$, because the difference $Z^k-Z^{k+1}$ cancels the error of
round $k$ exactly. The shifted iterates are then compared with the clipped barriers of the proof of
\cref{thm:grid}(b), as there. This gives a scalar recursion for an error envelope $e_k$, in which the only term not
multiplied by $h_k$ is the size of the current tail, which tends to zero.

\emph{Step 1: additive form and the tail.} Let $\tilde\xi^k_i=\varphi_i(\beta^k_i)\,\xi^k_i$ be the effective error along the
trajectory. Then
\[
\beta^{k+1}_i=\mathrm{pr}_i\bigl(x^k_i\bigr),\qquad
x^k_i=\beta^k_i+h_k\varphi_i(\beta^k_i)G_i(\beta^k)+h_k\tilde\xi^k_i .
\]
The error-free part is monotone: by \cref{lem:monotone-update} with $\eta=0$ and $\xi=0$, the map
$\beta\mapsto\beta_i+h_k\varphi_i(\beta_i)G_i(\beta)$ is weakly increasing on feasible strategies, in $\beta_i$ and in
$\beta_{i+1}$; the CFL condition for $\eta=0$ is exactly what this needs. Let
\[
Z^k_i=\sum_{j\ge k}h_j\tilde\xi^j_i\quad(i<N),\qquad Z^k_N=0,\qquad\|Z^k\|_1=\sum_{i<N}|Z^k_i| .
\]
The series converge by assumption, so $Z^k_i\to0$ and $\|Z^k\|_1\to0$; we only use $|Z^k_i|\le\|Z^k\|_1$. The tail satisfies
$Z^k_i=h_k\tilde\xi^k_i+Z^{k+1}_i$, so the error cancels in the shifted pre-projection value:
\begin{equation}\label{eq:D2-cancel}
x^k_i+Z^{k+1}_i=\beta^k_i+Z^k_i+h_k\varphi_i(\beta^k_i)G_i(\beta^k).
\end{equation}

\emph{Step 2: constants and the envelope.} Let $a>0$ be the minimum of $1/\ell$ and of the ratios $A_i/w_i$ over $i<N$ with
$w_i>0$. Then $\ell a\le1$ and $aw_i\le A_i$ for all $i<N$, so $\gamma(E)=aE$ satisfies the two conditions on $\gamma$ in
the proof of \cref{thm:grid}(b) for all $E\ge0$. Since $h_k\varphi_-\le\ell$, also $h_k\varphi_-a\le1$. On feasible strategies
the field $\beta\mapsto\varphi_i(\beta_i)G_i(\beta)$ is Lipschitz: if $\beta$ and $\beta'$ are feasible and
$|\beta_j-\beta'_j|\le\delta$ for $j=i,i+1$, then
\begin{equation}\label{eq:D2-lip}
\bigl|\varphi_i(\beta_i)G_i(\beta)-\varphi_i(\beta'_i)G_i(\beta')\bigr|\le K\delta,\qquad
K=\max_{i<N}\Bigl(\varphi_+\Bigl(\frac1\ell+\frac{2R}{\ell^2\Delta_i}\Bigr)+L_\varphi\Bigl(A_i+\frac R\ell\Bigr)\Bigr).
\end{equation}
Indeed, the floored differences $D,D'$ are at least $\ell\Delta_i$ and differ by at most $2\delta$, and $0\le\beta_i-\mu_i\le R$,
so $|G_i(\beta)-G_i(\beta')|\le\delta/\ell+2R\delta/(\ell^2\Delta_i)$; moreover $|G_i(\beta')|\le A_i+R/\ell$, $\varphi_i\le\varphi_+$,
and $\varphi_i$ is $L_\varphi$-Lipschitz by \condref{G1}. Let
\[
\Theta=K+\varphi_-a+\varphi_+/\ell .
\]
The three terms are the costs of the three comparisons below: $K$ for replacing the shifted barrier by the deterministic
one, $\varphi_+/\ell$ for the shift at the cap, and $\varphi_-a$ for the shift at the anchor. Define $e_0=R+\|Z^0\|_1$ and
\begin{equation}\label{eq:D2-envelope}
e_{k+1}=\max\bigl\{\|Z^{k+1}\|_1,\ (1-h_k\varphi_-a)e_k+h_k\Theta\|Z^k\|_1\bigr\}.
\end{equation}
We show by induction that
\begin{equation}\label{eq:D2-invariant}
|\beta^k_i+Z^k_i-\hat\beta_i|\le e_k\qquad\text{for all }k\text{ and }0\le i\le N.
\end{equation}
Then $|\beta^k_i-\hat\beta_i|\le e_k+\|Z^k\|_1$, and it remains to show $e_k\to0$ (Step~6).

For $k=0$, both $\beta^0_i$ and $\hat\beta_i$ lie in $[\mu_i,C]$, so $|\beta^0_i+Z^0_i-\hat\beta_i|\le R+\|Z^0\|_1$. For $i=N$ the
claim holds in every round, because $\beta^k_N=\hat\beta_N=C$ and $Z^k_N=0$. Fix a round $k$, assume
\eqref{eq:D2-invariant} for $k$, fix $i<N$, and write $h=h_k$, $E=e_k$, $Z=Z^k$ and $\varepsilon=\|Z^k\|_1$. Recall from the
proof of \cref{thm:grid}(b): below the cap, $\hat\beta_i<C$, we have $A_i\hat d_i=\Delta_i(\hat\beta_i-\mu_i)$ and
$\hat\beta_i-\mu_i\ge\ell A_i$; at the cap, $\hat\beta_i=C$, we have $w_i\le\ell A_i$. We also use that
$\varphi_-\le\varphi_i\le\varphi_+$ on feasible actions, so $G\le-x+y$ with $x,y\ge0$ implies
$\varphi_iG\le-\varphi_-x+\varphi_+y$, and $G\ge x\ge0$ implies $\varphi_iG\ge\varphi_-x$.

\emph{Step 3: shifted upper bound.} We show that, if $\hat\beta_i<C$,
\begin{equation}\label{eq:D2-upper}
\beta^k_i+Z_i+h\varphi_i(\beta^k_i)G_i(\beta^k)\le\hat\beta_i+(1-h\varphi_-a)E+h(K+\varphi_+/\ell)\varepsilon .
\end{equation}
Two barriers are involved. The \emph{shifted barrier} $\tilde U^+_j=\min\{C,\hat\beta_j+E-Z_j\}$ lies above $\beta^k$, by
\eqref{eq:D2-invariant} and $\beta^k_j\le C$, so it is feasible and monotonicity applies to it; but it carries the random
shift $Z$. The \emph{deterministic barrier} $U^+_j=\min\{C,\hat\beta_j+E\}$ of the proof of \cref{thm:grid}(b) has a known
field, and $|U^+_j-\tilde U^+_j|\le|Z_j|\le\varepsilon$. Monotonicity and \eqref{eq:D2-lip} give
\[
\beta^k_i+h\varphi_i(\beta^k_i)G_i(\beta^k)
\le\tilde U^+_i+h\varphi_i(\tilde U^+_i)G_i(\tilde U^+)
\le\tilde U^+_i+h\varphi_i(U^+_i)G_i(U^+)+hK\varepsilon .
\]
It remains to bound $\tilde U^+_i+Z_i+h\varphi_i(U^+_i)G_i(U^+)$ by $\hat\beta_i+(1-h\varphi_-a)E+h\varphi_+\varepsilon/\ell$.
Note $\tilde U^+_i+Z_i\le\hat\beta_i+E$.
\begin{itemize}
\item \emph{$\hat\beta_i+E\le C$.} Then $U^+_i=\hat\beta_i+E$ and $U^+_{i+1}-U^+_i\le\hat\beta_{i+1}-\hat\beta_i$, so the floored
difference at $U^+$ is at most $\hat d_i$, and
\[
G_i(U^+)\le A_i-\frac{\Delta_i(\hat\beta_i+E-\mu_i)}{\hat d_i}=-\frac{\Delta_iE}{\hat d_i}=-\frac{A_iE}{\hat\beta_i-\mu_i}\le-\frac{A_iE}{w_i}\le-aE .
\]
Hence $\varphi_i(U^+_i)G_i(U^+)\le-\varphi_-aE$, and the sum is at most $\hat\beta_i+E-h\varphi_-aE$.
\item \emph{$\hat\beta_i+E>C$.} Then $U^+_i=C\ge U^+_{i+1}$, the floor is active, and
$G_i(U^+)=A_i-w_i/\ell=A_i-(\hat\beta_i-\mu_i)/\ell-(C-\hat\beta_i)/\ell\le-(C-\hat\beta_i)/\ell$. Here the restoring force
is proportional to $C-\hat\beta_i$, not to $E$, so we measure the value part by
$F=\min\{C-\hat\beta_i+\varepsilon,E\}$: since $\tilde U^+_i\le C$ and $Z_i\le\varepsilon$, we have $\tilde U^+_i+Z_i\le\hat\beta_i+F$.
As $F-\varepsilon\le C-\hat\beta_i$ and $a\ell\le1$,
\[
G_i(U^+)\le-\frac{F-\varepsilon}\ell\le-aF+\frac\varepsilon\ell,
\qquad
\varphi_i(U^+_i)G_i(U^+)\le-\varphi_-aF+\varphi_+\frac\varepsilon\ell .
\]
The sum is at most $\hat\beta_i+(1-h\varphi_-a)F+h\varphi_+\varepsilon/\ell\le\hat\beta_i+(1-h\varphi_-a)E+h\varphi_+\varepsilon/\ell$,
because $F\le E$ and $1-h\varphi_-a\ge0$.
\end{itemize}
Adding $Z_i$ to the chain above gives \eqref{eq:D2-upper}.

\emph{Step 4: shifted lower bound.} We show that
\begin{equation}\label{eq:D2-lower}
\beta^k_i+Z_i+h\varphi_i(\beta^k_i)G_i(\beta^k)\ge\hat\beta_i-(1-h\varphi_-a)E-h(K+\varphi_-a)\varepsilon .
\end{equation}
Now $\tilde U^-_j=\max\{\mu_j,\hat\beta_j-E-Z_j\}\le\beta^k_j$ is feasible, $U^-_j=\max\{\mu_j,\hat\beta_j-E\}$ is the
deterministic lower barrier, and $|U^-_j-\tilde U^-_j|\le\varepsilon$. As in Step~3,
\[
\beta^k_i+h\varphi_i(\beta^k_i)G_i(\beta^k)\ge\tilde U^-_i+h\varphi_i(U^-_i)G_i(U^-)-hK\varepsilon,
\]
and the value part satisfies $\tilde U^-_i+Z_i\ge\hat\beta_i-E$ and $\tilde U^-_i+Z_i\ge\mu_i-\varepsilon$.
\begin{itemize}
\item \emph{$\hat\beta_i-E\ge\mu_i$.} Then $U^-_i=\hat\beta_i-E$ and $G_i(U^-)\ge aE$: below the cap, $U^-_{i+1}-U^-_i\ge\hat\beta_{i+1}-\hat\beta_i$
and the computation of the first case of Step~3 with reversed inequalities gives $G_i(U^-)\ge A_iE/w_i\ge aE$; at the cap,
the floor bounds the denominator below, so $G_i(U^-)\ge A_i-(U^-_i-\mu_i)/\ell=A_i-(w_i-E)/\ell\ge E/\ell\ge aE$. Hence the
sum is at least $\hat\beta_i-E+h\varphi_-aE$.
\item \emph{$\hat\beta_i-E<\mu_i$.} Then $U^-_i=\mu_i$ sits on the anchor, where $G_i(U^-)=A_i\ge aw_i\ge aD$ with
$D=\hat\beta_i-\mu_i\in[0,E)$; so $\varphi_i(U^-_i)G_i(U^-)\ge\varphi_-aD$. If $D+\varepsilon\le E$, we use
$\tilde U^-_i+Z_i\ge\mu_i-\varepsilon=\hat\beta_i-(D+\varepsilon)$, and the sum is at least
$\hat\beta_i-(1-h\varphi_-a)(D+\varepsilon)-h\varphi_-a\varepsilon\ge\hat\beta_i-(1-h\varphi_-a)E-h\varphi_-a\varepsilon$. If
$D+\varepsilon>E$, we use $\tilde U^-_i+Z_i\ge\hat\beta_i-E$ and $D>E-\varepsilon$, and the sum is at least
$\hat\beta_i-E+h\varphi_-a(E-\varepsilon)$. In both cases the shift costs at most $h\varphi_-a\varepsilon$.
\end{itemize}
With the Lipschitz error $hK\varepsilon$ this gives \eqref{eq:D2-lower}.

\emph{Step 5: projection and induction.} Let $\Psi=(1-h\varphi_-a)E+h\Theta\varepsilon$, so that $e_{k+1}=\max\{\|Z^{k+1}\|_1,\Psi\}$.
Since $K+\varphi_+/\ell\le\Theta$ and $K+\varphi_-a\le\Theta$, the right-hand side of \eqref{eq:D2-upper} is at most
$\hat\beta_i+\Psi$ and that of \eqref{eq:D2-lower} is at least $\hat\beta_i-\Psi$; by \eqref{eq:D2-cancel}, both inequalities bound
$x^k_i+Z^{k+1}_i$. The projection either keeps $x^k_i$ or moves it to a bound on the side of
$\hat\beta_i$:
\begin{itemize}
\item \emph{Upper bound.} If $\hat\beta_i=C$, then $\beta^{k+1}_i\le C=\hat\beta_i$. If $\hat\beta_i<C$ and $x^k_i\le\mu_i$, then
$\beta^{k+1}_i=\mu_i\le\hat\beta_i$. In both cases $\beta^{k+1}_i+Z^{k+1}_i-\hat\beta_i\le Z^{k+1}_i\le\|Z^{k+1}\|_1$. Otherwise
$\beta^{k+1}_i\le x^k_i$ and $\beta^{k+1}_i+Z^{k+1}_i-\hat\beta_i\le\Psi$ by \eqref{eq:D2-upper}.
\item \emph{Lower bound.} If $x^k_i\ge C$, then $\beta^{k+1}_i=C\ge\hat\beta_i$ and $\beta^{k+1}_i+Z^{k+1}_i-\hat\beta_i\ge-\|Z^{k+1}\|_1$.
Otherwise $\beta^{k+1}_i\ge x^k_i$ and the difference is at least $-\Psi$ by \eqref{eq:D2-lower}.
\end{itemize}
Hence $|\beta^{k+1}_i+Z^{k+1}_i-\hat\beta_i|\le e_{k+1}$, which completes the induction.

\emph{Step 6: the envelope tends to zero.} Put $\kappa=\varphi_-a>0$, so $0\le h_k\kappa\le1$. Fix $\theta>0$ and choose $k_1$ with
$\|Z^k\|_1<\theta$ and $\Theta\|Z^k\|_1\le\kappa\theta$ for all $k\ge k_1$. Let $k\ge k_1$. If $e_{k+1}>\theta$, the maximum in
\eqref{eq:D2-envelope} is attained by its second entry, so $e_{k+1}\le(1-h_k\kappa)e_k+h_k\kappa\theta$, that is,
$e_{k+1}-\theta\le(1-h_k\kappa)(e_k-\theta)$. Hence $q_k=\max\{e_k-\theta,0\}$ satisfies $q_{k+1}\le(1-h_k\kappa)q_k$, and
\[
q_k\le q_{k_1}\prod_{j=k_1}^{k-1}(1-h_j\kappa)\le q_{k_1}\exp\Bigl(-\kappa\sum_{j=k_1}^{k-1}h_j\Bigr)\to0,
\]
because $\sum_kh_k=\infty$. So $\limsup_ke_k\le\theta$ for every $\theta>0$, and $e_k\to0$. With \eqref{eq:D2-invariant} and
$\|Z^k\|_1\to0$ this gives $\beta^k\to\hat\beta$. The error $\|Z^{k+1}\|_1$ enters \eqref{eq:D2-envelope} through a maximum
and not additively, so it does not accumulate; this is why no rate for $h_k$ is needed.

Finally, if $\sum_kh_k|\xi^k_i|<\infty$, the series of the $h_k\tilde\xi^k_i$ converges absolutely, because the iterates are
feasible and $\varphi_i\le\varphi_+$ there; so the assumption of (a) holds.

\emph{(b)} We verify the assumption of (a) on almost every path. Since $\beta^0$ and $(h_k)$ are deterministic, the
iterate $\beta^k$ is a continuous function of $\xi^0,\dots,\xi^{k-1}$: the field $G_i$ is continuous because the floor
keeps the denominators positive, $\varphi_i$ is continuous on the feasible actions, and the projection is continuous.
Hence $\beta^k$ is $\mathcal F_k$-measurable, and $\varphi_i(\beta^k_i)$ is $\mathcal F_k$-measurable with values in
$[0,\varphi_+]$. So the effective error $\tilde\xi^k_i=\varphi_i(\beta^k_i)\xi^k_i$ is $\mathcal F_{k+1}$-measurable and
\[
\mathbb E[\tilde\xi^k_i\mid\mathcal F_k]=\varphi_i(\beta^k_i)\,\mathbb E[\xi^k_i\mid\mathcal F_k]=0,\qquad
\mathbb E[(\tilde\xi^k_i)^2]\le\varphi_+^2\sigma^2 .
\]
Thus the partial sums $\sum_{j<k}h_j\tilde\xi^j_i$ form a martingale. Its increments are orthogonal, so its second moment
is at most $\varphi_+^2\sigma^2\sum_jh_j^2<\infty$. By the martingale convergence theorem \citep{doob1953stochastic} the series
converges almost surely, and (a) applies to almost every path.

\emph{(c)} The iterates up to round $k_1$ depend only on the errors $\xi^k$ with $k<k_1$. Consider the event that
$|\xi^k_i|\le\eta$ for all $k<k_1$ and $i<N$. On this event, replace the errors from round $k_1$ on by zero. The truncated
errors satisfy $|\xi^k_i|\le\eta$ in every round, so \cref{thm:grid}(b) applies to them, and their iterates agree with the
actual iterates up to round $k_1$. Hence the bounds of \cref{thm:grid}(b) hold for all $k\le k_1$ on this event. By the
union bound over the $Nk_1$ pairs $(k,i)$, its complement has probability at most $Nk_1\varpi$.
\end{proof}

\begin{proof}[Proof of \cref{cor:rounds}]
\emph{Idea.} Part (a) instantiates \cref{thm:grid}: the floor bound $\ell\le1/(2(n-1)f_+)$ removes the floor bias, so
$\hat\beta=\check\beta$; the consistency estimate of \cref{thm:grid}(c) becomes explicit, of first order in the mesh; and the
second rate of \cref{thm:grid}(b) applies with $p=n-1$. Part (b) chooses the mesh, the step and the noise level so that
each of the three error terms is at most $\varepsilon/3$, and counts the rounds.

\emph{(a)} \emph{Step 1: the data.} By the proof of \cref{cor:auction}, $A=S^{n-1}$, $B=(n-1)S^{n-2}f$, $\mu(c)=c$ and $C=1$.
Hence $R=1$, $w_i=1-c_i$, $X=1-A$, $\Delta_i=A_i-A_{i+1}$ and $A_N=S(1)^{n-1}=0$. Since $f$ is a density on $[0,1]$, $f_-\le1\le f_+$.
For $\varphi\equiv1$ we have $\varphi_\pm=1$ and $L_\varphi=0$, so \eqref{eq:cfl} reads $h\le\ell$ and $h\le\ell^2\Delta_i$.

\emph{Step 2: no floor bias.} With $A_N=0$, the condition of \cref{thm:grid}(a) reads
$c_{i+1}-c_i\ge\ell\Delta_i+\ell(A_i-A_{i+1})=2\ell\Delta_i$. It holds because $|\fraction{\mathrm{d} A}{\mathrm{d} c}|=(n-1)S^{n-2}f\le(n-1)f_+$ gives
$\Delta_i\le(n-1)f_+(c_{i+1}-c_i)$, and $2\ell(n-1)f_+\le1$. This is where the bound on $\ell$ enters. So
$\hat\beta=\check\beta$.

\emph{Step 3: accuracy of the rest point.} Let $y_i=\check\beta_i-\bstar(c_i)$, so $y_N=0$. Subtracting
$(A_i+\Delta_i)\bstar(c_i)$ from the recursion $(A_i+\Delta_i)\check\beta_i=A_i\check\beta_{i+1}+\Delta_i\mu_i$ of
\eqref{eq:gridrest} with $\ell=0$ gives
\begin{equation}\label{eq:D2-errrec}
(A_i+\Delta_i)\,y_i=A_i\,y_{i+1}-\tau_i,\qquad
\tau_i=\Delta_i\bigl(\bstar(c_i)-\mu_i\bigr)-A_i\bigl(\bstar(c_{i+1})-\bstar(c_i)\bigr),
\end{equation}
as in the proof of \cref{thm:grid}(c). Thus $y_i$ is a convex combination of $y_{i+1}$ and $-\tau_i/\Delta_i$, with weights
$A_i/(A_i+\Delta_i)$ and $\Delta_i/(A_i+\Delta_i)$, so $|y_i|\le\max\{|y_{i+1}|,|\tau_i|/\Delta_i\}$,
and downward induction from $y_N=0$ gives $|y_i|\le\max_j|\tau_j|/\Delta_j$. We now compute the local error $\tau_i$. The
closed form $\bstar(c)=c+A(c)^{-1}\int_c^1A$ of the proof of \cref{cor:auction} and integration by parts, with $\fraction{\mathrm{d} A}{\mathrm{d} c}=-B$
and $A(1)=0$, give $A(c)\bstar(c)=\int_c^1B\mu$. Taking the difference at $c_i$ and $c_{i+1}$, and using $A_i=A_{i+1}+\Delta_i$
and $\Delta_i=\int_{c_i}^{c_{i+1}}B$,
\[
\tau_i=-\Delta_i\bigl(\bstar(c_{i+1})-\bstar(c_i)\bigr)+\int_{c_i}^{c_{i+1}}B\,(\mu-\mu_i).
\]
Both terms are bounded by slopes times the mesh $\delta$.
\begin{itemize}
\item \emph{Slope of $\bstar$.} By the proof of \cref{cor:auction}, $\fraction{\mathrm{d} \bstar}{\mathrm{d} c}=(n-1)fS^{-n}\int_c^1S^{n-1}$ for $c<1$. Since
$\int_c^1S^{n-1}\le\int_c^1S^{n-1}f/f_-=S(c)^n/(nf_-)$, we get $0\le \fraction{\mathrm{d} \bstar}{\mathrm{d} c}\le(n-1)f_+/(nf_-)\le f_+/f_-$. So the first
term lies in $[-\Delta_i(f_+/f_-)\delta,0]$.
\item \emph{Slope of $\mu$.} On the cell, $0\le\mu-\mu_i\le c_{i+1}-c_i\le\delta\le(f_+/f_-)\delta$. So the second term lies in
$[0,\Delta_i(f_+/f_-)\delta]$.
\end{itemize}
The two terms have opposite signs, so $|\tau_i|\le\Delta_i(f_+/f_-)\delta$, and therefore
$|\hat\beta_i-\bstar(c_i)|=|y_i|\le(f_+/f_-)\delta$.

\emph{Step 4: rates.} We apply the second bound of \cref{thm:grid}(b) with $p=n-1$, $a=f_-^{\,n-1}$ and $R=1$, and check its
side conditions. First, $S(c)=\int_c^1f\ge f_-(1-c)$, so $A_i\ge f_-^{\,n-1}w_i^{\,n-1}$. Second, $\ell\le1/(2(n-1)f_+)\le\frac12$ and
$f_-\le1$ give $\ell aR^{p-1}=\ell f_-^{\,n-1}\le1$. Third, $h\le\ell$ and $f_-\le1$ give $h\varphi_-apR^{p-1}=h(n-1)f_-^{\,n-1}\le(n-1)\ell\le1/(2f_+)\le1$.
For $n=2$ ($p=1$) the bound is $(1-hf_-)^k+\eta/f_-$, and for $n\ge3$ it is the maximum in the claim, since
$(p-1)\frac{\varphi_-a}2R^{p-1}=\frac{n-2}2f_-^{\,n-1}$.

\emph{Step 5: conclusion.} By the triangle inequality,
$\max_i|\beta^k_i-\bstar(c_i)|\le\max_i|\beta^k_i-\hat\beta_i|+\max_i|\hat\beta_i-\bstar(c_i)|$, and Steps~3 and~4 bound the two terms.

\emph{(b)} \emph{Step 1: the hypotheses of (a).} The uniform grid has mesh $\delta=1/N$, and $\ell=1/(2(n-1)f_+)$ is the
largest floor allowed in (a). For the CFL condition we need a lower bound on the cell weights. The map $s\mapsto s^{n-1}$
is superadditive on $[0,\infty)$ for $n\ge2$, and $S(c_i)-S(c_{i+1})=\int_{c_i}^{c_{i+1}}f\ge f_-/N$, so
\[
\Delta_i=S(c_i)^{n-1}-S(c_{i+1})^{n-1}\ge\bigl(S(c_i)-S(c_{i+1})\bigr)^{n-1}\ge(f_-/N)^{n-1}.
\]
Hence $h=\ell^2(f_-/N)^{n-1}\le\ell^2\Delta_i$, and $h\le\ell$ because $\ell\le\frac12$ and $f_-/N\le1$. So (a) applies.

\emph{Step 2: the three error terms.} We split $\varepsilon$ into three equal parts.
\begin{itemize}
\item \emph{Mesh.} The choice $N\ge3f_+/(f_-\varepsilon)$ gives $(f_+/f_-)\delta=f_+/(f_-N)\le\varepsilon/3$.
\item \emph{Noise.} For $n=2$, $\eta\le f_-\varepsilon/3$ gives $\eta/f_-\le\varepsilon/3$. For $n\ge3$, $\eta\le\frac12f_-^{\,n-1}(\varepsilon/3)^{n-1}$
gives $(2\eta/f_-^{\,n-1})^{1/(n-1)}\le\varepsilon/3$.
\item \emph{Transient.} Let $T_\varepsilon$ be the time at which the transient term of (a) reaches $\varepsilon/3$:
$T_\varepsilon=\log(3/\varepsilon)/f_-$ for $n=2$, and $T_\varepsilon=2((3/\varepsilon)^{n-2}-1)/((n-2)f_-^{\,n-1})$ for $n\ge3$. If $kh\ge T_\varepsilon$, then
$(1-hf_-)^k\le\e^{-f_-kh}\le\varepsilon/3$ for $n=2$, and $1+\frac{n-2}2f_-^{\,n-1}kh\ge(3/\varepsilon)^{n-2}$, that is,
$(1+\frac{n-2}2f_-^{\,n-1}kh)^{-1/(n-2)}\le\varepsilon/3$, for $n\ge3$.
\end{itemize}
Let $k_\varepsilon=\lceil T_\varepsilon/h\rceil$. For $k\ge k_\varepsilon$ the bound of (a) is at most
$\varepsilon/3+\varepsilon/3+\varepsilon/3=\varepsilon$ for $n=2$, and at most $\varepsilon/3+\varepsilon/3$ for $n\ge3$, where the noise and transient
terms enter through a maximum.

\emph{Step 3: counting rounds.} We have $k_\varepsilon\le T_\varepsilon/h+1$ and $1/h=\ell^{-2}(N/f_-)^{n-1}$ with $\ell^{-2}=4(n-1)^2f_+^2$.
Since $f_+/(f_-\varepsilon)\ge1$, $N<3f_+/(f_-\varepsilon)+1\le4f_+/(f_-\varepsilon)$, so $N/f_-\le4f_+/(f_-^2\varepsilon)$. For $n=2$,
\[
\frac{T_\varepsilon}h\le\frac{\log(3/\varepsilon)}{f_-}\cdot4f_+^2\cdot\frac{4f_+}{f_-^2\varepsilon}=16\Bigl(\frac{f_+}{f_-}\Bigr)^3\frac{\log(3/\varepsilon)}\varepsilon .
\]
For $n\ge3$, $T_\varepsilon\le2\cdot3^{n-2}\varepsilon^{-(n-2)}/((n-2)f_-^{\,n-1})$, and
\[
\frac{T_\varepsilon}h\le\frac{2\cdot3^{n-2}}{(n-2)f_-^{\,n-1}\varepsilon^{n-2}}\cdot4(n-1)^2f_+^2\cdot\frac{4^{n-1}f_+^{\,n-1}}{f_-^{\,2(n-1)}\varepsilon^{n-1}}
=\frac{2\cdot3^{n-2}4^n(n-1)^2f_+^{\,n+1}}{(n-2)\,f_-^{\,3(n-1)}}\,\varepsilon^{-(2n-3)} .
\]
Adding $1$ gives the bounds on $k_\varepsilon$.
\end{proof}

\end{document}